\documentclass[aps,pra,superscriptaddress,twocolumn]{revtex4-2}
\usepackage{graphicx,amsmath,amssymb,amsfonts,latexsym,color,dcolumn,bm}

\newcommand{\beq}{\begin{equation}}
\newcommand{\eeq}{\end{equation}}
\newcommand{\bea}{\begin{eqnarray}}
\newcommand{\eea}{\end{eqnarray}}

\providecommand{\bra}[1]{\langle #1 \rvert}
\providecommand{\ket}[1]{\lvert #1 \rangle}

\usepackage{amsfonts,amssymb}
\usepackage{dsfont}
\usepackage{physics}
\usepackage{tocvsec2}
\usepackage{stmaryrd} 

\usepackage{appendix}

\definecolor{blue(pigment)}{rgb}{0.2, 0.2, 0.6}

\newcommand{\mohamed}[1]{\textcolor{black}{#1}}
\newcommand{\nicolas}[1]{\textcolor{blue(pigment)}{#1}}

\usepackage{tikz}

\usepackage{soul} 

\usepackage{xcolor}
\usepackage[normalem]{ulem} 

\usepackage{amsmath}
\usepackage{bbold}
\usepackage{hyperref}
\hypersetup{
     colorlinks=true,      
    linkcolor=blue,        
    citecolor=blue,        
    filecolor=magenta, 
    urlcolor=black          
}

\usepackage{comment}

\hypersetup{
  colorlinks=true,
  linkcolor=blue,
  urlcolor=blue,
  citecolor=blue
} 

\begin{document}

\author{Mohamed Meguebel}
\email{mohamed.meguebel@telecom-paris.fr}
\affiliation{Telecom Paris, Institut Polytechnique de Paris, 19 Place Marguerite Perey, 91120 Palaiseau, France}
\author{Maxime Federico}
\affiliation{Télécom Paris, Institut Polytechnique de Paris, 19 Place Marguerite Perey, 91120 Palaiseau, France}
\author{Simone Felicetti}
\affiliation{Institute for Complex Systems, National Research Council (ISC-CNR), Via dei Taurini 19, 00185 Rome, Italy}
\affiliation{Physics Department, Sapienza University, P.le A. Moro 2, 00185 Rome, Italy}
\author{Nadia Belabas}
\affiliation{Centre for Nanosciences and Nanotechnology, CNRS, Université Paris-Saclay,
UMR 9001, 10 Boulevard Thomas Gobert, 91120, Palaiseau, France}
\author{Nicolas Fabre} \email{nicolas.fabre@telecom-paris.fr}
\affiliation{Telecom Paris, Institut Polytechnique de Paris, 19 Place Marguerite Perey, 91120 Palaiseau, France}

\date{\today}
\begin{abstract}
Light-matter interactions with quantum dots have been extensively studied to harness key quantum properties of photons, such as indistinguishability and entanglement. In this theoretical work, we exploit the atomic-like four-level structure of a quantum dot coupled to a waveguide to model a shaping frequency entangling gate (FrEnGATE) for single photons. Our approach is based on the identification of input frequencies and an atomic level structure for which frequency-dependent one-photon transitions are adiabatically eliminated, while frequency-dependent two-photon transitions are resonantly enhanced. The frequency entanglement performance of the gate is analyzed using a Schmidt decomposition for continuous variables, revealing a trade-off between entanglement generation efficiency and entanglement quality. We further demonstrate the use of the  FrEnGATE for the generation of entangled frequency qudit states. 
\end{abstract}
\vskip2pc

\title{Generation of frequency entanglement with an effective quantum dot-waveguide two-photon quadratic interaction
}
\maketitle

\section{Introduction}

Photonic quantum technologies~\cite{BlablaphotonicQIslussarenko2019photonic, ContextBlablaphotonicQIObrieno2009photonic} rising from the second quantum revolution~\cite{BabbleQuantumtechnologiesdowling2003quantum,BabbleQuantumtechnologiesacin2018quantum,BabbleQuantumtechnologiesdeutsch2020harnessing}, include a range of application from quantum communication~\cite{BabbleQuantumcommunicationgisin2007quantum,BabbleQuantumcommunicationvajner2022quantum} to quantum metrology~\cite{Babblequantummetrologygiovannetti2006quantum,Babblequantummetrologygiovannetti2011advances} and computation~\cite{ContextDVrailencodingkok2007review,Blablaphotoniccomputingbartolucci2023fusion,Blablaphotonicquantumcomputingromero2024photonic}.  Quantum information can be encoded into discrete~\cite{ContextphotonicDVQInielsen2000quantum,ContextphotonicDVQIhorodecki2021quantum} or continuous variables~\cite{ContextphotonicCVQIweedbrook2012gaussian,ContextCVQIserafini2023quantum}, and some encodings pursue an intermediate hybrid approach~\cite{ContextphotonichybridQIandersen2014hybrid}. Examples of discrete variables for photonic quantum system are polarization~\cite{ContextphotonDVpolarizationsansoni2014integrated} or the rail~\cite{ContextDVrailencodingkok2007review,ContextphotonicDVrailencodingwu2013nogotheorem} encoding for qubit-based photonic quantum information. For a $d$-dimensional computational basis defining qudit-based photonic quantum information, frequency-comb~\cite{ContextphotonicDVquditsfrequencycomnbernhard2013shaping,ContextphotonicDVquditsfrequencycomblu2019quantum,ContextphotonicTFGridstateshenry2024parallelization} or orbital angular momentum of light~\cite{ContextphotonicDVquditsOAMgarcia2011universal,ContextphotonicDVquditsOAMperumangatt2017quantum,ContextphotonicDVquditsOAMandrews2011structured} encodings can be listed out. When it comes to continuous variables, although the most common form of encoding relies on the electric field quadratures~\cite{ContextphotonicCVquadraturesbraunstein2005quantum,ContextphotonicCVModesandstatestrepsfabre2020modes}, other proposals have been investigated. For instance, continuous variables corresponding to transverse spatial degrees of freedom of photons~\cite{ContextphootnicCVtransversespatialtasca2011continuous} or time-frequency continuous variables~\cite{ContextphotonicCVtimefrequencyfabre2022time,ContextphotonicCVtimefrequencydescamps2024gottesman}. The latter aims at harnessing continuous degrees of freedom of \textit{time-of-arrival} and \textit{frequency} in the single-photon subspace unlike field quadrature-based quantum information which relies on the number of photons statistical distribution in a given mode~\cite{ContextphotonicCVModesandstatestrepsfabre2020modes}. The time-frequency continuous variables formalism defines auxiliary modes which are any possible single-photon degrees of freedom that are not the time-frequency variables and do not couple to them. These auxiliary modes act as photon carriers and do not encode any quantum information. Provided that no two photons occupy the same complete set of auxiliary modes -- they may share individual modes, but not the entire configuration -- the canonical commutation relation (CCR) of the \textit{time-of-arrival} $\hat \tau$ and \textit{frequency} $\hat \omega$ operators is the same~\cite{ContextphotonicCVtimefrequencyfabre2022time} as that of \textit{position} $\hat x$ and \textit{momentum} $\hat p$ for the  position-momentum field quadratures. This entails a Heisenberg algebra for both and the same mathematical descriptions therein -- from the phase-space description to the universal set of gates \cite{ContextphotonicCVquadraturesbraunstein2005quantum,ContextphotonicCVtimefrequencyfabre2022time}. Notwithstanding these mathematical similarities, it should be emphasized that in these two cases the information is encoded in different physical degrees of freedom. While the quadratures revolve around the photon-number statistical distribution in a given mode, the time-frequency continuous variables encode quantum information continuously in the single-photon subspace. As such, non-Gaussian operations are difficult to implement in the former case~\cite{ContextCVnonGaussianMattia}, particularly for higher-order processes and multimode gates. These implementations typically employ offline measurements~\cite{ContextCVnonGaussianmultimodeoffline} or non-linear interactions~\cite{ContextCVnonGaussiannonlinearzavatta2004quantum}. In contrast, time-frequency continuous degrees of freedom benefit from the versatility of waveshapers, that allow performing any single-photon gate \cite{PhysRevA.94.063842,ContextphotonicCVtimefrequencyfabre2022time}. Nevertheless, while entangling modes within the field quadrature formalism is well established \cite{yoshikawa_generation_2016,doi:10.1126/science.aay2645,doi:10.1126/science.aay4354}, generating time-frequency entanglement --  or entanglement in any other degree of freedom --  between single photons remains notoriously challenging~\cite{Contextphotonphotoninteractionchang2014quantum}. A prevalent approach in linear optical quantum computing involves a probabilistic scheme that relies on post-selection, utilizing linear optical elements and measurements of ancillary photons.  \cite{ContextprobabilisticpostselectionKLMknill2001scheme,ContextDVrailencodingkok2007review,Blablaphotoniccomputingbartolucci2023fusion}. \\ 
In contrast, \textit{measurement-free} -- \textit{i.e.} without post-selection -- time-frequency entanglement naturally arises in the photon pairs generated via spontaneous parametric down-conversion (SPDC) and spontaneous four-wave mixing (SFWM). However, these processes have significant drawbacks. (i) SPDC and SFWM do not directly entangle preexisting photons but rather produce entangled pairs. (ii) The photon pair generation rate is inefficient and putting SPDC in cascade is not scalable for generating larger entangled states. (iii) These non-linear processes also generate multiphoton states and are thus imperfect sources of true
frequency-correlated photon pairs. To circumvent these pitfalls, other schemes should and have been investigated. \\ Namely, Le Jeannic \textit{et al.} proposed in~\cite{ArticleracineHannale2022dynamical} an experimental implementation employing a quantum dot (QD) embedded in a photonic crystal waveguide which scatters~\cite{Scatteringtheoryfan2010input} time-correlated photon pairs. Nonetheless, the light-matter interaction considered in~\cite{ArticleracineHannale2022dynamical} is dictated by a linear Hamiltonian in bosonic creation and annihilation operators, which can limit the gate fidelity of controlled-phase gate \cite{Articleracinenogonysteen2017limitations}. In~\cite{ArticleracineSimonealushi2023waveguide}, Alushi \textit{et al.} provided a waveguide quantum electrodynamics (WQED) framework based on a quadratic Hamiltonian in the bosonic creation and annihilation operators. Using scattering theory calculations, the authors demonstrated how this quadratic Hamiltonian allowed overcoming the aforementioned no-go theorem~\cite{Articleracinenogonysteen2017limitations}. Although possible physical implementations were suggested, the derivations of the quadratic Hamiltonian were based on a generic toy model, without considering a specific device’s requirements. In this work, we leverage their waveguide QED toolbox within a realistic physical system to establish frequency entanglement of initially separable single photons.\\ 

 We design a photonic scheme where a QD is embedded in a single-mode, linear, and nonmagnetic waveguide for which, under certain conditions that permit an adiabatic elimination, the light-matter interaction in the weak coupling regime can be shown to yield an \textit{ab initio} quadratic Hamiltonian identical to the \textit{ad hoc} Hamiltonian considered by Alushi \textit{et al.}~\cite{ArticleracineSimonealushi2023waveguide}. The proposed photonic system can be employed to entangle single photons in their continuously distributed frequencies. The light-matter interaction model is founded upon an adiabatic elimination procedure~\cite{AdiabaticeliminationLambdasystembrion2007adiabatic,Adiabaticeliminationbeyondpaulisch2014beyond,Adiabaticeliminationmultiphotonmaity2024adiabatic,AdiabaticeliminationHeisenbergRouchonle2023heisenberg,Adiabaticeliminationgonzalez2024tutorial,Adiabaticeliminationsubspacefinkelstein2020adiabatic}. Adiabatic elimination is typically carried out within either an isolated (closed system) framework or an open system framework. In the isolated system approach \cite{AdiabaticeliminationLambdasystembrion2007adiabatic,Adiabaticeliminationbeyondpaulisch2014beyond}, states that are far off-resonance -- \textit{i.e}., those with large frequency detunings compared to that of the states of interest -- are effectively discarded from the dynamics. In contrast, the open system derivation \cite{AdiabaticeliminationHeisenbergRouchonle2023heisenberg,Adiabaticeliminationgonzalez2024tutorial,Adiabaticeliminationsubspacefinkelstein2020adiabatic} considers a system coupled to an external bath whose degrees of freedom are adiabatically eliminated, leading to an effective evolution for the system that incorporates dissipation and decoherence. Our approach employs frequency-dependent \textit{joint} transition operators, where atomic transitions are coupled to the continuously defined bosonic creation and annihilation operators of the field. This formalism encapsulates the interplay between the atomic and photonic degrees of freedom in a frequency-resolved manner. Within this framework, frequency-dependent \textit{joint} one-photon transition operators are adiabatically eliminated in favor of frequency-dependent \textit{joint} two-photon transition operators described in the Heisenberg picture.  Moreover, the present model notably addresses the requirement of having at most one photon per auxiliary mode configuration--here being the polarization, as mentioned in \cite{ContextphotonicCVtimefrequencyfabre2022time}. Similarly to Le Jeannic \textit{et al.}~\cite{ArticleracineHannale2022dynamical}, the frequency-correlated photon pairs are effectively produced without post-selection~\cite{ContextprobabilisticpostselectionKLMknill2001scheme,ContextBlablaphotonicQIObrieno2009photonic}, \text{i.e.} \textit{measurement-free}. By applying a Schmidt decomposition to analyze the continuous-frequency entanglement, we further highlight a trade-off between the efficiency of entanglement generation and the quality of the entanglement. Furthermore, we demonstrate that the shaping frequency entangling gate (FrEnGATE) can be used for the generation of frequency qudit states~\cite{ContextphotonicDVquditsfrequencycomnbernhard2013shaping,ContextphotonicTFquditreimer2016generation,ContextphotonicTFqudtistateskues2017chip,ContextphotonicTFquditstateslu2018quantum,maltese_generation_2020,ContextphotonicCVtimefrequencyNicolasGridstates,ContextphotonicTFquditsstateslu2023frequency,yamazaki_linear_2023,ContextphotonicTFGridstateshenry2024parallelization}. \mohamed{At an intuitive level, the protocol can be understood as follows. Two initially independent photons propagate through a waveguide embedding a quantum dot, where an effective quadratic light-matter interaction induces frequency entanglement between the outgoing photons, eliminating the need for post-selection or auxiliary photons. This protocol thus offers a direct and scalable route to photon-photon frequency entanglement.} \\

The article is organized as follows. In Sec.~\ref{Sec. Quantum dot-waveguide
effective quadratic two-photon interaction}, we cover the light-matter interaction system and the calculation of the effective two-photon quadratic Hamiltonian based on the \textit{joint} one-photon operator adiabatic elimination. Then, in Sec.~\ref{Sec. TiFrEnGa}, we make use of the methods from Alushi \textit{et al.}~\cite{ArticleracineSimonealushi2023waveguide} to show how photon-photon frequency entanglement can be generated without auxiliary photons and measurement-free, effectively outputting a two-photon Gaussian distribution along the sum and difference of frequencies. The entanglement generation and efficiency are scrutinized in Sec.~\ref{Sec. Entanglement, fidelity and efficiency investigation} before applying our method to frequency qudit states.

\section{Quantum dot-waveguide
    effective quadratic two-photon interaction model}
\label{Sec. Quantum dot-waveguide
    effective quadratic two-photon interaction}

\subsection{Physical system}
\subsubsection{Waveguide}
The physical system that we consider to implement the photonic frequency entangling gate  is a QD embedded in a single-mode, linear and nonmagnetic waveguide. The waveguide electric field quantization follows that of~\cite{ArticleracineMullertrivedi2020generation} and is detailed in \mohamed{Supplement 1, Section II}. We extend the analysis by taking into account the polarization degree of freedom along with the waveguide's time-reversal symmetry, yielding the following expression for the electric field operator
\begin{equation}
    \hat{\boldsymbol{E}}(\boldsymbol{r}) = \sum_{\sigma}\sum_{\mu \in \{\pm\}}\int_{\mathbb R}d\omega\;(\boldsymbol{\mathcal{E}}_{\sigma \mu}(\boldsymbol \rho, \beta(\omega)) e^{i\beta(\omega)\mu z}\hat a_{\sigma \mu}(\omega) + \text{h.c}),
\end{equation}
where $\sigma$, $\mu$, $\boldsymbol{\rho}$, $\beta(\omega)$, $\boldsymbol{\mathcal{E}}_{\sigma\mu} \left(\boldsymbol{r},\beta\right) = \boldsymbol{\mathcal{E}}_{\sigma \mu}(\boldsymbol \rho, \beta(\omega))e^{i\beta(\omega)z}$ correspond, respectively, to the polarization, the direction of propagation--either rightward $\mu=+$ or leftward $\mu=-$ along the $z$-axis aligned with the QD's growth axis -, the transverse coordinate, the wavevector -- depending on the frequency $\omega$ through the dispersion relation -- and the propagating modes, solutions to the wave equation 
\begin{equation}
\label{Equation wave equation}
    \boldsymbol \nabla \times \boldsymbol \nabla \times \boldsymbol{\mathcal{E}}_{\sigma\mu} \left(\boldsymbol{r},\beta\right) - \frac{\omega(\beta)^2}{c^2}\epsilon(\boldsymbol r,\mu\beta)\boldsymbol{\mathcal{E}}_{\sigma\mu} \left(\boldsymbol{r},\beta\right) = \boldsymbol 0,
\end{equation}
where $\epsilon(\boldsymbol r, \mu\beta)$ is the permittivity of the material. The operators $\hat a_{\sigma \mu}(\omega)$ are bosonic operators fulfilling the standard commutation relation 
\begin{equation}
\label{Equation standard commutation relation}
     [\hat a_{\sigma \mu}(\omega), \hat a_{\sigma' \mu'}^\dagger(\omega')] = \delta(\omega-\omega')\delta_{\sigma \sigma'}\delta_{\mu \mu'}.
\end{equation}
Then, the free Hamiltonian for the waveguide quantized field reads
\begin{equation}
    \hat H_{\text{free,WG}}= \sum_{\sigma}\sum_{\mu\in\{\pm\}}\int_{\mathbb R}d\omega \; \hbar \omega \hat a_{\sigma \mu}^\dagger(\omega)\hat a_{\sigma \mu}(\omega),
\end{equation}
discarding the zero-point energy.

\subsubsection{Quantum dot}
\begin{figure*}
    \centering
    \includegraphics[width=\linewidth]{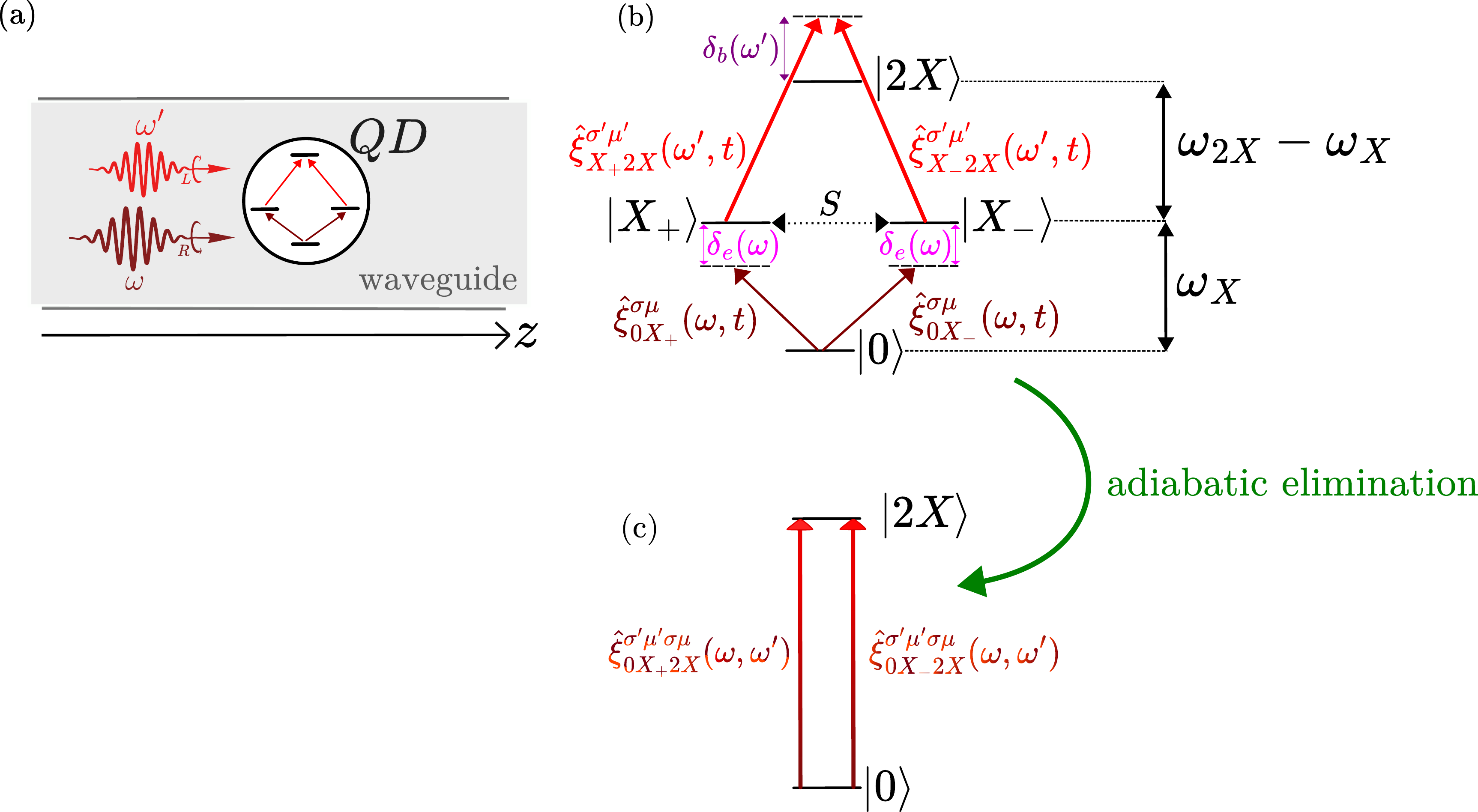}
    \caption{\mohamed{(a) Sketch of the QD embedded in the waveguide.} (b) QD's four-level structure in the circular polarization basis. The energy structure is displayed \textit{before} the one-photon joint operators $\hat \xi_{0X_\pm}^{\sigma \mu}(\omega,t) = \left(\ket{X_\pm}\bra{0}\otimes \hat a_{\sigma\mu}(\omega)\right)(t)$ and $\hat \xi_{X_\pm 2X}^{\sigma'\mu'}(\omega',t) = \left(\ket{2X}\bra{X_\pm}\otimes\hat a_{\sigma\mu}(\omega)\right)(t)$ adiabatic eliminations.  The one-photon transition detunings are $\delta_e(\omega) \equiv \omega-\omega_{X}$ and $\delta_b(\omega')\equiv\omega'-(\omega_{2X}-\omega_X)$. The excitonic transitions $\ket{0}\leftrightarrow \ket{X_\pm}$ can be driven by a right-handed and left-handed circularly polarized photon, respectively. The biexcitonic transitions $\ket{X_\pm}\leftrightarrow \ket{2X}$ can be triggered by a left-handed and right-handed circularly polarized photon, respectively. The fine-structure splitting (FSS) $S$ couples the two excitonic states $\ket{X_\pm}$ together. (c) Effective two-level QD's energy structure \textit{after} the adiabatic elimination. This adiabatic elimination switches from four one-photon joint transitions operators (a) to two two-photon joint transitions operators $\hat \xi_{0X_\pm2X}^{\sigma'\mu'\sigma\mu}(\omega',\omega)$ (b) driving the transition $\ket{0}\rightarrow \ket{2X}$ either through $\ket{X_+}$ or $\ket{X_-}$. This effectively results in a two-level atomic system. The two-photon joint transition operators $\hat \xi_{0X_\pm2X}^{\sigma'\mu'\sigma\mu}(\omega',\omega)$ in (b) are time-independent during the interaction time verifying the adiabatic elimination conditions. Note that the additional $X_\pm$ indices in these joint two-photon transition operators were introduced to indicate the origin of $\hat \xi_{0X_\pm2X}^{\sigma'\mu'\sigma\mu}(\omega',\omega)$. They both read $\hat \xi_{0X_\pm2X}^{\sigma'\mu'\sigma\mu}(\omega',\omega) = \ket{2X}\bra{0}\otimes \hat a_{\sigma'\mu'}(\omega')\hat a_{\sigma\mu}(\omega)$ regardless of the transition path.}
    \label{Figure one-photon transitions adiabatic elimination}
\end{figure*}

The QD is modeled as a four-level system with a ground state, two excitonic states and one biexcitonic state (see Fig.~\ref{Figure one-photon transitions adiabatic elimination} (a)). 
This four-level atomic-like structure can be represented in two different bases -- the linear and circular bases -- with different optical selection rules~\cite{QDgeneralgywat2010spins,Articleracineostfeldt2022demand,Articleracinegonzalez2023entanglement}. In the first basis, identically linearly polarized photons can drive the QD from its ground state to its biexcitonic state. In the circular basis, the optical selection rules require the involved photons to be circularly polarized and orthogonal, see  Fig.~\ref{Figure one-photon transitions adiabatic elimination} (a). The QD's free Hamiltonian can be expressed in the circular basis as 
\begin{equation}
    \begin{split}
        \hat H_{\text{free, QD}} &= \hbar\omega_{2X}\ket{2X}\bra{2X} \\ 
         &+ \hbar\omega_X\left(\ket{X_+}\bra{X_+}+\ket{X_-}\bra{X_-} \right) \\ 
         &+ \hbar S\left(\ket{X_+}\bra{X_-}+\ket{X_-}\bra{X_+}\right),
    \end{split}
\end{equation}
with $S$ the so-called fine-structure splitting (FSS) due to the QD's asymmetry~\cite{QDgeneralgywat2010spins,QDlodahl2015interfacing,QDgeneralollivier2021quantum}. The FSS scale typically ranges from 0 to 100 GHz in III–V semiconductor QDs~\cite{QDlodahl2015interfacing,QDFSSfognini2018universal} and several strategies to suppress it have been investigated~\cite{QDFSSmano2010self,QDFSSwang2012eliminating,QDFSSfognini2018universal,QDFSSlettner2021strain}. In the linear basis, the FSS lifts the degeneracy of the excitonic levels while it induces a coupling between the two excitonic states in the circular basis~\cite{Articleracinegonzalez2023entanglement}. Unless mentioned otherwise, we work in the QD's circular basis throughout.

\subsection{Effective two-photon quadratic Hamiltonian}
\label{subSection Effective two-photon quadratic Hamiltonian}
\subsubsection{Jaynes-Cummings Hamiltonian}
The light-matter interaction is described within the Goeppert-Mayer gauge and the dipole approximation in such a way that the interaction Hamiltonian is
\begin{equation}
    \begin{split}
        \hat H_{\text{int}} &= -\hat{\boldsymbol{d}}_{0X_+}\cdot \hat{\boldsymbol{E}}(\boldsymbol{0})-\hat{\boldsymbol{d}}_{0X_-}\cdot \hat{\boldsymbol{E}}(\boldsymbol{0}) \\
        &-\hat{\boldsymbol{d}}_{X_+2X}\cdot \hat{\boldsymbol{E}}(\boldsymbol{0}) 
       - \hat{\boldsymbol{d}}_{X_-2X}\cdot \hat{\boldsymbol{E}}(\boldsymbol{0}),
    \end{split}
\end{equation}
in the QD's and field's circular basis with $\hat{\boldsymbol d}_{..}$ the dipole moment operators of the different transitions and with the QD's position set at the reference origin of the waveguide $\boldsymbol r = \boldsymbol 0$, without loss of generality. For instance, $\hat{\boldsymbol d}_{0X_+} = \boldsymbol{d}_{0X_+}\ket{X_+}\bra{0}+\text{h.c}$ where $\ket{0}$ \mohamed{denotes} the QD's ground state \mohamed{from now on}. The first term $-\hat{\boldsymbol{d}}_{0X_+}\cdot \hat{\boldsymbol{E}}(\boldsymbol{0})$ can be expressed as 
\begin{equation}
\label{Equation first interaction term dipolar approximation}
    \begin{split}
        &-\hat{\boldsymbol{d}}_{0X_+}\cdot \hat{\boldsymbol E}(\boldsymbol 0) = 
        -\sum_{\sigma \in \{R,L\}} \sum_{\mu \in \{\pm\}} \int_{\mathbb R}d\omega\; \\ 
        &\bigg(\boldsymbol{d}_{0X_+}\cdot \boldsymbol{\mathcal E}_{\sigma \mu}(\boldsymbol 0, \beta(\omega)) \ket{X_+}\bra{0}\otimes \hat a_{\sigma \mu}(\omega) \\
        &+  \boldsymbol{d}_{0X_+}\cdot \boldsymbol{\mathcal E}_{\sigma \mu}^*(\boldsymbol 0, \beta(\omega)) \ket{X_+}\bra{0}\otimes \hat a_{\sigma \mu}^\dagger(\omega) + \text{h.c}\bigg),
    \end{split}
\end{equation}
in the field circular basis $\sigma = R,L$ where $R$ and $L$ stand for right-handed and left-handed circular polarizations, respectively. One can define the frequency-dependent coupling terms 
\begin{equation}
\label{Equation one-photon coupling term}
    g_{0X_+}^{\sigma  \mu}(\omega) \equiv -\frac{\boldsymbol{d}_{0X_+}\cdot \boldsymbol{\mathcal E}_{\sigma \mu}(\boldsymbol 0, \beta(\omega))}{\hbar} 
\end{equation}
\begin{equation}
    f_{0X_+}^{\sigma \mu}(\omega) \equiv -\frac{\boldsymbol{d}_{0X_+}\cdot \boldsymbol{\mathcal E}_{\sigma \mu}^*(\boldsymbol 0, \beta(\omega))}{\hbar}.
\end{equation} 
Up to first order in these coupling terms and in $S$ -- which acts as the coupling term in the circular basis -- the operators $\ket{X_+}\bra{0}\otimes \hat a_{\sigma \mu}(\omega)$ and $\ket{X_+}\bra{0}\otimes \hat a_{\sigma \mu}^\dagger(\omega)$ evolve in the Heisenberg picture with frequencies $\omega+\omega_X$ and $|\omega-\omega_X|$, respectively. Assuming the weak coupling regime, where all the coupling terms in Eq.~\eqref{Equation first interaction term dipolar approximation} are small compared to $\omega + \omega_X$, one can apply the standard rotating-wave approximation (RWA) and discard the non-resonant terms associated with the coupling term $f_{0X_+}^{\sigma \mu}(\omega)$. Therefore,  the first term of the interaction Hamiltonian is:
\begin{equation}
    \begin{split}
        &-\hat{\boldsymbol{d}}_{0X_+}\cdot \hat{\boldsymbol E}(\boldsymbol 0) = 
        \sum_{\sigma \in \{R,L\}} \sum_{\mu \in \{\pm\}} \int_{\mathbb R}d\omega\;\\
        & \cross \bigg(\hbar g_{0X_+}^{\sigma \mu}(\omega) \ket{X_+}\bra{0}\otimes \hat a_{\sigma \mu}(\omega)+\text{h.c}\bigg).
    \end{split}
\end{equation}
Following the same procedure for $\hat{\boldsymbol{d}}_{0X_-}\cdot \hat{\boldsymbol{E}}(\boldsymbol 0)$, $\hat{\boldsymbol{d}}_{X_+2X}\cdot \hat{\boldsymbol{E}}(\boldsymbol{0})$ and $\hat{\boldsymbol{d}}_{X_-2X}\cdot \hat{\boldsymbol{E}}(\boldsymbol{0})$, the interaction Hamiltonian reads 
\begin{align}
\label{Equation dipolar and RWA Hamiltonian before adiabatic elimination}
        \hat{H}_{\text{int}} &= \hbar\sum_{\sigma \in \{R,L\}}\sum_{\mu\in\{\pm\}}\int_{\mathbb R}d\omega\;\bigg(g_{0X_+}^{\sigma \mu}(\omega)\ket{X_+}\bra{0}\otimes \hat a_{\sigma \mu}(\omega)\nonumber \\
        &+g_{X_+2X}^{\sigma \mu}(\omega)\ket{2X}\bra{X_+}\otimes \hat a_{\sigma \mu}(\omega)\nonumber\\ 
        &+g_{0X_-}^{\sigma \mu}(\omega)\ket{X_-}\bra{0}\otimes \hat a_{\sigma \mu}(\omega)\nonumber\\
        &+g_{X_-2X}^{\sigma \mu}(\omega)\ket{2X}\bra{X_-}\otimes \hat a_{\sigma \mu}(\omega)+\text{h.c}\bigg),
\end{align}
which is a sum of four one-photon Hamiltonians. In order to retrieve a quadratic Hamiltonian analogous to that of Alushi \textit{et al}.~\cite{ArticleracineSimonealushi2023waveguide}, we would like to have two-photon Hamiltonians instead. With this aim, we resort to an adiabatic elimination to obtain an effective two-level system characterized by frequency-dependent two-photon transitions.

\subsubsection{Adiabatic elimination}
\label{subsubsection Adiabatic elimination}
  Inspired by the stimulated-Raman adiabatic passage (STIRAP)~\cite{STIRAPoriginalgaubatz1990population,STIRAPshore2017picturing}, we perform an adiabatic elimination~\cite{AdiabaticeliminationLambdasystembrion2007adiabatic,Adiabaticeliminationbeyondpaulisch2014beyond,Adiabaticeliminationmultiphotonmaity2024adiabatic,AdiabaticeliminationHeisenbergRouchonle2023heisenberg,Adiabaticeliminationgonzalez2024tutorial,Adiabaticeliminationsubspacefinkelstein2020adiabatic,adiabaticeliminationatomicphysicslarson2021jaynes,adiabaticeliminationatomicphysicspuri1988quantum,adiabaticeliminationatomicphysicsalexanian1995unitary,adiabaticeliminationatomicphysicswu1996effective} to suppress rapidly evolving one-photon processes in favor of two-photon ones. To this end, we invoke the transition operators time-evolution Heisenberg equation as it was done in~\cite{ArticleracineRamanpassagegerry1990dynamics,adiabaticeliminationatomicphysicspuri1988quantum,adiabaticeliminationatomicphysicsalexanian1995unitary,adiabaticeliminationatomicphysicswu1996effective}. Nonetheless, unlike~\cite{ArticleracineRamanpassagegerry1990dynamics,adiabaticeliminationatomicphysicspuri1988quantum,adiabaticeliminationatomicphysicsalexanian1995unitary,adiabaticeliminationatomicphysicswu1996effective}, we manipulate \textit{joint} light-matter operators. In our case, they encompass both matter and light while~\cite{ArticleracineRamanpassagegerry1990dynamics,adiabaticeliminationatomicphysicsalexanian1995unitary,adiabaticeliminationatomicphysicswu1996effective} considered bare matter operators only. Puri \textit{et al.}~\cite{adiabaticeliminationatomicphysicspuri1988quantum} did evoke the physical differences that rise from considering the evolution of the joint transition operators, namely for spontaneous photon-number-dependent Stark shifts. Nevertheless, their work applies to a single-mode cavity and not to a continuum for which the bookkeeping of each frequency-dependent joint one-photon transition operator is paramount. This \textit{joint} operator terminology is not to be confused with \textit{dressed} states from quantum optics' atom-field \textit{dressed} states or polaritons~\cite{adiabaticeliminationatomicphysicslarson2021jaynes} which refers to coherent superpositions of light-matter states. The joint one-photon transition operators are written as $\ket{j}\bra{i}\otimes \hat a_{\sigma \mu}(\omega)$ depicting a transition from state $\ket{i}$ to $\ket{j}$ by absorbing a photon. For instance, the $\hat \xi_{0X_+}^{\sigma \mu}(\omega) \equiv \ket{X_+}\bra{0}\otimes \hat a_{\sigma \mu}(\omega)$ joint operator corresponds to the $\ket{0}\rightarrow \ket{X_+}$ QD excitonic transition by annihilation of a field excitation at frequency $\omega$, polarization $\sigma$ and with a direction of propagation $\mu$. The corresponding Heisenberg picture time-evolution of this joint one-photon transition operator is given by
\begin{equation}
\label{Equation of motion Heisenberg picture}
    \frac{d\hat \xi_{0X_+}^{\sigma \mu}(\omega,t)}{dt} = \frac{[\hat \xi_{0X_+}^{\sigma \mu}(\omega,t),\hat H]}{i\hbar}.
\end{equation}
It should be noted that, here, the time $t$ is the dynamical time of the Hamiltonian evolution and not the \textit{time-of-arrival} defined in the time-frequency continuous variables formalism~\cite{ContextphotonicCVtimefrequencythesefabre2020quantum,ContextphotonicCVtimefrequencyfabre2022time}. Up to second order in the FSS parameter $S$ and in the coupling terms $g_{..}^{..}(.)$, the effective interaction Hamiltonian can be expressed as 
\begin{equation}
\label{Equation effective Hamiltonian after adiabatic elimination}
    \begin{split}
        &\hat H_{\text{int}} = \left(\hat H_{\text{int}} \right)_{\ket{0}\leftrightarrow \ket{X_+}\leftrightarrow \ket{2X}} +\left(\hat H_{\text{int}} \right)_{\ket{0}\leftrightarrow \ket{X_-}\leftrightarrow \ket{2X}},
    \end{split}
\end{equation}
where we have separated the two interaction branches which read 
\begin{equation}
    \label{Equation general effective interaction Hamiltonian X_pm branch}
     \begin{split}
        &\left(\hat H_{\text{int}} \right)_{\ket{0}\leftrightarrow \ket{X_\pm}\leftrightarrow \ket{2X}} =\hbar\sum_{\substack{\sigma'\in \{R,L\} \\ \mu' \in \{\pm\}}} \sum_{\substack{\sigma \in \{R,L\} \\ \mu \in \{\pm\}}}\int_{\mathbb R^2}d\omega' d\omega\; \\ &\bigg(g_{X_\pm}^{\sigma'\mu'\sigma \mu }(\omega',\omega)\ket{2X}\bra{0}\otimes \hat a_{\sigma'\mu'}(\omega')\hat a_{\sigma \mu}(\omega) + \text{h.c}\bigg).
    \end{split}
\end{equation}
The calculated two-photon coupling terms are
\begin{equation}
\begin{split}
\label{Equation two-photon coupling term X_pm branch}
    g_{X_\pm}^{\sigma' \mu'\sigma \mu }(\omega',\omega) &= g_{X_\pm2X}^{\sigma'\mu'}(\omega')g_{0X_\pm}^{\sigma \mu}(\omega) \\&\times\bigg[\frac{1}{\omega-\omega_X}-\frac{1}{\omega'-(\omega_{2X}-\omega_X)}\bigg],
\end{split}
\end{equation}
where the one-photon detunings $\delta_e(\omega) \equiv \omega-\omega_X$ and $\delta_{b}(\omega') \equiv \omega'-(\omega_{2X}-\omega_X)$ arises. The detailed calculations are included in \mohamed{Supplement 1, Section III.A}. It should be pointed out that the FSS contribution has been eliminated by neglecting coupling terms of third or higher-order in the interaction Hamiltonian. Although a second-order term involving $S$ does emerge from the time-evolution of the joint one-photon transition operator, it is suppressed within the adiabatic elimination.  This adiabatic elimination regime is valid as long as: (i) $\delta_e(\omega)$ and $\delta_b(\omega')$ are large in absolute value compared to the $\ket{0}\leftrightarrow \ket{2X}$ two-photon detuning $|\omega+\omega'-\omega_{2X}|$, and to the photons' bandwidths. This condition ensures that the two-photon joint transition operators undergo a considerably slower evolution than the one-photon joint transition operators. \mohamed{(ii) The one-photon detunings are significantly larger than the one-photon coupling terms. This dispersive condition ensures that the excitonic levels remain negligibly populated.}. (iii) The interaction time $t$ is such that $1/t$ is much smaller than the one-photon detunings and much larger than the photons' bandwidths and two-photon detunings. This guarantees that the one-photon joint transition operators have averaged to zero, whereas the two-photon joint transition operators have undergone minimal evolution and can thus be regarded as constant. \\
\mohamed{In Supplement 1, Section III.B we provide rigorous bounds on the contributions of fast-oscillating operators to the effective Hamiltonian, and in Supplement 1, Section III.C we report numerical simulations of biexciton decay both inside and outside the adiabatic-elimination regime. When the adiabatic conditions are satisfied, one-photon transitions are strongly suppressed, the exciton population remains small (peak value $\approx 7\%$) and the biexciton decays preferentially via the two-photon channel, supporting the validity of the effective two-photon Hamiltonian derived above. We also observe that the excitonic population exhibits rapid oscillations before settling to a small residual plateau. However, the population remains below 10\%  and reaches a population plateau $\approx 1.8\%$ level throughout and can be safely neglected in the dispersive regime. Since this effect has no significant impact on the biexciton decay pathway, it is not included in the effective Hamiltonian Eq.~\eqref{Equation general effective interaction Hamiltonian X_pm branch}.} \\
The effective interaction Hamiltonian Eq.~\eqref{Equation effective Hamiltonian after adiabatic elimination} describes the QD adiabatic passage from its ground to its biexcitonic state by annihilation of two photons at frequencies $\omega$ and $\omega'$, polarizations $\sigma$ and $\sigma'$ and with directions of propagation $\mu$ and $\mu'$, and reciprocally for the relaxation from the biexcitonic to the ground state. As expected, employing the STIRAP-like method took us from four one-photon interaction Hamiltonians to two two-photon interaction Hamiltonians. Whether one interaction regime or the other occurs depends on the interaction timescale relative to the system spectral characteristics. Therefore, one can use this time criterion to selectively filter the photons that are effectively going to be frequency-entangled with this adiabatic elimination process. This one-photon operators adiabatic elimination is outlined in Fig.~\ref{Figure one-photon transitions adiabatic elimination}. 
The two-photon coupling terms Eq.~\eqref{Equation two-photon coupling term X_pm branch} are non-separable in the two-photon frequencies $\omega$ and $\omega'$. As it is discussed in Sec.~\ref{Sec. TiFrEnGa}, this is paramount for the  FrEnGATE. One way to fathom this is to think of the two-photon interaction with the QD as a spectral reshaping. Since the two-photon coupling term is non-separable in the two photon frequencies, the reshaping is also non-separable, as is the resulting emitted two-photon distribution. In addition to their non-separability, the coupling terms are also in general non-symmetric, that is to say $g_{X_\pm}^{\sigma'\mu' \sigma \mu}(\omega',\omega) \neq g_{X_\pm}^{\sigma \mu \sigma'\mu'}(\omega,\omega')$. This will become clearer in the next section, which addresses the optical selection rules. At this stage, one can intuitively understand this ordering by observing that the indices are arranged such that the leftmost ones correspond to the biexcitonic photon, while the rightmost ones pertain to the excitonic photon. This follows from the proportionality relation $g_{\pm}^{\sigma'\mu' \sigma \mu}(\omega',\omega) \propto g_{X_\pm 2X}^{\sigma'\mu'}(\omega')g_{0X_\pm}^{\sigma \mu}(\omega)$. Indeed, if the single-photon coupling terms were different, the two-photon coupling terms associated with the two excitation pathways $\ket{0}\leftrightarrow\ket{X_\pm}\leftrightarrow\ket{2X}$  would necessarily differ as well. The retrieved Hamiltonian Eq.~\eqref{Equation effective Hamiltonian after adiabatic elimination} constructed from Eq.~\eqref{Equation general effective interaction Hamiltonian X_pm branch} is exactly of the same form as the one suggested by Alushi \textit{et al.}~\cite{ArticleracineSimonealushi2023waveguide} in their waveguide QED framework 
\begin{equation}
\label{Equation Simone's Hamiltonian}
    \begin{split}
        &\hat H_{\text{int,Alushi}} = \hbar\sum_{\mu',\mu\in\{\pm\}}\int_{\mathbb R^2}d\omega'd\omega\;\\
        &\left(g^{\mu'\mu}(\omega',\omega)\hat \sigma_+ \otimes \hat a_{\mu'}(\omega')\hat a_{\mu}(\omega) + \text{h.c}\right),
    \end{split}
\end{equation}
where $\hat \sigma_{+}$ is a general raising operator associated to the two-level system under consideration and where the polarization has not been taken into account. Unlike Eq.~\eqref{Equation Simone's Hamiltonian}, the Hamiltonian Eq.~\eqref{Equation effective Hamiltonian after adiabatic elimination} that we derived has been constructed based on a specific photonic system, ensuring a concrete physical foundation thus laying the groundwork for experimental implementations. More specifically, the coupling terms Eq.~\eqref{Equation two-photon coupling term X_pm branch} have an explicit form hence providing greater insights as to how to engineer the light-matter system. Besides, one may notice that we have yet to specify the optical selection rules. The derivations detailed in \mohamed{Supplement 1, Section III.A} are thus sufficiently general to be applied to photonic systems analogous to the one investigated here with other auxiliary modes and not only the polarization and direction of propagation. 
\subsubsection{Optical selection rules}
In this section, we restrict the effective interaction Hamiltonian Eq.~\eqref{Equation effective Hamiltonian after adiabatic elimination} to fulfill the QD's optical selection rules~\cite{QDlodahl2015interfacing,QDgeneralgywat2010spins}. Fixing the quantization axis along the photons' propagation axis $z$, $\sigma$-polarized photons propagating in the $\mu=-$ direction are seen as $\sigma^\perp$-polarized by the QD. Hence, $R$-polarized photons can drive the $\ket{0}\leftrightarrow \ket{X_+}$ and $\ket{X_-}\leftrightarrow \ket{2X}$ transitions if they propagate with $\mu = +$ and the $\ket{0}\leftrightarrow \ket{X_-}$ and $\ket{X_+}\leftrightarrow \ket{2X}$ if they travel with $\mu=-$. The reasoning is identical for $L$-polarized photons. The two interaction paths Eq.~\eqref{Equation general effective interaction Hamiltonian X_pm branch} thus read 
\begin{equation}
\label{Equation optical selection rules interaction Hamiltonian X_+ branch}
    \begin{split}
  & \left(\hat H_{\text{int}} \right)_{\ket{0}\leftrightarrow \ket{X_+}\leftrightarrow \ket{2X}} = \hbar\int_{\mathbb R^2}d\omega' d\omega\ket{2X}\bra{0}\otimes\\ &\bigg(g_{X_+}^{L+R+}(\omega',\omega)\hat a_{L+}(\omega')\hat a_{R+}(\omega)\\
   &+g_{X_+}^{R-L-}(\omega',\omega)\hat a_{R-}(\omega')\hat a_{L-}(\omega) \\ 
   &+g_{X_+}^{R-R+}(\omega',\omega)\hat a_{R-}(\omega')\hat a_{R+}(\omega)\\
   &+g_{X_+}^{L+L-}(\omega',\omega)\hat a_{L+}(\omega')\hat a_{L-}(\omega)+\text{h.c}\bigg)
    \end{split}
\end{equation}
and 
\begin{equation}
\label{Equation optical selection rules interaction Hamiltonian X_- branch}
    \begin{split}
  & \left(\hat H_{\text{int}} \right)_{\ket{0}\leftrightarrow \ket{X_-}\leftrightarrow \ket{2X}} = \hbar\int_{\mathbb R^2}d\omega' d\omega\ket{2X}\bra{0}\otimes\\ &\bigg(g_{X_-}^{R+L+}(\omega',\omega)\hat a_{R+}(\omega')\hat a_{L+}(\omega)\\
   &+g_{X_-}^{L-R-}(\omega',\omega)\hat a_{L-}(\omega')\hat a_{R-}(\omega) \\ 
   &+g_{X_-}^{L-L+}(\omega',\omega)\hat a_{L-}(\omega')\hat a_{L+}(\omega)\\
   &+g_{X_-}^{R+R-}(\omega',\omega)\hat a_{R+}(\omega')\hat a_{R-}(\omega)+\text{h.c}\bigg).
    \end{split}
\end{equation}
By grouping the two interaction branches Hamiltonian Eq.~\eqref{Equation optical selection rules interaction Hamiltonian X_+ branch} and Eq.~\eqref{Equation optical selection rules interaction Hamiltonian X_- branch} together and by rearranging the $\omega$ and $\omega'$ dummy indices, one obtains
\begin{equation}
\label{Equation Hamiltonian where I've grouped the two interaction paths together}
    \begin{split}
        &\hat H_{\text{int}}=\hbar\int_{\mathbb R^2}d\omega'd\omega\ket{2X}\bra{0}\otimes\bigg(g^{++}(\omega',\omega)\hat a_{L+}(\omega')\hat a_{R+}(\omega)\\
        &+g^{--}(\omega',\omega)\hat a_{R-}(\omega')\hat a_{L-}(\omega)+g^{-+}(\omega',\omega)\hat a_{R-}(\omega')\hat a_{R+}(\omega) \\
        &g^{+-}(\omega',\omega)\hat a_{L+}(\omega')\hat a_{L-}(\omega) +\text{h.c}\bigg ),
    \end{split}
\end{equation}
where we have defined the two-photon coupling terms comprising both interaction paths for each pair of propagation directions $(\mu,\mu')$ as
\begin{equation}
\begin{split}
\label{Equation coupling terms after regrouping the two interaction paths}
    &g^{++}(\omega',\omega) = g_{X_+}^{L+R+}(\omega',\omega)+g_{X_-}^{R+L+}(\omega,\omega') \\
    &g^{--}(\omega',\omega) = g_{X_+}^{R-L-}(\omega',\omega) + g_{X_-}^{L-R-}(\omega,\omega') \\ 
    &g^{-+}(\omega',\omega) = g_{X_+}^{R-R+}(\omega',\omega) + g_{X_-}^{R+R-}(\omega,\omega') \\ 
    &g^{+-}(\omega',\omega) = g_{X_+}^{L+L-}(\omega',\omega) + g_{X_-}^{L-L+}(\omega,\omega').
\end{split}
\end{equation}
We recall that the indices are ordered such that  the  leftmost ones correspond  to  the  biexciton  photon, while the rightmost ones relate to the  excitonic photon. For instance, the coupling term $g_{X_+}^{L+R+}(\omega',\omega)$ depicts a transition where the biexcitonic photon is $L$-polarized and propagating rightward at frequency $\omega'$ and the excitonic photon is $R$-polarized and propagating rightward at frequency $\omega$.  Eq.~\eqref{Equation optical selection rules interaction Hamiltonian X_+ branch} and Eq.~\eqref{Equation optical selection rules interaction Hamiltonian X_- branch} each contribute eight terms to the total Hamiltonian, whereas the rearranged Hamiltonian in Eq.~\eqref{Equation Hamiltonian where I've grouped the two interaction paths together} contains only four. This can be understood by the fact that photons do not carry any information regarding what transitions -- excitonic or biexcitonic -- they have driven. Their degrees of freedom are solely the frequency, polarization and direction of propagation, as seen in the commutation relation Eq.~\eqref{Equation standard commutation relation}. Whether the transition is $\ket{0}\leftrightarrow \ket{X_\pm}$ or $\ket{X_\pm}\leftrightarrow \ket{2X}$ is imposed by the QD's optical selection rules and is conveyed through the coupling terms. In other words, there is meaning in referring to \textit{excitonic} and \textit{biexcitonic} photons only from the QD's perspective. Therefore, the Hamiltonian in Eq.~\eqref{Equation Hamiltonian where I've grouped the two interaction paths together} erases the information regarding the $\ket{0}\leftrightarrow \ket{2X}$ interaction path -- $\ket{0}\leftrightarrow \ket{X_+}\leftrightarrow \ket{2X}$ or $\ket{0}\leftrightarrow \ket{X_-}\leftrightarrow \ket{2X}$ and only retains the degrees of freedom relevant from the photons' perspective, namely the frequency, the polarization and the direction of propagation; thus merging the two interaction paths into one.

\subsubsection{Two-photon coupling terms}
\label{Subsubsection two-photon coupling term}
In this section, we focus on the two-photon coupling terms given in Eq.~\eqref{Equation coupling terms after regrouping the two interaction paths}. As explained, the two-photon Hamiltonians Eq.~\eqref{Equation optical selection rules interaction Hamiltonian X_+ branch} and Eq.~\eqref{Equation optical selection rules interaction Hamiltonian X_- branch} for the two interaction paths can be combined into one global Hamiltonian Eq.~\eqref{Equation Hamiltonian where I've grouped the two interaction paths together} where the details of the transition branch have been removed. Let us assume that 
\begin{equation}
\begin{split}
\label{Equation product of the two one-photon coupling terms}
    &g_{X_+2X}^{L+}(\omega')g_{0X_+}^{R+}(\omega) = g_{X_-2X}^{R+}(\omega)g_{0X_-}^{L+}(\omega') \\ 
    &g_{X_+2X}^{R-}(\omega')g_{0X_+}^{L-}(\omega) = g_{X_-2X}^{L-}(\omega)g_{0X_-}^{R-}(\omega') \\
    &g_{X_+2X}^{R-}(\omega')g_{0X_+}^{R+}(\omega) = g_{X_-2X}^{R+}(\omega)g_{0X_-}^{R-}(\omega') \\ 
    &g_{X_+2X}^{L+}(\omega')g_{0X_+}^{L-}(\omega) = g_{X_-2X}^{L-}(\omega)g_{0X_-}^{L+}(\omega'),
\end{split}
\end{equation}
meaning that the interaction paths have the same strength. The magnitudes of the electric dipole moments are independent of the excitonic state $\ket{X_\pm}$~\cite{QDgeneralgywat2010spins,QDlodahl2015interfacing}: $ ||\boldsymbol d_{0X_+}|| =  ||\boldsymbol d_{0X_-}||$ and $||\boldsymbol d_{X_+2X}|| =  ||\boldsymbol d_{X_-2X}||$. Therefore, given the expression of the one-photon coupling term Eq.~\eqref{Equation one-photon coupling term}, this approximation is legitimate provided that the norm of the propagation mode $\boldsymbol{\mathcal{E}}_{\sigma \mu}(\boldsymbol 0,\beta(\omega))$ -- solution to the wave equation Eq.~\eqref{Equation wave equation} -- does not depend on either the polarization nor the direction of propagation; this is valid provided there is propagation isotropy. Consequently, the product of the two one-photon coupling terms in Eq.~\eqref{Equation product of the two one-photon coupling terms} is considered equal to $D/\hbar^2\times u(\omega')u(\omega)$
where $D=||\boldsymbol d_{0X_\pm}||\times||\boldsymbol d_{X_\pm 2X}||$ and $u(\omega)$ is the magnitude of the propagation mode which depends on the frequency. The two-photon coupling terms~\eqref{Equation coupling terms after regrouping the two interaction paths} in the global Hamiltonian~\eqref{Equation Hamiltonian where I've grouped the two interaction paths together} are thus equal and read 
\begin{equation}
\begin{split}
\label{Equation two-photon coupling term after assuming same strength for each path}
    g^{\mu'\mu}(\omega',\omega) &= \frac{D}{\hbar^2}u(\omega')u(\omega)\times\bigg[\frac{1}{\omega-\omega_X}-\frac{1}{\omega'-(\omega_{2X}-\omega_X)} \\
    &+ \frac{1}{\omega'-\omega_X}-\frac{1}{\omega-(\omega_{2X}-\omega_X)}\bigg].
\end{split}
\end{equation}
The poles in Eq.~\eqref{Equation two-photon coupling term after assuming same strength for each path} can be disregarded because the adiabatic elimination calculation assumed the off-resonance condition for the one-photon transitions. The biexcitonic level has a characteristic frequency equal to $\omega_{2X} = 2\omega_{X}-\delta_X$ where $\delta_X$ denotes the binding frequency which is in general positive and of the order of $10^{-2}\omega_X$~\cite{QDgeneralgywat2010spins}. This energy shift is due to the energy needed to separate an exciton into its individual electron and hole charge carriers. In the telecommunication wavelengths for instance, type III-V QDs can be engineered~\cite{QDTelecomkors2018telecom,Telecomsinglephotonsourcemuller2018quantum,QDSinglephotonSourcesarakawa2020progress} to have $\omega_{X}\approx 2\pi\times 190$  THz, that is a corresponding wavelength $\lambda_{X}\approx$ 1550 nm. The two-photon coupling term can be rewritten as 
\begin{equation}
\begin{split}
\label{Equation two-photon coupling term after assuming same strength for each path taking delta_X}
    g^{\mu'\mu}(\omega',\omega) &= \frac{D}{\hbar^2}u(\omega')u(\omega)\times\bigg[\frac{1}{\omega-\omega_X}-\frac{1}{\omega'-(\omega_{X}-\delta_X)} \\
    &+ \frac{1}{\omega'-\omega_X}-\frac{1}{\omega-(\omega_{X}-\delta_X)}\bigg].
\end{split}
\end{equation}
\begin{figure}
    \centering
    \includegraphics[width=1\linewidth]{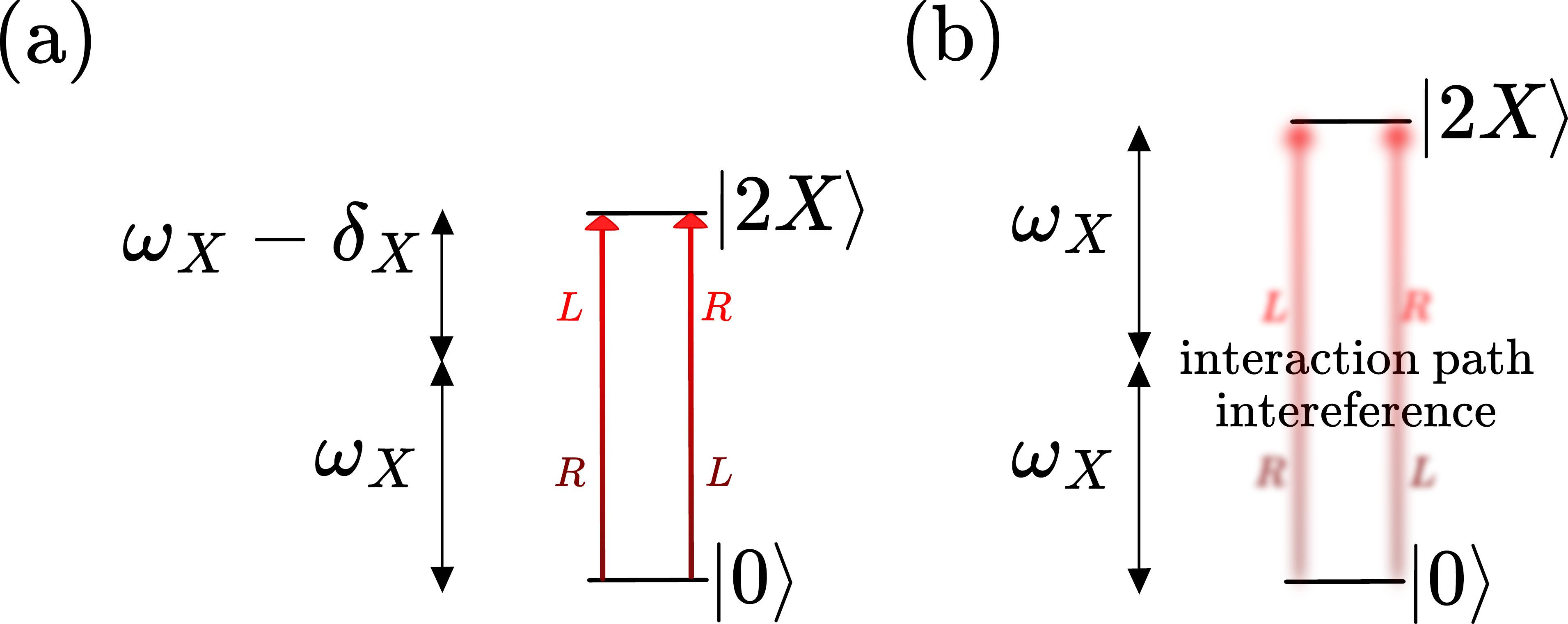}
    \caption{Effective two-level system where the interaction can be driven through two interaction paths: one where the first photon is $R$-polarized (respectively $L$-polarized) with a frequency far-detuned from $\omega_X$ and the second photon is $L$-polarized (respectively $R$-polarized) with a frequency far-detuned from $\omega_X-\delta_X$. (a) For $\delta_X \neq 0$, the two transition paths are distinguishable and do not interefere leading to a non-zero two-photon coupling term. (b) For $\delta_X=0$, the two interaction paths are now indistinguishable and destructively interfere. This entails a two-photon coupling term equal to zero thus preventing the $\ket{0}\leftrightarrow \ket{2X}$ transition.}
    \label{Figure interaction path interferences}
\end{figure}
Notably, it appears that the two-photon coupling term vanishes if $\delta_X$ is equal to zero, that is to say that the excitonic $\ket{0}\leftrightarrow \ket{X_\pm}$ and biexcitonic $\ket{X_\pm}\leftrightarrow \ket{2X}$ transitions have the same energies. In this case, the two indistinguishable interaction paths $\ket{0}\leftrightarrow\ket{X_\pm}\leftrightarrow \ket{2X}$ interfere destructively, therefore preventing the $\ket{0}\leftrightarrow \ket{2X}$ two-photon interaction emerging from the adiabiatic elimination described previously, see Fig.~\ref{Figure interaction path interferences}. Two-photon interaction path interferences were described in other photonic systems. For instance, in~\cite{PhotonblockadePhysRevA.83.021802} the pump-driven cavity-cavity photon blockade effect is interpreted as quantum interferences between different excitation paths. Building on the above considerations, the theoretical engineering of the two-photon coupling term in Eq.~\eqref{Equation two-photon coupling term after assuming same strength for each path taking delta_X} can be realized by shaping the magnitude of the propagation mode 
$u$ across the two photons' bandwidths, effectively leveraging the waveguide dispersion properties.

\section{Shaping frequency entangling gate}
\label{Sec. TiFrEnGa}
This section details how the effective two-photon quadratic interaction Hamiltonian -- with Eq.~\eqref{Equation Hamiltonian where I've grouped the two interaction paths together} or without selection rules Eq.~\eqref{Equation effective Hamiltonian after adiabatic elimination} -- can be exploited to generate frequency correlations between single photons. The overall methodology and computations -- based on scattering theory~\cite{Scatteringtheoryfan2010input,Scatteringtheorysakurai2020modern} -- follow those presented in~\cite{ArticleracineSimonealushi2023waveguide}. The additional time-scale constraints required for the adiabatic elimination regime to hold are discussed.

\subsection{Scattering theory and Markovian approximation}
In this section, we use the WQED toolbox developed in Alushi \textit{et al.}~\cite{ArticleracineSimonealushi2023waveguide}. The main calculation steps are detailed in \mohamed{Supplement 1, Section IV}.

\subsubsection{Wigner-Weisskopf ansatz}
The present scattering theory calculation is performed within the Schrödinger picture. The Hamiltonian~\eqref{Equation Hamiltonian where I've grouped the two interaction paths together} preserves the number of weighted excitations. The matter-field dynamics can thus be encapsulated in a vector state $\ket{\psi(t)}$ with a fixed number of weighted excitations  
\begin{equation}
 \label{Equation Wigner-Weisskopf ansatz} 
\begin{split}
   &\ket{\psi(t)} = C_{2X}(t)\hat \xi_{2X0}\ket{\boldsymbol 0}+\\
   &\sum_{\substack{\sigma'\in \{R,L\} \\ \mu' \in \{\pm\}}} \sum_{\substack{\sigma \in \{R,L\} \\ \mu \in \{\pm\}}}\int_{\mathbb R^2}d\omega' d\omega\; C^{\sigma'\sigma }_{\mu'\mu}(\omega',\omega;t)\hat a_{\sigma'\mu'}^\dagger(\omega')\hat a_{\sigma \mu}^\dagger(\omega)\ket{\boldsymbol 0},
\end{split}
\end{equation}
where $C_{2X}(t)$ and $C^{\sigma'\sigma }_{\mu'\mu}(\omega',\omega;t)$ are the probability amplitudes to have the system respectively in the QD's biexcitonic state without photons and in the QD's ground state with two photons at frequency $\omega$ and $\omega'$, polarization $\sigma$ and $\sigma'$, and direction of propagation $\mu$ and $\mu'$, respectively. $C^{\sigma'\sigma }_{\mu'\mu}(\omega',\omega;t)$ is also referred to as the joint spectral amplitude (JSA) and its squared modulus as the joint spectral intensity (JSI). The state $\ket{\boldsymbol 0} \equiv \ket{0}\otimes\ket{\text{vac}}$ represents the global vacuum state for both light and matter. We define the bare transition operator $\hat \xi_{2X0} = \hat \xi_{02X}^\dagger \equiv \ket{2X}\bra{0}$. The complete QD-waveguide isolated system is encapsulated within the Wigner-Weisskopf ansatz Eq.~\eqref{Equation Wigner-Weisskopf ansatz}, whose dynamics is governed by the Schrödinger equation $i\hbar \partial_t \ket{\psi(t)} = \hat H\ket{\psi(t)}$ where $\hat H$ is the total Hamiltonian. 
\subsubsection{Markovian and weak coupling approximation}
In the following, we switch to the collective variables  $\omega_\Sigma = \omega+\omega'$ and $\omega_\Delta = \omega-\omega'$. This is motivated by the forthcoming Markovian approximation and the action of the FrEnGATE which as shown in Sec.~\ref{Subsection ShaTiFrEnGa},  effectively reshapes the reference axes of the two-photon distribution from the individual frequencies $\omega$ and $\omega'$ to the collective frequencies variables $\omega_\Sigma$ and $\omega_\Delta$. The expressions of the probability amplitudes $C_{2X}(t)$ and $C^{\sigma'\sigma} _{\mu'\mu}(\omega_\Sigma,\omega_\Delta;t)$ are related to the memory function 
\begin{equation}
    K(\tau) \equiv \int_{\mathbb R}d\omega_\Sigma\; e^{-i\omega_\Sigma\tau}\left( \frac{1}{2}\sum_{\mu',\mu\in\{\pm\}}\int_{\mathbb R} d\Delta\; |g^{\mu' \mu}(\omega_\Sigma,\omega_\Delta)|^2 \right) 
\end{equation}

of the waveguide seen as a bath which the QD is coupled to.  Assuming that the coupling strength is negligible compared to the biexcitonic energy $\omega_{2X}$, the probability amplitude $C_{2X}(\tau)$ can be approximated as $C_{2X}(\tau) = e^{-i\omega_{2X}\tau} S_{2X}(\tau)$. The term $e^{-i\omega_{2X}\tau}$ represents the rapid oscillation due to the biexcitonic energy and $S_{2X}(\tau)$ is a slowly varying function of time. This is the weak coupling approximation. Further, supposing that the coupling terms $g^{\mu'\mu}(\omega_\Sigma, \omega_\Delta)$ are almost constant with respect to the collective frequency $\omega_\Sigma$, the memory kernel $K(\tau)$ can be taken as sharply peaked at $\tau=0$. This implies that the bath does not retain information about the QD's previous states. This is the Markovian approximation. Within these two approximations, the scattered two-photon probability amplitude can be computed as 

\begin{widetext}
\begin{equation}
\label{Equation general scattered output states}
\begin{split}
   C^{\sigma'\sigma} _{\mu'\mu}(\omega_\Sigma,\omega_\Delta;t_1) &= e^{-i\omega_\Sigma(t_1-t_0)}\bigg[C^{\theta'\theta} _{\nu'\nu}(\omega_\Sigma,\omega_\Delta;t_0)\delta_{\sigma'\theta'}\delta_{\mu'\nu'}\delta_{ \sigma \theta }\delta_{\mu \nu} \\
   &-\frac{\pi\left(g^{\mu'\mu}(\omega_\Sigma,\omega_\Delta)\right)^*}{\frac{\Gamma}{2}+i(\omega_{2X}-\omega_\Sigma)}\sum_{\nu',\nu\in\{\pm\} }\int_{\mathbb R}d\omega_\Delta'\;g^{\nu'\nu}(\omega_\Sigma, \omega_\Delta')C _{\mu'\mu}(\omega_\Sigma,\omega_\Delta;t_0) \bigg],
\end{split}
\end{equation}
\end{widetext}
where $\Gamma$ labels the QD's decay rate and $t_0$ and $t_1$ the scattered times fulfilling $(t_1-t_0)\gg 1/\Gamma$ for the QD to have fully decayed by the time the two-photon output state is formally reached. The polarization degrees of freedom indices are omitted whenever there is a two-photon coupling term as it implicitly dictates them due to the selection rules. The two-photon output state consists of two contributions: one term represents photons that did not interact with the QD and are only time-shifted, while the other term corresponds to photons that interacted with the QD, as indicated by the coupling terms.
\begin{figure*}
    \centering
    \includegraphics[width=1\linewidth]{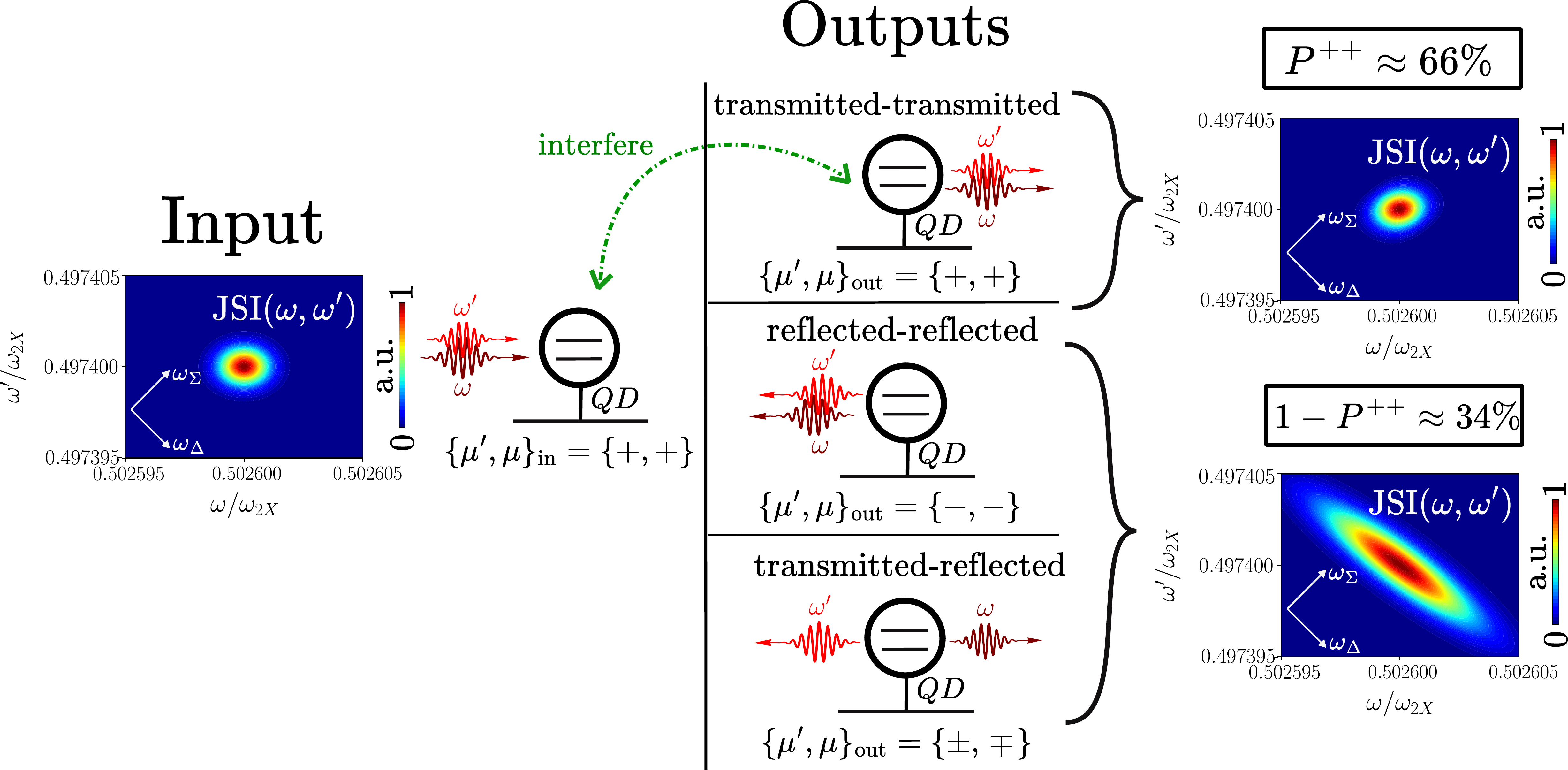}
    \caption{Scattering channels for the two input photons propagating rightward, \text{i.e.}. $\{\mu',\mu\}_{\text{in}} = \{+,+\}$, with an initial Gaussian joint spectral amplitude (left panel) along both collective variables $\omega_\Sigma = \omega+\omega'$ and $\omega_{\Delta} = \omega-\omega'$ of width $\alpha = 10^{-6}\omega_{2X}$ and centers $\omega_e =0.5026\omega_{2X}$, $\omega_b=0.4974\omega_{2X}$.  The coupling term is taken as Gaussian of width $\beta=4\alpha$. The decay rate is chosen as $\Gamma = 10^{-5}\omega_{2X}\gg\alpha$. There are two kinds of outputs (right panels), the one for which the output photons' auxiliary modes are identical to that of the input (upper right panel), \text{i.e.}., $\{\mu',\mu\}_{\text{in}} = \{\mu',\mu\}_{\text{out}}$ and those for which they are not (lower right panel), \text{i.e.}., $\{\mu',\mu\}_{\text{in}} \neq \{\mu',\mu\}_{\text{out}}$ . In the first case, of probability $P^{++}\approx 66$\% for the chosen physical parameters, the input photons can interfere with the entangled photons scattered out by the QD. This results in an output two-photon distribution that is weakly modified with respect to the input two-photon distribution. In the second case $\{\mu',\mu\}_{\text{in}} \neq \{\mu',\mu\}_{\text{out}}$, which occurs with probability $1-P^{++}\approx 34$\% for the chosen physical parameters, only the entangled photons scattered out by the QD contribute to the two-photon distribution. This produces an output two-photon distribution that is reshaped along both collective variables. Along $\omega_\Sigma$, the JSA is filtered by the Lorentzian profile originating from the Markovian approximation. The two-photon coupling term shapes the JSA along the $\omega_\Delta$ collective variable. The exact values of the JSIs have been renormalized for readability. They are presented in arbitrary units (a.u.) on a linear scale.}
    \label{Figure scattering channels}
\end{figure*}
\subsection{Shaping frequency entangling gate }
\label{Subsection ShaTiFrEnGa}

\subsubsection{Gaussian input and coupling} 
\label{Subsubsection Gaussian input and coupling}
In this section, we outline how to adjust the system's physical parameters to ensure that the scattered output states, described by Eq.~\eqref{Equation general scattered output states}, exhibit frequency correlations. More specifically, we aim at recovering Gaussian distributions along both collective variable $\omega_\Sigma$ and $\omega_\Delta$. Shaping the distribution this way
is mainly motivated by the fact that Gaussianity is well characterized for several quantum information protocols~\cite{ContextphotonicCVQIweedbrook2012gaussian,GaussianQIlaudenbach2018continuous}. Let us consider a Gaussian input two-photon state for rightward incoming photons with polarizations $R$ and $L$ and center frequencies $\omega_e$ and $\omega_b$
\begin{equation}
\label{Equation Gaussian input two-photon distribution}
\begin{split}
    C^{\theta'\theta}_{\nu'\nu}(\omega_\Sigma,\omega_\Delta;t_0) &= \frac{\delta_{\theta R}\delta_{\nu +}\delta_{\theta' L}\delta_{\nu' +}}{\sqrt{2\pi\alpha ^2}}\\
    &\times e^{-\frac{\left[\omega_\Sigma- (\omega_e+\omega_b)\right]^2}{4\alpha^2}}e^{-\frac{\left[\omega_\Delta-(\omega_e-\omega_b)\right]^2}{4\alpha^2}},
\end{split}
\end{equation}
normalized to one with $C_{2X}(t_0)=0$. This distribution is isotropic in the two-frequency plane given that the standard deviations along the two collective variables are identical and equal to $\alpha$. Let us further assume in the same line as~\cite{ArticleracineSimonealushi2023waveguide} that the coupling terms can be engineered as discussed in Sec.~\ref{Subsubsection two-photon coupling term} to a Gaussian of the form 
\begin{equation}
\label{Equation Gaussian two-photon coupling term}
    g^{\mu'\mu}(\omega_\Delta) = \sqrt{\frac{\gamma^{\mu'\mu}}{\pi}}\left(\frac{1}{2\pi\beta^2}e^{-\frac{\left(\omega_\Delta-(\omega_e-\omega_b)\right)^2}{\beta^2}}\right)^{1/4},
\end{equation}
where $\gamma^{\mu'\mu}$ is an emission rate taken isotropic that is $\gamma^{\mu'\mu}=\Gamma/4$. This coupling-term is therefore assumed to be independent of the sum of frequencies $\omega_\Sigma$ which aligns with the Markovian approximation previously applied. Based on the structure of the two-photon coupling term Eq.~\eqref{Equation two-photon coupling term after assuming same strength for each path}, we formally discuss in \mohamed{\text{Supplement 1, Section VI}} the required form of the propagation mode $u(\omega)$ across the two photons' bandwidths to achieve the desired Gaussian coupling Eq.~\eqref{Equation Gaussian two-photon coupling term} using a standard numerical optimization method. More advanced algorithms such as simulated annealing and softwares like COMSOL can be used to refine the waveguide design~\cite{Waveguideengineeringpressl2018semi} and achieve the desired mode profile. Putting the input Gaussian two-photon distribution Eq.~\eqref{Equation Gaussian input two-photon distribution} and the Gaussian two-photon coupling term Eq.~\eqref{Equation Gaussian two-photon coupling term} together in the expression of the scattered output states for $\{\mu',\mu\} = \{+,+\}$
\begin{widetext}
\begin{equation}
\label{Equation Scattered output with same auxiliary degrees of freedom}
    \begin{split}
        C_{++}^{LR}(\omega_\Sigma,\omega_\Delta;t_1) &= e^{-i\omega_\Sigma(t_1-t_0)}\bigg[C_{++}^{LR}(\omega_\Sigma,\omega_\Delta;t_0) \\
        &-
  \sqrt{\frac{1}{16\pi(\alpha^2+\beta^2)}}\frac{\Gamma }{\frac{\Gamma}{2}+i(\omega_{2X}-\omega_\Sigma)}e^{-\frac{\left[\omega_\Sigma- (\omega_e+\omega_b)\right]^2}{4\alpha^2}}e^{-\frac{\left(\omega_\Delta-(\omega_e-\omega_b)\right)^2}{4\beta^2}}\bigg] 
    \end{split}
\end{equation}
and for $\{\mu',\mu\} \neq \{+,+\}$
\begin{equation}
\label{Equation Scattered output without same auxiliary degrees of freedom}
\begin{split}
    C_{\mu'\mu}^{\sigma'\sigma}(\omega_\Sigma,\omega_\Delta;t_1) =-e^{-i\omega_\Sigma(t_1-t_0)}\sqrt{\frac{1}{16\pi(\alpha^2+\beta^2)}}\frac{\Gamma }{\frac{\Gamma}{2}+i(\omega_{2X}-\omega_\Sigma)} 
     e^{-\frac{\left[\omega_\Sigma- (\omega_e+\omega_b)\right]^2}{4\alpha^2}}e^{-\frac{\left(\omega_\Delta-(\omega_e-\omega_b)\right)^2}{4\beta^2}},
\end{split}
\end{equation}
\end{widetext}
where the polarizations $\sigma$ and $\sigma'$ match the optical selection rules for each direction of propagation channel.
 An example of the corresponding squared moduli -- or JSI -- are plotted in Fig.~\ref{Figure scattering channels}. It appears that while the outputs Eq.~\eqref{Equation Scattered output without same auxiliary degrees of freedom} exhibit frequency correlations or anticorrelations depending on the ratios of the different widths $\Gamma,\alpha$ and $\beta$, the output Eq.~\eqref{Equation Scattered output with same auxiliary degrees of freedom} displays a two-photon distribution similar to the input although slightly distorted along the two collective variables $\omega_\Sigma$ and $\omega_\Delta$. The latter can be interpreted as an interference pattern arising from the interaction of the photons that did not interact with the QD and those which did. This interference phenomenon does not manifest when considering photons with different auxiliary degrees of freedom as the input photons. The direction of propagation and polarization hence define a coherence condition for the intereferences to manifest in the present scattering derivation. 

\subsubsection{Photon-bandwidth-limiting filtering}
\begin{figure}
    \centering
        \includegraphics[width=\linewidth]{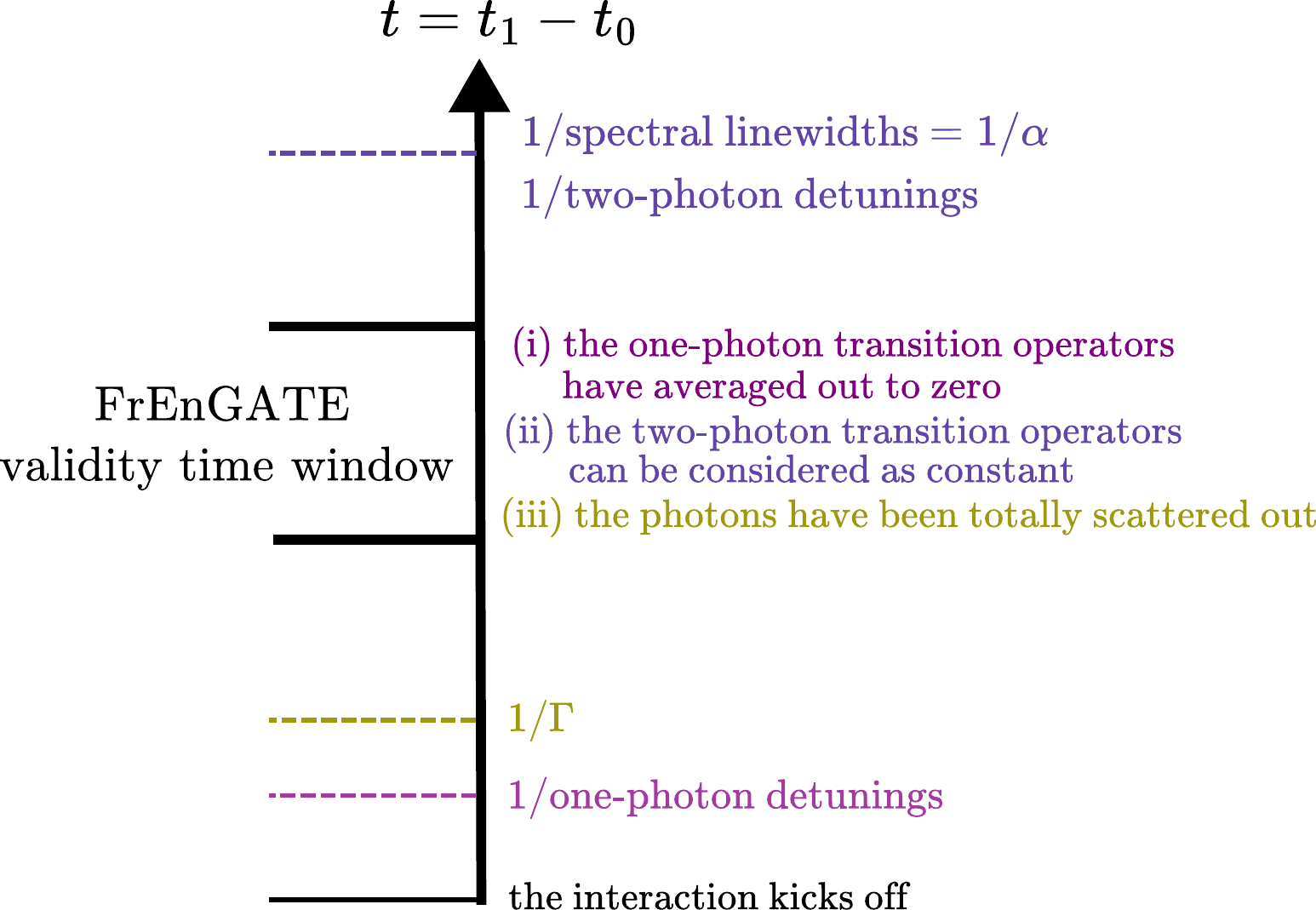}
        \caption{Time-scale constraints for the FrEnGATE. The adiabatic elimination of one-photon transitions requires that the interaction time be much longer than the inverse of the one-photon detunings, ensuring that the corresponding joint operators average out to zero and can be neglected (condition (i)). To treat the two-photon joint operators as effectively time-independent during the scattering process, the time must also remain much shorter than the inverse of the two-photon detunings and the spectral linewidths (condition (ii)). Finally, for the scattering theory approach to be valid, the time must exceed $1/\Gamma$ ensuring that the QD has fully decayed to its ground state by the end of the interaction (condition (iii)).}
        \label{Figure time scales}
\end{figure}

From the scattered output amplitudes' equations~\eqref{Equation Scattered output with same auxiliary degrees of freedom} and~\eqref{Equation Scattered output without same auxiliary degrees of freedom}, it stands out that the term corresponding to the photons that have interacted with the QD is composed of a Gaussian dependency along the difference of frequencies $\omega_\Delta$ dictated by the Gaussian form of the coupling term Eq.~\eqref{Equation Gaussian two-photon coupling term} and two competing dependencies along the sum of frequencies $\omega_\Sigma$ -- a Gaussian one imposed by the input two-photon distribution and a Lorentzian one enforced by the Markovian approximation. The latter acts as a frequency filter preventing photons to be emitted by the QD if they do not match the two-photon resonance condition $\omega_\Sigma= \omega_{2X}$ -- up to the decay rate $\Gamma$. Nonetheless, this resonance condition does not impose any constraint on the difference of frequencies $\omega_\Delta$. The Gaussian two-photon input along $\omega_\Sigma$ on the other hand forces the scattered output photons to have the same frequency distribution as the input thus preventing sums of frequencies to be generated by the QD. These two filterings are centered respectively at $\omega_{2X}$ and $\omega_e+\omega_b$ with the latter being close to $\omega_{2X}$ with respect to the one-photon detunings under the adiabatic elimination we performed in Sec.~\ref{subsubsection Adiabatic elimination}. The Lorentzian and Gaussian filterings can be considered centered at the same frequency provided $|\omega_{2X}-(\omega_e+\omega_b)| \ll \Gamma,\alpha$. Therefore, in the limit regime for which $\Gamma \gg \alpha$, the frequency filtering along the collective variable $\omega_\Sigma$ is governed by the input two-photon distribution, resulting in a non-interefered output Eq.~\eqref{Equation Scattered output without same auxiliary degrees of freedom} two-photon JSI now reading 
\begin{equation}
\begin{split}
\label{Equation output JSI without interferences}
    |C_{\mu'\mu}^{\sigma'\sigma}(\omega_\Sigma,\omega_\Delta;t_1)|^2 &= \frac{1}{4\pi(\alpha^2+\beta^2)} \\ &\times e^{-\frac{\left(\omega_\Delta-(\omega_e-\omega_b) \right)^2}{2\beta^2}}e^{-\frac{\left[\omega_\Sigma- (\omega_e+\omega_b)\right]^2}{2\alpha^2}},
\end{split}
\end{equation}
for $\{\mu',\mu\} \neq \{+,+\}$. This two-photon output distribution Eq.~\eqref{Equation output JSI without interferences} is Gaussian in both collective variables $\omega_\Sigma$ and $\omega_\Delta$. Remarkably, and unlike Alushi \textit{et al.}, the present photonic microscopic model actually requires this $\Gamma \gg \alpha$ condition to be met. Indeed, for the scattering calculations, one considered $(t_1-t_0)\Gamma\gg 1$. In contrast, for the adiabatic elimination resulting in the effective quadratic Hamiltonian Eq.~\eqref{Equation effective Hamiltonian after adiabatic elimination} to hold, the interaction time was assumed to be such that $(t_1-t_0)|\overline {\delta_e}| \gg 1$, $(t_1-t_0)|\overline{\delta_b}| \gg 1$ and $(t_1-t_0)\alpha \ll 1$ with $\overline {\delta_e}$ and $\overline {\delta_b}$ the mean value of the one-photon transition detunings previously discussed. This ensures that the one-photon transition operators adiabatically average out while the two-photon transition operators remain constant in time. The $(t_1-t_0)\Gamma\gg 1$ and $(t_1-t_0)\alpha \ll 1$ conditions inevitably entail $\Gamma \gg \alpha$. The different time scales are sketched in Fig.~\ref{Figure time scales}. \mohamed{Neglecting the broadband-induced, Stark shift permits relaxing the strict interaction-time constraint $(t_1-t_0)\alpha\ll 1$. However, the filtering hierarchy $\Gamma\gg\alpha$ remains necessary to ensure the two-photon output is Gaussian in both collective variables.}
At the telecom band, the decay rate $\Gamma$ is typically of the order of a few GHz~\cite{QDdecayratekettler2016neutral,QDdecayratetelecomzeuner2021demand} which would require single-photon sources' linewidths in the telecom band to be narrower than a few hundred MHz. Although telecom band single-photon sources' linewidths are generally broader~\cite{Telecomsinglephotonsourcetakemoto2007optical,Telecomsinglephotonsourcebelhassen2018chip,Telecomsinglephotonsourceyu2023telecom,Telecomsinglephotonsourcecraddock2023high}, several studies have successfully demonstrated devices with narrower linewidths~\cite{Telecomsinglephotonsourcemuller2018quantum,ContextphotonicTFGridstateshenry2024parallelization,QDFWMSinglephotonsourcepsiquantum2025manufacturable}. In that regard, the time $(t_1-t_0)$ should typically be within the $0.1-1$ ns range. Furthermore, the spontaneous emission decay rate may be significantly enhanced by increasing the Purcell factor. The described scattering process thus operates as a frequency reshaping performed by the light-matter interaction coupling term which reshapes the two-photon distribution by modifying its width along the difference of frequencies $\omega_\Delta$ from $\alpha$ to $\beta$. As it will be discussed shortly, the ratio $\beta/\alpha$ is directly related to the resulting frequency entanglement. Namely, the two photons are maximally frequency-correlated for $\beta/\alpha \rightarrow +\infty$ (see Fig.~\ref{Figure trade-off probability of success VS quality of entanglement}, panel (c)) and frequency-anticorrelated for $\beta/\alpha \rightarrow 0$ (see Fig.~\ref{Figure trade-off probability of success VS quality of entanglement}, panel (a)). There are no frequency correlations and the two-photon output distribution is separable when $\beta=\alpha$ (see Fig.~\ref{Figure trade-off probability of success VS quality of entanglement}, panel (b)). 

\section{Entanglement and efficiency investigation of the frequency-entanglement device}
\label{Sec. Entanglement, fidelity and efficiency investigation}
In this section, the entanglement properties are quantitatively analyzed revealing a trade-off between the generation success probability and the entanglement quality. By employing our overall methodology, the FrEnGATE is then shown to apply for the generation of frequency-entangled frequency qudit states.

\subsection{Efficiency of the frequency-entanglement generation}
As previously discussed in Sec.~\ref{Subsubsection Gaussian input and coupling}, out of the four scattering outputs depending on the photons' directions of propagation, the one for which the output photons' auxiliary degrees of freedom match that of the input photons exhibits interference patterns that make it unsuitable for the frequency-entanglement generation. Consequently, the efficiency of the entanglement generation is directly given by the probability $P^{\text{success}}$ of having the photons scattered out with different auxiliary degrees of freedom as the input photons. Here for instance, since we considered input photons incoming moving to the right, \text{i.e.}, $\{\mu',\mu\} =\{+,+\}$,  $P^{\text{success}} = 1-P^{++} = P^{--}+P^{+-}+P^{-+}$ where 
\begin{equation}
\label{Equation general success probability formula}
   P^{\mu'\mu} = \frac{\int_{\mathbb R^2}d\omega_\Sigma d\omega_\Delta\;|C_{\omega_\Sigma \omega_\Delta}^{\sigma'\mu'\sigma\mu}(t_1)|^2}{\int_{\mathbb R^2}d\omega_\Sigma d\omega_\Delta\;|C_{++}^{LR}(\omega_\Sigma,\omega_\Delta;t_0)|^2}.
\end{equation}
The previously discussed frequency filtering imposed by the Lorentzian emission profile explains why this probability is maximized when $\Gamma \gg \alpha$. For an isotropic process $\gamma^{\mu'\mu} = \Gamma/4$, in the $\Gamma\gg\alpha$ regime and for an input two-photon amplitude normalized to one, the probability of success can be expressed as a function of the ratio $\beta/\alpha$ using Eq.~\eqref{Equation output JSI without interferences} 
\begin{equation}
\label{Equation probability of success}
    P^{\text{success}} = \frac{3\beta/\alpha}{2(1+\beta^2/\alpha^2)}.
\end{equation}
 The success probability of outputting a suitable two-photon distribution -- that is to say without interferences -- is at most equal to 75\% for $\alpha=\beta$. Nevertheless, $\alpha=\beta$ corresponds to a separable two-photon JSI Eq.~\eqref{Equation output JSI without interferences}. Indeed, the closer the ratio $\beta/\alpha$ is to one, the less frequency-entangled the photons are. They are frequency-correlated for $\beta\gg \alpha$ and frequency-anticorrelated for $\alpha\gg\beta$. Subsequently, there is a trade-off between the success probability of producing frequency-entangled photons and the amount of entanglement that is generated. All appears as if the two-photon output distribution opposed the reshaping along $\omega_\Delta$ to ensure that no frequencies along this collective variable are created -- as it does along $\omega_\Sigma$. This probability of success is achieved without requiring off-line measurements of auxiliary photons, unlike photon entanglement protocols based on the KLM scheme~\cite{ContextprobabilisticpostselectionKLMknill2001scheme,ContextBlablaphotonicQIObrieno2009photonic}. Moreover, by exploiting chiral waveguides~\cite{chiralQOpticslodahl2017chiral} to preferentially channel non‑interfering scattering events, the success probability can be increased even further. Note that one could have utilized the QD's spontaneous emission by first pumping it to its biexcetionic state and then let it spontaneously decay. Provided that the photons are collected within the time window in which the adiabatic elimination regime is valid, frequency-entanglement is generated with a Lorentzian dependence of width the decay rate $\Gamma$ along the sum of frequencies $\omega_\Sigma = \omega+\omega'$ and a Gaussian dependence of width $\beta$ along the difference of frequencies $\omega_{\Delta} = \omega-\omega'$. Nevertheless, since there would be no initial Gaussian dependence to draw from, the distribution along $\omega_\Sigma$ will always be Lorentzian and not Gaussian in the Markovian approximation previously discussed. \\

\subsection{Schmidt decomposition of the output spectral two-photon state}
\begin{figure*}
    \centering
        \includegraphics[width=\linewidth]{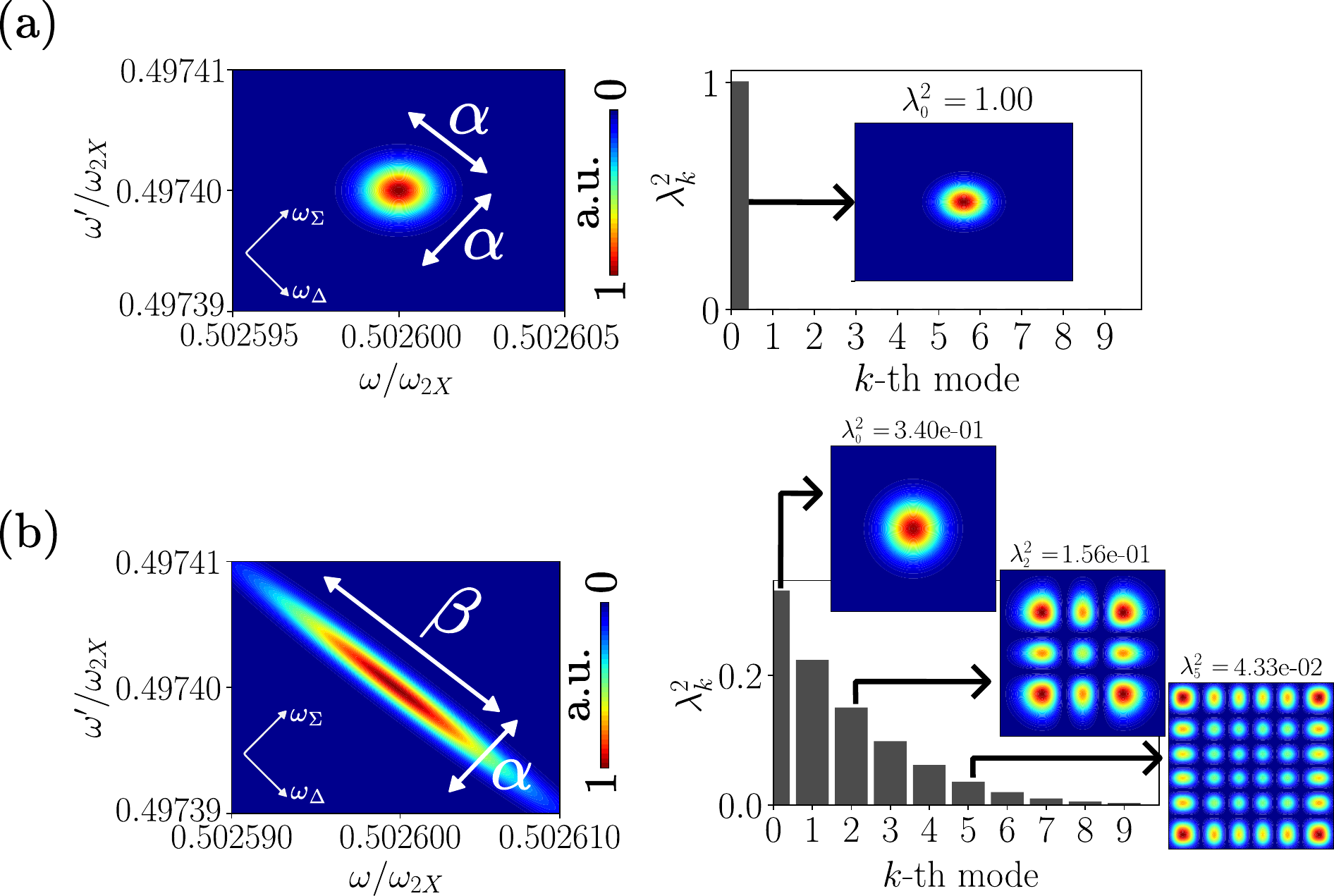}
        \caption{(a) Modes structure for the input two-photon distribution Gaussian, in both $\omega_\Sigma = \omega+\omega'$ and $\omega_\Delta=\omega-\omega'$ of width $\alpha=10^{-6}\omega_{2X}$ and centers $\omega_e =0.5026\omega_{2X}$, $\omega_b=0.4974\omega_{2X}$. The normalized entanglement entropy and the Schmidt number are computed to be $S_{\text{entropy}}=0.00$ and $K=1.00$ respectively. (b) Modes structure for the output two-photon distribution Gaussian in both $\omega_\Sigma = \omega+\omega'$ and $\omega_\Delta=\omega-\omega'$ of width $\alpha=10^{-6}\omega_{2X}$, centers $\omega_e =0.5026\omega_{2X}$, $\omega_b=0.4974\omega_{2X}$ and $\beta=10\alpha$, respectively. The decay rate is taken as $\Gamma = 10^{-5}\omega_{2X}\gg\alpha$. The normalized entanglement entropy and the Schmidt number are computed to be $S_{\text{entropy}}=0.39$ and $K=4.72$ respectively. The first, third and fifth Schmidt modes' JSIs have been plotted. The exact values of the JSIs have been renormalized for readability. They are presented in arbitrary units (a.u.) on the same linear scale in $\omega/\omega_{2X}$ and $\omega'/\omega_{2X}$ as the (b) left panel. We formally indicate the standard deviations $\alpha$ and $\beta$ although it should be noted that the full widths at half maximum (FWHMs) are rigorously equal to 2$\alpha\sqrt{2\ln2}$ and 2$\beta\sqrt{2\ln2}$, respectively. }
        \label{Figure Schmidt decomposition and number of modes for input}
\end{figure*}

\begin{figure*}
    \centering
    \includegraphics[width=0.7\linewidth]{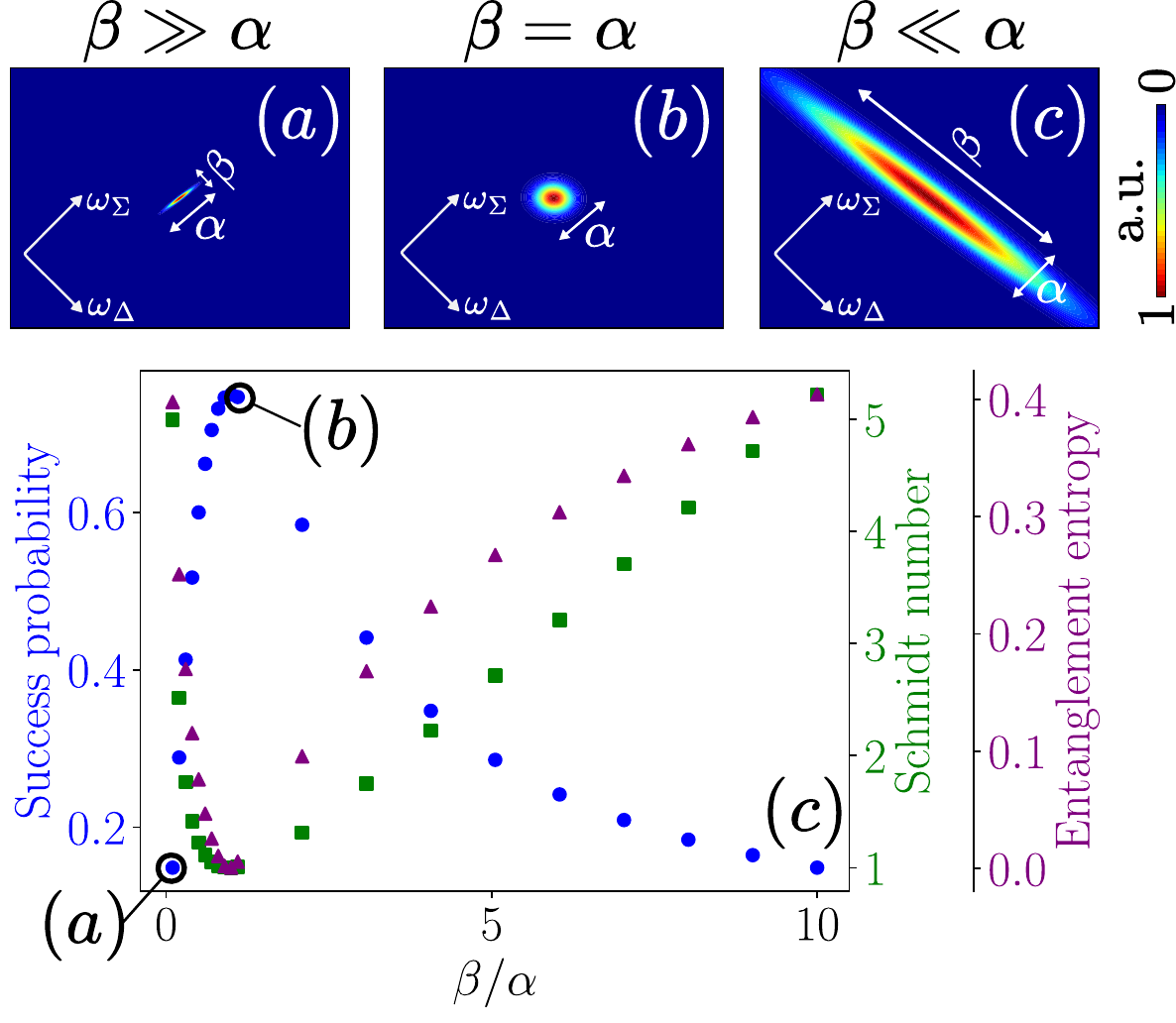}
    \caption{Trade-off between the probability of successfully outputting a non-interfered entangled two-photon distribution and the quality of the entanglement measured with the Schmidt number and normalized entanglement entropy with respect to the ratio $\beta/\alpha$ of the Gaussian coupling width $\beta$ to the two-photon input width $\alpha$. The normalized entanglement entropy is given with respect to its maximum value. The Gaussian coupling width $\beta$ varies from 0.1$\alpha$ to 10$\alpha$ with $\alpha = 10^{-6}\omega_{2X}$. The decay rate is taken as $\Gamma = 10^{-5}\omega_{2X}\gg\alpha$. The illustrative JSIs (panels (a), (b), (c)) are all on the same linear scale in $\omega/\omega_{2X}$ and $\omega'/\omega_{2X}$. The panel (a) depicts a regime where $\beta \ll \alpha$ (frequency-anti-correlated photons), the panel (c) where $\beta \gg \alpha$ (frequency-correlated photons) and the panel (b) where $\beta=\alpha$ (no frequency correlations). We formally indicate the standard deviations $\alpha$ and $\beta$ although it should be noted that the FWHMs are rigorously equal to 2$\alpha\sqrt{2\ln2}$ and 2$\beta\sqrt{2\ln2}$ respectively. }
    \label{Figure trade-off probability of success VS quality of entanglement}
\end{figure*}

In this subsection, we evaluate the entanglement of the two-photon output state Eq.~\eqref{Equation Scattered output without same auxiliary degrees of freedom}, postselected on scattering events where the auxiliary degrees of freedom differ from those of the input state. To proceed, we perform a Schmidt decomposition for continuous variables bipartite systems~\cite{CVSchmidtdecompositionParkerPhysRevA.61.032305,CVSchmidtdecompositionlucaslamata2005dealing,bogdanov_schmidt_2006}. Further details are provided in \mohamed{\text{Supplement 1, Section V}}. The probability amplitude $C_{\mu'\mu}^{\sigma'\sigma }(\omega',\omega;t_1)$ of the pure renormalized output state
\begin{equation}
\label{Equation two-photon scattered output state before Schmidt decomposition}
    \ket{\psi_{\mu'\mu}^{\sigma'\sigma}(t_1)} = \int_{\mathbb R^2}d\omega'd\omega\; C_{\mu'\mu}^{\sigma'\sigma }(\omega',\omega;t_1)\hat a_{\sigma'\mu'}^\dagger(\omega')\hat a_{\sigma \mu}^\dagger(\omega)\ket{\text{vac}}
\end{equation}
is expressed by means of a discrete expansion 
\begin{equation}
    \ket{\psi_{\mu'\mu}^{\sigma'\sigma}(t_1)} = \sum_{k=1}^{l} \lambda_k^{\sigma'\mu'\sigma\mu} \hat b^{(2)\dagger}_{\sigma'\mu'}(k)\hat b^{(1)\dagger}_{\sigma\mu}(k)\ket{\text{vac}},
\end{equation}
where we truncated the number of modes to $l$ to perform the numerical singular-value decomposition. 
The operators $\hat b^{(1)\dagger}_{\sigma\mu}(k)$ and $\hat b^{(2)\dagger}_{\sigma'\mu'}(k)$ are bosonic creation operators which can be expressed as 
\begin{equation}
    \begin{split}
        &\hat b^{(1)\dagger}_{\sigma\mu}(k) = \int_{\mathbb R}d\omega\;\Theta_k^{(1)}(\omega)\hat a_{\sigma\mu}^\dagger(\omega) \\ 
        &\hat b^{(2)\dagger}_{\sigma'\mu'}(k) = \int_{\mathbb R}d\omega'\;\Theta_k^{(2)}(\omega')\hat a_{\sigma'\mu'}^\dagger(\omega').
    \end{split}
\end{equation}
These operators create photons in the so-called Schmidt modes labelled by the $k$ index. The functions $\Theta_k^{(1)}(\omega)$ and $\Theta_k^{(2)}(\omega')$ are linear combinations of a set of orthogonal functions. These orthogonal functions are here chosen to be the Hermite-Gauss orthogonal functions because of the Gaussianity of the two-photon distribution.  The $\lambda_k^{\sigma'\mu'\sigma\mu}$ Schmidt coefficients are real, non-negative, unique up to reordering and obey $\sum_{k=1}^{\text{min}(m_0,n_0)}\left(\lambda_k^{\sigma'\mu'\sigma\mu}\right)^2=1$ provided $C_{\mu'\mu}^{\sigma'\sigma }(\omega',\omega;t_1)$ is normalized to one.  From the Schmidt coefficients, two measures of entanglement can be computed~\cite{ContextphotonicCVquadraturesbraunstein2005quantum}: (i) the entanglement entropy $S_{\text{entropy}}^{\sigma'\mu'\sigma\mu} = -\sum_{k=1}^{\text{min}(m_0,n_0)}(\lambda_k^{\sigma'\mu'\sigma\mu} )^2\log((\lambda_k^{\sigma'\mu'\sigma\mu} )^2)$ -- corresponding to the von Neumann entropy of the reduced density matrix obtained by tracing out either one of the photons -- (ii) the Schmidt number $K^{\sigma'\mu'\sigma\mu} = 1/\sum_{k=1}^{\text{min}(m_0,n_0)}(\lambda_k^{\sigma'\mu'\sigma\mu} )^4$. Namely, the Schmidt number gives the effective number of modes in the Schmidt decomposition. In the following, we drop the Schmidt number and entropy of entanglement labeling for readability. In Fig.~\ref{Figure Schmidt decomposition and number of modes for input}, we plot the input and output two-photon joint-spectral intensity (JSI) and the modes structure for $\beta=10\alpha$ and in the $\Gamma\gg \alpha$ regime. For the input, since the two-photon distribution is separable in the individual frequencies $\omega$ and $\omega'$, the entanglement entropy is equal to 0 and the Schmidt number to 1. On the other hand, the output two-photon distribution is non-separable in $\omega$ and $\omega'$ as long as $\beta\neq\alpha$. For $\beta=10\alpha$, the Schmidt number is approximately equal to 5 and the entanglement entropy is close to 36\% of its maximum value.
In Fig.~\ref{Figure trade-off probability of success VS quality of entanglement}, we display for ten values of $\beta/\alpha$ ranging from 0.1 to 10 the evolution of the entanglement generation success probability Eq.~\eqref{Equation probability of success}, the Schmidt number $K$ and the normalized entanglement entropy $S_{\text{entropy}}$. Note that the computed Schmidt number evolution follows its analytical expression for a doubly Gaussian JSA~\cite{Schmidtnumberwrtratiozielnicki2018joint}. As previously discussed, a trade-off can be observed between the quality of the entanglement -- estimated by the Schmidt number and the entanglement entropy -- and success probability of the entanglement generation. \mohamed{Naturally, the latter is evaluated within the adiabatic elimination regime detailed in Sec.II.B, where one-photon transition channels are strongly suppressed. In realistic experimental conditions where this regime may not be perfectly realized, some reduction in entanglement generation is expected due to residual one-photon transition events. Nonetheless, as shown by the numerical simulations in \text{Supplement 1.III.C}, the effective two-photon Hamiltonian Eq.~(13-14) provides an excellent approximation of the dynamics within the adiabatic elimination regime, indicating that the Schmidt number and entropy of entanglement reported correctly reflect the entanglement properties of the frequency-entangled photon pair.} \\
An illustrative summary of typical physical values related to the output two-photon distributions is displayed in Tab.~\ref{Tab illustrative table for two values of beta alpha} for two different values of the $\beta/\alpha$ ratio. This section shows how the FrEnGATE can produce Gaussian frequency-entangled biphoton states with Gaussian profiles and with high quality and efficiency. The resulting biphoton state can exhibit a Schmidt number of approximately 5, a normalized entanglement entropy of 0.4, and 98\% ellipticity, with an entanglement generation efficiency reaching up to 15\%-- significantly exceeding what is typically achievable with SPDC or SFWM. 
\begin{table}
\centering
\begin{tabular}{|l|l|l|}
\hline
                                                                                       & \multicolumn{1}{c|}{$\beta=10\alpha$} & $\beta=30\alpha$               \\ \hline
\begin{tabular}[c]{@{}l@{}}Entanglement generation \\ probability success\end{tabular} & 15\%                                     & 5\%                \\ \hline
Normalized entanglement entropy                                                          & 0.4                                  & 0.6                          \\ \hline
Schmidt number                                                                         & 5                                  & 12                             \\ \hline
Ellipticity                                                                            & 98\%     & 99.8\% \\ \hline
Frequency bandwidth along $\omega_\Sigma$                                              & 2.4 GHz                                & 790 MHz                      \\ \hline
Frequency bandwidth along $\omega_\Delta$                                              & 24 GHz                           & 24 GHz                     \\ \hline
\end{tabular}
\caption{Characteristic physical values related to the output two-photon distribution for two values of the $\beta/\alpha$ ratio. The Gaussian coupling term width is fixed at $\beta = 10^{-5}\omega_{2X} \approx 2\pi\times 3.8$ GHz. These frequency bandwidths along both collective variables can be compared to Le Jeannic \textit{et al.}~\cite{ArticleracineHannale2022dynamical} who found frequency bandwidths of the order of a few GHz.}
\label{Tab illustrative table for two values of beta alpha}
\end{table}
\subsection{\nicolas{FrEnGATE} with frequency qudit states}
\label{Subsection FrEnGATE with frequency qudit states}
\begin{figure*}
    \centering
    \includegraphics[width=1\linewidth]{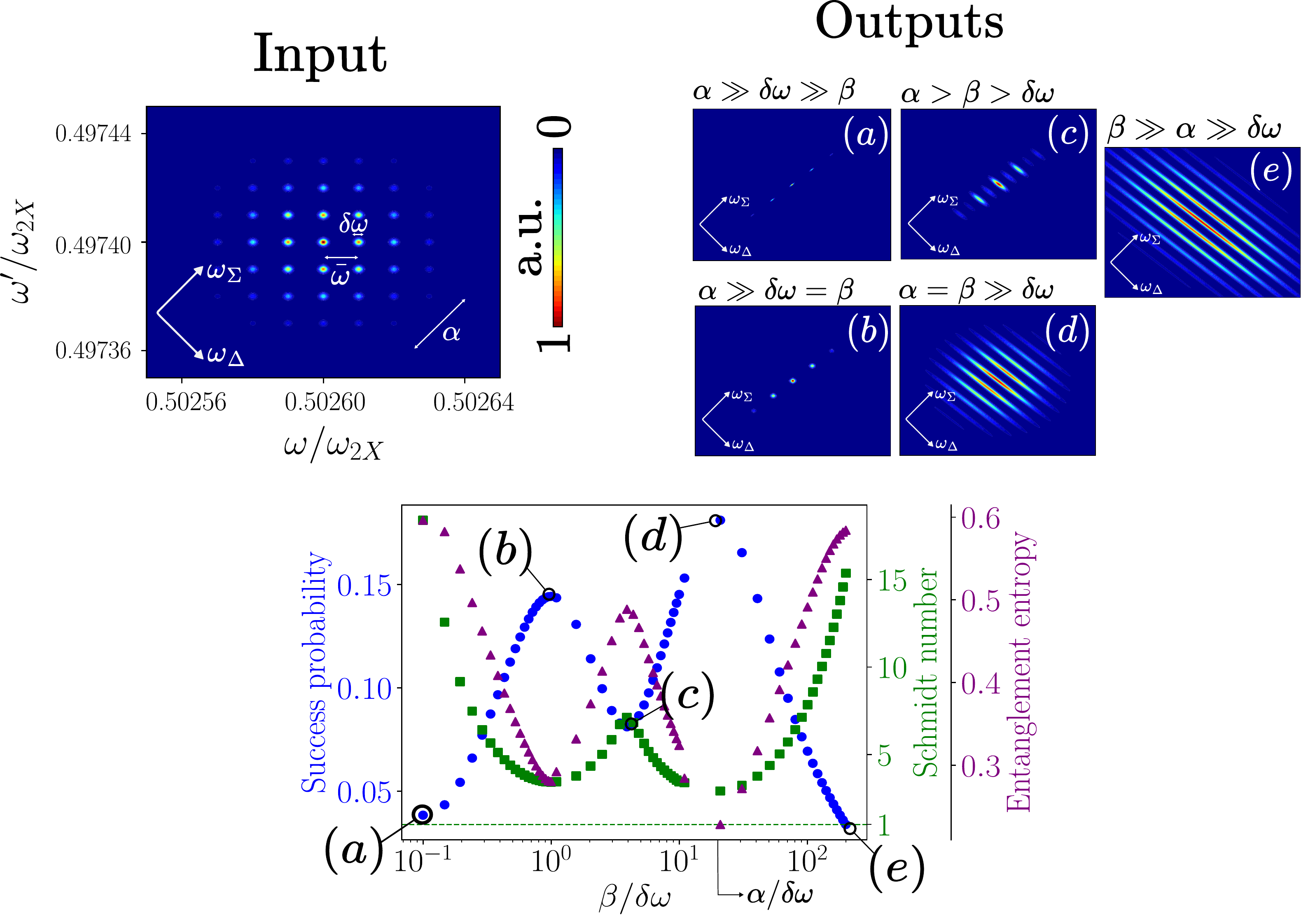}
    \caption{(bottom) Trade-off between the probability of successfully outputting a non-interfered entangled two-photon distribution and the quality of the entanglement measured with the Schmidt number and normalized entanglement entropy with respect to the ratio $\beta/\delta \omega$ of the Gaussian coupling width $\beta$ and $\delta \omega$ the input grid state peaks' width. The $\beta/\delta \omega$-axis is displayed in a logarithmic scale. The Gaussian coupling width $\beta$ varies from 0.1$\delta \omega$ to 45$\delta \omega$ with $\delta \omega = 10^{-6}\omega_{2X}$. The entanglement entropy is given with respect to its maximum value. The Gaussian envelope input is taken as $\alpha = 2\times 10^{-5}\omega_{2X}$ and the FSR as $10^{-5}\omega_{2X}$. The decay rate is taken as $\Gamma = 2\times 10^{-5}\omega_{2X}\gg\alpha$. The illustrative output JSIs panels (a), (b), (c), (d) and (e) (top right corner) and the input JSI (top left corner) are all on the same linear scale in $\omega/\omega_{2X}$ and $\omega'/\omega_{2X}$. The input JSA is separable with a normalized entropy of entanglement equal to 0 and a Schmidt number equal to 1.}
    \label{Figure gridstates success probability VS S and K}
\end{figure*}

In this subsection, we cover the action of the FrEnGATE when the input state is a two-photon frequency qudit states~\cite{ContextphotonicTFquditsstateslu2023frequency,ContextphotonicTFquditstateslu2018quantum,ContextphotonicTFqudtistateskues2017chip,ContextphotonicTFquditreimer2014integrated,ContextphotonicTFquditreimer2016generation,ContextphotonicTFquditxie2015harnessing,ContextphotonicDVquditsfrequencycomnbernhard2013shaping,ContextphotonicTFGridstateshenry2024parallelization,ContextphotonicCVtimefrequencyNicolasGridstates,FSRgridstatesjang2024programmable,FSRgridstateschao2010octave,FSRgridstatesrielander2017frequency}. Such frequency grid states are single-photon states characterized by discrete frequency components arranged in a comb-like structure and modulated by an overall envelope. An isotropic frequency grid state in the $(\omega,\omega')$ plane can be generated starting with the isotropic Gaussian two-photon distribution $C_{++}^{LR}(\omega_\Sigma,\omega_\Delta;t_0)$ (see Eq.~\eqref{Equation Gaussian two-photon coupling term}) of standard deviation $\alpha$ and applying a frequency filter such as an optical cavity. This filtering process operates on the individual frequencies $\omega$ and $\omega'$ shaping the desired grid structure in the frequency domain $C_{++}^{LR}(\omega_\Sigma,\omega_\Delta;t_0) \rightarrow C_{++}^{LR}(\omega_\Sigma,\omega_\Delta;t_0)f_{\text{filter}}(\omega)f_{\text{filter}}(\omega')$ where the filtering function $f_{\text{filter}}$ is assumed to be identical for both frequencies. This filtering function is typically an Airy function that can be approximated as a sum of Gaussians in the high-finesse regime of the optical cavity: $f_{\text{filter}}(\omega) = \sum_{n\in \mathbb Z}T_{n}(\omega)$ where $T_n(\omega) = \exp(-(\omega-n\overline \omega)^2/(2\delta\omega^2))$. This high-finesse regime holds for a free spectral range (FSR) much larger than the width of the peaks, \textit{i.e.} $\overline \omega \gg \delta \omega$. For the frequency peaks to be distinct, the envelope standard deviation $\alpha$ should be larger than the FSR. Standard frequency comb FSRs~\cite{ContextphotonicTFquditsstateslu2023frequency} below 100 GHz are usually of the order of 10-20 GHz~\cite{FSRgridstatesjang2024programmable,ContextphotonicTFGridstateshenry2024parallelization,ContextphotonicCVtimefrequencyNicolasGridstates,ContextphotonicTFquditstateslu2018quantum} but lower values close to 1 GHz~\cite{FSRgridstateschao2010octave,FSRgridstatesrielander2017frequency} have also been reported. Given the adiabatic elimination constraint imposing the decay rate $\Gamma$ to be large with respect to $\alpha$, we take $\overline \omega = 10^{-5}\omega_{2X} \approx 2\pi\times 1$ GHz, $\alpha = 2\times 10^{-5}\omega_{2X} \approx 2\pi \times$ 2 GHz and $\delta \omega = 10^{-6}\omega_{2X} \approx 2\pi \times 100\;\text{MHz}\;\ll \overline \omega$ for the high-finesse regime to hold. This requires $\Gamma$ to be larger than $2\times 2\pi \times 10$ GHz which is a bit higher than the standard biexcitonic decay rates~\cite{QDdecayratekettler2016neutral,QDdecayratetelecomzeuner2021demand} but not too far off.  The frequency-filtered distribution $C_{++}^{LR}(\omega_\Sigma,\omega_\Delta;t_0)f_{\text{filter}}(\omega)f_{\text{filter}}(\omega')$ can be used as the FrEnGATE input. The non-interfered two-photon output distribution can be analytically calculated as 
\begin{widetext}
\begin{equation}
\begin{split}
\label{Equation output of input grid state}
      C_{\mu'\mu}^{\sigma'\sigma }(\omega',\omega;t_1)=-e^{-i\omega_\Sigma(t_1-t_0)}\sqrt{\frac{\delta \omega^2}{16\pi\left(\alpha^2\beta^2+\delta \omega^2(\alpha^2+\beta^2)\right)}}e^{-\frac{[\omega_\Delta-(\omega_e-\omega_b)]^2}{4\beta^2}} \frac{\Gamma}{\frac{\Gamma}{2}+i(\omega_{2X}-\omega_\Sigma)}e^{-\frac{[\omega_\Sigma-(\omega_e+\omega_b)]^2}{4\alpha^2}}\sum_{n,m\in \mathbb Z}f_{nm}(\omega_\Sigma),
\end{split}
\end{equation}
\end{widetext}
for $\{\mu,\mu'\} \neq \{+,+\}$. The computation details are given in \mohamed{\text{Supplement 1, Section VII}}. The comb structure is now along the sum of frequencies $\omega_\Sigma$ with frequency peaks expressed as 
\begin{equation}
    f_{nm}(\omega_\Sigma) = e^{D_{nm}}e^{-\frac{1}{4\delta \omega^2}\left[\omega_\Sigma-\overline \omega(n+m) \right]^2},
\end{equation}
where 
\begin{equation}
    D_{nm} = -\frac{1}{4}\frac{\left(\alpha^2+\beta^2\right)\left((\omega_e-\omega_b)-\overline \omega(n-m) \right)^2}{\alpha^2\beta^2+\delta \omega^2(\alpha^2+\beta^2)}.
\end{equation}
As for Eq.~\eqref{Equation Scattered output without same auxiliary degrees of freedom}, there is a filtering competition along $\omega_\Sigma$ between the input two-photon Gaussian distribution and the Lorentzian emission profile. In the $\Gamma \gg \alpha$ regime enforced by the adiabatic elimination regime and assuming $\Gamma, \alpha \gg |\omega_{2X}-(\omega_e+\omega_b)|$, the two-photon distribution Eq.~\eqref{Equation output of input grid state} corresponds to a frequency comb where frequency peaks are approximately $\overline \omega$ apart and lie along the $\omega_\Sigma$ collective variable, with a Gaussian envelope of standard deviation $\alpha$. The frequency peaks standard deviation $\beta$ along the difference of frequencies $\omega_\Delta$ is enforced by the Gaussian coupling term. Employing the continuous-variable Schmidt decomposition we display in Fig.~\ref{Figure gridstates success probability VS S and K} the normalized entanglement entropy and the Schmidt number along with the probability Eq.~\eqref{Equation general success probability formula} of successfully outputting the non-interfered two-photon state Eq.~\eqref{Equation output of input grid state} as functions of the ratio $\beta/\delta\omega$.
Our analysis reveals five regimes. The two two-photon states with the lowest Schmidt number and entanglement entropy emerge either due to the separability within the peaks (Fig.~\ref{Figure gridstates success probability VS S and K}~(b)) when $\beta = \delta \omega$ or to the apparent separability of the envelope when $\beta = \alpha$ (Fig.~\ref{Figure gridstates success probability VS S and K}~(d)). Although these two-photon states exhibit the lowest Schmidt number and entanglement entropy, neither metric reaches the separable limit -- Schmidt number equal to 1 and entropy of entanglement equal to 0 --, so they remain partially entangled. These partially-entangled two-photon states have the highest success probability, up to approximately 19\% for Fig.~\ref{Figure gridstates success probability VS S and K}~(d).  Second, when $\beta \ll \delta \omega \ll \alpha$ (Fig.~\ref{Figure gridstates success probability VS S and K}~(a)), the two-photon output state is the most frequency-anti-correlated because both the peaks and the envelope are stretched along the $\omega_\Sigma$ variable. For $\beta \gg \alpha \gg \delta \omega$ (Fig.~\ref{Figure gridstates success probability VS S and K}~(e)), the two-photon output state appears as the most frequency-correlated because both the peaks and the envelope are stretched along the $\omega_\Sigma$ variable. In the regime $\beta \gg \alpha \gg \delta \omega$ and for a FSR $\alpha \gtrsim \overline \omega$, one can engineer the initial distribution in such a way for the output state to correspond to a cat state with elliptical lobes. The Fig.~\ref{Figure gridstates success probability VS S and K}~(a) and Fig.~\ref{Figure gridstates success probability VS S and K}~(e) regimes have the lowest success probabilities, dropping to around 1\%. Last, the regime where $\alpha > \beta > \delta \omega$ (Fig.~\ref{Figure gridstates success probability VS S and K}~(c)) is an intermediate regime where the photons are not as frequency-entangled as in the $\beta \ll \delta \omega \ll \alpha$ and $\beta \gg \alpha \gg \delta \omega$ regimes but exhibit higher frequency correlations than the two other regimes (Fig.~\ref{Figure gridstates success probability VS S and K}~(b) and Fig.~\ref{Figure gridstates success probability VS S and K}~(d)). Furthermore, this intermediate regime has an intermediate success probabilities as compared to the other regimes, reaching close to 8\%.  These different regimes show the interplay between the two-photon envelope and frequency peaks' apparent separability. Furthermore, Fig.~\ref{Figure gridstates success probability VS S and K} confirms the trade-off between the generation success probability and the entanglement previously discussed: the higher the Schmidt number and entanglement entropy, the lower the generation success probability.  \\
\mohamed{In Table~\ref{Tab illustrative table of different parameters}, we give a snapshot of the QD-waveguide parameters considered throughout this manuscript. These values can be adapted to different QD/waveguide photonic platforms provided that (i) the adiabatic-elimination conditions are satisfied, and (ii) the hierarchy $\Gamma \gg \alpha$ holds so that the Lorentzian reemission profile does not become the limiting frequency filter.}

\begin{table}
\centering
\begin{tabular}{|l|l|l|}
\hline
                                                                                 Parameters      & \multicolumn{1}{c|}{CV (Sec.~\ref{Subsection ShaTiFrEnGa})} & Qudits (Sec.~\ref{Subsection FrEnGATE with frequency qudit states})               \\ \hline
\begin{tabular}[c]{@{}l@{}}$\omega_X$~\cite{QDTelecomkors2018telecom,Telecomsinglephotonsourcemuller2018quantum,QDSinglephotonSourcesarakawa2020progress} \end{tabular} & $2\pi\times 190\;\text{THz}$                                     & $2\pi\times 190\;\text{THz}$           \\ \hline
$\delta_X$~\cite{QDgeneralgywat2010spins}                                                        & $10^{-2}\omega_X$                                  & $10^{-2}\omega_X$                         \\ \hline
$S$~\cite{QDlodahl2015interfacing,QDFSSfognini2018universal}                                                                        & 0-100 GHz                              & 0-100 GHz                          \\ \hline
$\alpha$~\cite{Telecomsinglephotonsourcetakemoto2007optical,Telecomsinglephotonsourcebelhassen2018chip,Telecomsinglephotonsourceyu2023telecom,Telecomsinglephotonsourcecraddock2023high}                                                                        & $2\pi\times 200\; \text{MHz}$    & $2\pi\times 2 \; \text{GHz}$ \\ \hline
$\Gamma$~\cite{QDdecayratekettler2016neutral,QDdecayratetelecomzeuner2021demand}                                          &  $2\pi\times 1\;\text{GHz}$                                & $2\pi\times 20\;\text{GHz}$                    \\ \hline
FSR~\cite{FSRgridstatesjang2024programmable,ContextphotonicTFGridstateshenry2024parallelization,ContextphotonicCVtimefrequencyNicolasGridstates,ContextphotonicTFquditstateslu2018quantum,FSRgridstateschao2010octave,FSRgridstatesrielander2017frequency}                                        & N/A                         & $2\pi\times 1\;\text{GHz}$                   \\ 
\hline
\end{tabular}
\caption{Parameters of the QD–waveguide system considered in the FrEnGATE protocol for both the continuous-variable (see Fig.~\ref{Figure trade-off probability of success VS quality of entanglement}) frequency states and frequency qudit states (see Fig.~\ref{Figure gridstates success probability VS S and K}).}
\label{Tab illustrative table of different parameters}
\end{table}

\section{Conclusion}
The present theoretical work first provides an \textit{ab initio} effective two-level quadratic two-photon Hamiltonian in which the frequency is continuously defined. The derivation is based on a Heisenberg-picture adiabatic elimination of frequency-dependent joint one-photon transition operators reducing to frequency-dependent two-photon transition operators. In contrast, adiabatic elimination in atomic physics -- for instance in cavity systems -- is typically conducted in the interaction-Schrödinger picture, where the populations of irrelevant states are suppressed on average without explicitly keeping track of the frequency dependence of the interaction. Starting from a generic four-level system interacting with a bosonic field continuum, such one-photon transition-operator adiabatic elimination yields a two-photon quadratic Hamiltonian where the field is coupled to an effective two-level atomic-like system, \textit{e.g.} a QD embedded in a waveguide. This Hamiltonian is effective in that it emerges under several physical considerations: (i) Provided the one-photon detunings are large compared to the one-photon coupling terms \mohamed{and the FSS}. (ii) The interaction time is large with respect to the inverse of the one-photon detuning, ensuring that the one-photon transition operators have averaged to zero. (iii) The interaction time is small compared to the inverse of the photon linewidths and of the two-photon detunings, which entails that the two-photon transition operators are roughly constant and can be reinjected in the Schrödinger picture evolution. The computed two-photon coupling terms are frequency-dependent and non-separable in the two photons' frequencies. Alushi \textit{et al.}'s methodology~\cite{ArticleracineSimonealushi2023waveguide} based on a scattering theory computation in the Markovian approximation is then applied to analyze the appearance of frequency-entanglement in the output field. Under the Markovian approximation, the initial two-photon input is reshaped with a Lorentzian-dependence along the collective variables $\omega_\Sigma = \omega+\omega'$ determined by the biexcitonic decay rate $\Gamma$. When the frequency-dependent two-photon coupling term is specifically designed along the collective variable $\omega_\Delta$, the initial two-photon input distribution follows the profile of the coupling function along this variable. Further, considering a separable, isotropic, Gaussian two-photon input of width $\alpha$, and assuming a Gaussian two-photon coupling term along $\omega_\Delta$, the $\Gamma\gg\alpha$ regime yields a frequency-entangled two-photon output with Gaussian distributions along both collective variables $\omega_\Sigma$ and $\omega_\Delta$. A trade-off between the frequency-entanglement generation efficiency and its quality -- assessed with the Schmidt number and the entanglement entropy -- is showcased using a Schmidt decomposition for frequency continuous variables. For example, the frequency-entanglement generation efficiency reaches up to 15\% for a frequency-entangled biphoton state with a Schmidt number of around 5,  a normalized entanglement entropy of 0.4 and an ellipticity of 98\%. This frequency-entanglement generation efficiency is much higher than what can be achieved with SPDC~\cite{ContextSPDCphotongeneraitonefficicneybock2016highly,ContextSPDCshi2024efficient} or SFWM~\cite{ContextSFWMphotongeneffewang2024progress} and is expected to reach even higher values by controlling the waveguide chirality. The FrEnGATE can also be implemented with time–frequency qudit states, where the trade-off between frequency-entanglement and success probability emerges again. Depending on the ratio of the Gaussian coupling’s standard deviation to the spacing between spectral peaks, one can identify five distinct regimes of frequency correlation. This versatility shows that the FrEnGATE applies equally well in both continuous-variable and discrete-variable (qudit) time–frequency settings.
\\ Importantly, the FrEnGATE is \textit{measurement-free} as it does not require any auxiliary photons, post-selection or intense pump to perform frequency entanglement, by contrast with fusion-based methods~\cite{Blablaphotoniccomputingbartolucci2023fusion} for instance. Nevertheless, in the time-frequency continuous variable framework, the "true" frequency entangling gate has been shown~\cite{ContextphotonicCVtimefrequencythesefabre2020quantum,fabre_photonic_2024} to be quartic in the bosonic creation and annihilation operators, that is \textit{a priori} complicated to carry out experimentally. This quartic Hamiltonian amounts to a rotation of the JSI and acts as a non-linear frequency beam-splitter. This non-linear frequency beam-splitter is a proper unitary gate and is paramount for quantum information protocols, such as universal quantum computing \cite{ContextphotonicCVtimefrequencyfabre2022time}. While obtaining an Hamiltonian which is quartic in the bosonic creation and annihilation operators is particularly relevant within the time-frequency variable framework for time-frequency entanglement, such a Hamiltonian also holds potential for quadrature-based variables. Indeed, quartic operations in bosonic creation and annihilation operators would correspond to non-Gaussian gates, a topic of considerable interest in quantum information processing research. The FrEnGATE -- a quadratic Hamiltonian in the bosonic creation and annihilation operators -- paves the way for both theoretical analyses and experimental realizations of the time–frequency entangling gate in photonic platforms.\\

\section*{ACKNOWLEDGMENT}
\mohamed{Mohamed Meguebel acknowledges the support of the Program QuanTEdu-France n° ANR-22-CMAS-0001 France 2030}. Maxime Federico acknowledges funding from European Union’s Horizon Europe research and
innovation programme under the project Quantum Secure Network Partnership (QSNP, grant agreement No 101114043). Simone Felicetti acknowledges financial support from National Recovery and Resilience Plan (PNRR) Extended Partnership (MUR) project PE0000023-NQSTI, financed by the European Union-- Next Generation EU and from the foundation Compagnia di San Paolo, grant vEIcolo no. 121319. We acknowledge discussions with Hélène Ollivier, Eva Maria Gonzalez Ruiz, Hanna le Jeannic, Arne Keller and Pérola Milman for the completion of this manuscript.
\section*{Disclosures}
The authors declare no conflicts of interest.
\section*{Data availability}
Data underlying the results presented in this paper are not publicly available at this time but may be obtained
from the authors upon reasonable request.

\onecolumngrid

\bibliography{biblio}

\end{document}


\author{Mohamed Meguebel}
\email{mohamed.meguebel@telecom-paris.fr}
\affiliation{Telecom Paris, Institut Polytechnique de Paris, 19 Place Marguerite Perey, 91120 Palaiseau, France}
\author{Maxime Federico}
\affiliation{Telecom Paris, Institut Polytechnique de Paris, 19 Place Marguerite Perey, 91120 Palaiseau, France}
\author{Simone Felicetti}
\affiliation{Institute for Complex Systems, National Research Council (ISC-CNR), Via dei Taurini 19, 00185 Rome, Italy}
\affiliation{Physics Department, Sapienza University, P.le A. Moro 2, 00185 Rome, Italy}
\author{Nadia Belabas}
\affiliation{Centre for Nanosciences and Nanotechnology, CNRS, Universite Paris-Saclay,
UMR 9001, 10 Boulevard Thomas Gobert, 91120, Palaiseau, France}
\author{Nicolas Fabre} \email{nicolas.fabre@telecom-paris.fr}
\affiliation{Telecom Paris, Institut Polytechnique de Paris, 19 Place Marguerite Perey, 91120 Palaiseau, France}

\date{\today}
\pacs{}
\vskip2pc 
 
\title{Generation of frequency entanglement with an effective quantum dot-waveguidetwo-photon quadratic interaction:  supplemental document}

\maketitle

\onecolumngrid

\section{Introduction}
This supplemental document provides computational details corresponding to various sections of the main text. It is organized as follows. Section~\ref{Appendix Field quantization in dielectrics} addresses the quantization of the waveguide field. Section~\ref{Appendix adiabatic elimination} details the Heisenberg-picture adiabatic elimination procedure. Section~\ref{Appendix scattering theory derivation} summarizes the waveguide quantum electrodynamics derivation toolbox introduced by Alushi \textit{et al.}~\cite{ArticleracineSimonealushi2023waveguide}. Section~\ref{Appendix Schmidt decomposition for continous variables} outlines the Schmidt decomposition method for continuous variables using discrete orthogonal function bases. Section~\ref{Appendix Optimization of the propagation mode} sketches a numerical optimization of the waveguide spectral mode profile. Finally, Section~\ref{Appendix Grid states} provides the calculation steps leading to the analytical form of the scattered frequency-qudit two-photon output state.

\section{Field quantization in a dielectric waveguide}
\label{Appendix Field quantization in dielectrics}
In this section, we go over the field quantization in a lossless, nonmagnetic and single mode waveguide inspired from~\cite{ArticleracineMullertrivedi2020generation} where we additionnaly account for the polarization and the waveguide time-reversal symmetry. Other quantization procedures in a dielectric waveguide~\cite{Dielectricquantizationhuttner1992quantization,Dielectricquantizationgruner1996green,Dielectricquantizationdung1998three} and more specifically its interaction with an ensemble of point charges~\cite{Dielectricquantizedinteractiondung2002intermolecular,Dielectricquantizedinteractiondung2002resonant,Dielectricquantizedinteractionwubs2004multiple} and a quantum dot~\cite{QDinteractionWGGreenyao2010chip} can be found in the references therein. A field propagating in a single direction, rightward along the $z$-axis, the quantized field can be expressed as
\begin{equation}
    \hat{\boldsymbol{E}}(\boldsymbol{r}) = \sum_{\sigma}\int_{\mathbb R^+}d\beta\;\mathcal E_{\sigma}(\beta)\boldsymbol{E}_\sigma \left(\boldsymbol{r},\beta\right)\hat a_{\sigma}(\beta) + \text{h.c},
\end{equation}
where $\sigma$ is the polarization, $\beta$ the wavenumber, $\mathcal E_\sigma(\beta)$ a normalization constant 
\begin{equation}
    \mathcal{E}_{\sigma}(\beta) = \sqrt{\frac{\hbar \omega(\beta)}{2\epsilon_0\int_{\Gamma}\epsilon(\boldsymbol\rho,\beta)|\boldsymbol E_{\sigma}(\boldsymbol \rho,\beta)|^2d^2\boldsymbol \rho}}
\end{equation}
and $\boldsymbol{E}_\sigma \left(\boldsymbol{r},\beta\right) = \boldsymbol{E}_\sigma \left(\boldsymbol{\rho},\beta\right)e^{i\beta z}$ the propagating modes, solution to the wave equation 
\begin{equation}
    \boldsymbol \nabla \times \boldsymbol \nabla \times \boldsymbol{E}_\sigma \left(\boldsymbol{r},\beta\right) - \frac{\omega(\beta)^2}{c^2}\epsilon(\boldsymbol r,\beta)\boldsymbol{E}_\sigma \left(\boldsymbol{r},\beta\right) = \boldsymbol 0,
\end{equation}
where $\boldsymbol \rho$ is the transverse coordinate and $\omega(\beta)$ the 1D dispersion relation $\beta^2 = n_{\text{eff}}^2(\omega)\frac{\omega}{c}$ with $n_{\text{eff}}(\omega)$ the effective refractive index. The terme 'effective' encapsulates the idea that the mode propagates as if it were in a homogeneous medium whose refractive index is the average seen by the field. This effective index $n_{\text{eff}}$ does not explicity vary with the transverse coordinate $\boldsymbol r$ because all the spatial variations in the dielectric function are contained in the mode profile~\cite{WaveguideEMteich2007fundamentals,WaveguideEMbenisty2022introduction}. These propagating modes constitute an orthogonal basis in such a way that 
\begin{equation}
    \epsilon_0 \int d^3\boldsymbol r \; \epsilon (\boldsymbol{r},\beta) \boldsymbol{E}^*_\sigma \left(\boldsymbol{r},\beta\right)\cdot \boldsymbol{E}_{\sigma'} \left(\boldsymbol{r},\beta'\right) = \delta(\beta-\beta')\delta_{\sigma \sigma '}.
\end{equation}
The quantized electric field thus reads 
\begin{equation}
    \hat{\boldsymbol{E}}(\boldsymbol{r}) = \sum_{\sigma}\int_{\mathbb R^+}d\beta\;\mathcal E_{\sigma}(\beta)\boldsymbol{E}_\sigma \left(\boldsymbol{\rho},\beta\right)e^{i\beta z}\hat a_{\sigma}(\beta) + \text{h.c}.
\end{equation}
From now on, we will group together $\mathcal E_{\sigma}(\beta)$ and $\boldsymbol{E}_\sigma \left(\boldsymbol{\rho},\beta\right)$ in a single term $\boldsymbol{\mathcal{E}}_{\sigma}(\boldsymbol \rho, \beta)$ for readability. Taking into account the leftward propagating field along the $z$-axis 
\begin{equation}
\begin{split}
    \hat{\boldsymbol{E}}(\boldsymbol{r}) = \sum_{\sigma}\int_{\mathbb R^+}d\beta\;\bigg(\boldsymbol{\mathcal{E}}_{\sigma}(\boldsymbol \rho, \beta)e^{i\beta z}\hat a_{\sigma}(\beta)+\boldsymbol{\mathcal{E}}_{\sigma}(\boldsymbol \rho, -\beta)e^{-i\beta z}\hat a_{\sigma}(-\beta)+\text{h.c}\bigg).
\end{split}
\end{equation}
From here, we invoke the time-reversal symmetry~\cite{Timereversalsymmetryarntzenius2009time} of the waveguide which enforces 
\begin{equation}
    \boldsymbol{\mathcal{E}}_{\sigma}(\boldsymbol \rho, -\beta) = \boldsymbol{\mathcal{E}}_{\sigma}^*(\boldsymbol \rho, \beta),
\end{equation}
which entails 
\begin{equation}
    \hat{\boldsymbol{E}}(\boldsymbol{r}) = \sum_{\sigma}\sum_{\mu \in \{\pm\}}\int_{\mathbb R^+}d\beta\;\boldsymbol{\mathcal{E}}_{\sigma \mu}(\boldsymbol \rho, \beta) e^{i\beta\mu z}\hat a_{\sigma \mu}(\beta) + \text{h.c},
\end{equation}
with 
\begin{align}
    \boldsymbol{\mathcal{E}}_{\sigma \mu}(\boldsymbol \rho, \beta) &= \boldsymbol{\mathcal{E}}_{\sigma}(\boldsymbol \rho, \mu \beta) \\ 
    \hat a_{\sigma \mu}(\beta) &= \hat a_{\sigma}(\beta \mu).
\end{align}
The previously introduced bosonic operators fulfill the commutation relations 
\begin{equation}
    [\hat a_{\sigma \mu}(\beta), \hat a_{\sigma' \mu'}^\dagger(\beta')] = \delta(\beta-\beta')\delta_{\sigma \sigma'}\delta_{\mu \mu'}.
\end{equation}
One can toggle to the frequency $\omega(\beta)$ representation as 
\begin{equation}
\label{Appendice quantized field before real line extension}
    \hat{\boldsymbol{E}}(\boldsymbol{r}) = \sum_{\sigma}\sum_{\mu \in \{\pm\}}\int_{\omega_c}^{+\infty}d\omega\;\boldsymbol{\mathcal{E}}_{\sigma \mu}(\boldsymbol \rho, \beta(\omega)) e^{i\beta(\omega)\mu z}\hat a_{\sigma \mu}(\omega) + \text{h.c},
\end{equation}
where we have defined 
\begin{equation}
    \hat a_{\sigma \mu}(\omega) = \frac{\hat a_{\sigma \mu}\left(\beta(\omega)\right)}{\sqrt{v_G(\omega)}},
\end{equation}
where $v_G(\omega)$ correspond to the waveguide group velocity at frequency $\omega$ and $\omega_c$ to the waveguide cutoff frequency respectively. The latter is assumed to be far below the photons' frequency distributions in such a way that one can expand the integral in Eq.~\eqref{Appendice quantized field before real line extension} to the entire real line
\begin{equation}
    \hat{\boldsymbol{E}}(\boldsymbol{r}) = \sum_{\sigma}\sum_{\mu \in \{\pm\}}\int_{\mathbb R}d\omega\;\boldsymbol{\mathcal{E}}_{\sigma \mu}(\boldsymbol \rho, \beta(\omega)) e^{i\beta(\omega)\mu z}\hat a_{\sigma \mu}(\omega) + \text{h.c},
\end{equation}
with the commutation relation expressed with the frequency degree of freedom 
\begin{equation}
     [\hat a_{\sigma \mu}(\omega), \hat a_{\sigma' \mu'}^\dagger(\omega')] = \delta(\omega-\omega')\delta_{\sigma \sigma'}\delta_{\mu \mu'}.
\end{equation}
The previously defined bosonic operators act as 
\begin{align}
    &\hat a_{\sigma\mu}(\omega)\ket{\text{vac}} = 0,\;\;\;\;\hat a_{\sigma \mu}(\omega)\ket{\omega'}_{\sigma'\mu'} = \delta_{\sigma\sigma'}\delta(\omega-\omega')\ket{\text{vac}}
    \\ 
    &\hat a_{\sigma\mu}^\dagger(\omega)\ket{\text{vac}} = \ket{\omega}_{\sigma\mu},\;\;\;\; \hat a_{\sigma \mu}^\dagger(\omega)\ket{\omega'}_{\sigma'\mu'} = \ket{\omega',\omega}_{\sigma\mu,\sigma'\mu'},
\end{align}
where $\ket{\text{vac}}$ is the vacuum state and $\ket{\omega}_{\sigma\mu}$ the state corresponding to one photon at polarization $\sigma$ traveling in the direction of propagation $\mu$. The field free Hamiltonian reads 
\begin{equation}
\begin{split}
    \hat H_{\text{free, field}} = \sum_{\sigma}\sum_{\mu \in \{\pm\}} \int_{\mathbb R}d\omega\; \hbar \omega \hat a_{\sigma \mu}^\dagger(\omega)\hat a_{\sigma\mu}(\omega).
\end{split}
\end{equation}
For the time-frequency continuous variables framework~\cite{ContextphotonicCVtimefrequencyfabre2022time,ContextphotonicCVtimefrequencydescamps2024gottesman} to mathematically resemble that of electric field quadratures, it is additionally required that no two photons can occupy the same auxiliary modes. For instance, they can share the same propagation direction as long as they have different polarizations.
\section{Transition operators adiabatic elimination}
\label{Appendix adiabatic elimination}
In this section, we provide full computation for the adiabatic elimination that leads to the effective interaction Hamiltonian in Sec. II.B.2 of the main text
\begin{equation}
\label{Appendix Equation effective Hamiltonian after adiabatic elimination}
    \begin{split}
        &\hat H_{\text{int}} = \left(\hat H_{\text{int}} \right)_{\ket{0}\leftrightarrow \ket{X_+}\leftrightarrow \ket{2X}} +\left(\hat H_{\text{int}} \right)_{\ket{0}\leftrightarrow \ket{X_-}\leftrightarrow \ket{2X}},
    \end{split}
\end{equation}
where we divide the two interaction paths 
\begin{equation}
\label{Appendix Equation general effective interaction Hamiltonian X_pm branch}
    \begin{split}
        &\left(\hat H_{\text{int}} \right)_{\ket{0}\leftrightarrow \ket{X_\pm}\leftrightarrow \ket{2X}} = \hbar \sum_{\substack{\sigma'\in \{R,L\} \\ \mu' \in \{\pm\}}} \sum_{\substack{\sigma \in \{R,L\} \\ \mu \in \{\pm\}}}\int_{\mathbb R^2}d\omega' d\omega\; \bigg(g_{X_\pm}^{\sigma'\mu'\sigma \mu }(\omega',\omega)\ket{2X}\bra{0}\otimes \hat a_{\sigma'\mu'}(\omega')\hat a_{\sigma \mu}(\omega) + \text{h.c}\bigg),
    \end{split}
\end{equation}
where we incorporate the photon-number-dependent Stark shifts energy renormalization in the natural frequencies. The calculated two-photon coupling terms are shown to be equal to
\begin{equation}
\begin{split}
\label{Appendix Equation two-photon coupling term X_pm branch}
    g_{X_\pm}^{\sigma' \mu'\sigma \mu }(\omega',\omega) = g_{X_\pm2X}^{\sigma'\mu'}(\omega')g_{0X_\pm}^{\sigma \mu}(\omega)\times\bigg[\frac{1}{\omega-\omega_X}-\frac{1}{\omega'-(\omega_{2X}-\omega_X)}\bigg].
\end{split}
\end{equation}
\subsection{\mohamed{Derivation of the effective Hamiltonian}}
\label{Appendix Derivation of the effective Hamiltonian}
Consider the Heisenberg picture transition operator $\hat \xi_{0X_+}^{\sigma \mu}(\omega,t) \equiv \left( \ket{X_+}\bra{0}\otimes \hat a_{\sigma \mu}(\omega)\right)(t)$ dynamical equation
\begin{equation}
\label{Appendix Equation equation of motion}
     \frac{d \hat{\xi}_{0X_+}^{\sigma \mu}(\omega,t)}{dt} = \frac{1}{i\hbar}\left[\hat{\xi}_{0X_+}^{\sigma \mu }(\omega,t),\hat H_H(t)\right],
\end{equation}
where the Hamiltonian $\hat H_H(t)$ in the Heisenberg picture is equal to the Hamiltonian $\hat H = \hat H_{\text{free}}+\hat H_{\text{int}}$ previously defined 
\begin{align}
\label{Appendix Equation free Hamiltonian}
    \frac{\hat H_{\text{free}}}{\hbar} &= \omega_{2X}\ket{2X}\bra{2X} + \omega_X\left(\ket{X_+}\bra{X_+}+\ket{X_-}\bra{X_-} \right) + S\left(\ket{X_+}\bra{X_-}+\ket{X_-}\bra{X_+}\right)+ \sum_{\substack{\sigma \in \{R, L\} \\ \mu \in \{\pm\}}}\int_{\mathbb{R}}d\omega\; \omega \hat a_{\sigma \mu}^\dagger(\omega)\hat a_{\sigma \mu}(\omega) \nonumber \\
    &\quad  \\
\label{Appendix Equation interaction Hamiltonian before adiabatic elimination}
    \frac{\hat{H}_{\text{int}}}{\hbar} &= \sum_{\substack{\sigma \in \{R, L\} \\ \mu \in \{\pm\}}}\int_{\mathbb{R}}d\omega\;\bigg( 
    g_{0X_+}^{\sigma \mu}(\omega)\ket{X_+}\bra{0}\otimes \hat a_{\sigma \mu}(\omega)+ g_{X_+2X}^{\sigma \mu}(\omega)\ket{2X}\bra{X_+}\otimes \hat a_{\sigma \mu}(\omega) + g_{0X_-}^{\sigma \mu}(\omega)\ket{X_-}\bra{0}\otimes \hat a_{\sigma \mu}(\omega) \nonumber \\
    &\quad + g_{X_-2X}^{\sigma \mu}(\omega)\ket{2X}\bra{X_-}\otimes \hat a_{\sigma \mu}(\omega) + \text{h.c.} \bigg)
\end{align}
in the Schrödinger picture because the latter is time-independent. This joint operator characterizes the QD transition from its ground state $\ket{0}$ to its excitonic state $\ket{X_+}$ by absorbing a photon of frequency $\omega$, polarization $\sigma$ and direction of propagation $\mu$. The photons driving the excitonic transitions $\ket{0}\leftrightarrow\ket{X_\pm}$ and the biexcitonic transitions $\ket{X_\pm}\leftrightarrow\ket{2X}$ are referred to as \textit{excitonic} and \textit{biexcitonic} photons respectively. It is important to note that the time $t$ represents the dynamical time in the Hamiltonian evolution and not to the \textit{time-of-arrival} defined in the context of time-frequency continuous variables. Within the free Hamiltonian dynamics, this dynamical time simply amounts to an offset of the time-of-arrival while its action is \textit{a priori} no longer nontrivial when accounting for the interaction. One can compute the following 
\begin{equation}
\label{Eq Appendix first dynamical equation for the excitonic joint operators}
\begin{split}
    \hat{\xi}_{0X_+}^{\sigma \mu}(\omega,t)\frac{\hat H}{\hbar} &= \left(\hat{\xi}_{0X_+}^{\sigma \mu}(\omega)\sum_{\substack{\sigma' \in \{R, L\} \\ \mu' \in \{\pm\}}}\int_{\mathbb R}d\omega'\;\omega'\hat a_{\sigma'\mu'}^\dagger(\omega') \hat a_{\sigma' \mu'}(\omega')\right)(t) \\ 
    &+\left(\sum_{\substack{\sigma' \in \{R, L\} \\ \mu' \in \{\pm\}}}\int_{\mathbb R}d\omega'\;\left( g_{0X_+}^{\sigma'\mu'*}(\omega')\ket{X_+}\bra{X_+}\otimes\hat{a}_{\sigma \mu}(\omega)\hat{a}^\dagger_{\sigma '\mu'}(\omega')+g_{0X_-}^{\sigma'\mu'*}(\omega')\ket{X_+}\bra{X_-}\otimes \hat{a}_{\sigma \mu }(\omega)\hat {a}_{\sigma '\mu'}^\dagger(\omega')\right)\right)(t)
\end{split}
\end{equation}
and
\begin{equation}
\label{Eq Appendix second dynamical equation for the excitonic joint operators}
    \begin{split}
        \frac{\hat H}{\hbar} \hat{\xi}_{0X_+}^{\sigma \mu }(\omega,t) &= \omega_{X}\hat{\xi}_{0X_+}^{\sigma \mu}(\omega,t) + S\hat{\xi}_{0X_-}^{\sigma \mu}(\omega,t) + \left(\sum_{\substack{\sigma' \in \{R, L\} \\ \mu' \in \{\pm\}}}\int_{\mathbb R}d\omega'\;\omega'\hat{a}_{\sigma' \mu'}^\dagger(\omega')\hat{a}_{\sigma' \mu'}(\omega')\hat{\xi}_{0X_+}^{\sigma \mu }(\omega)\right)(t) \\ 
        &\left(\sum_{\substack{\sigma' \in \{R, L\} \\ \mu' \in \{\pm\}}}\int_{\mathbb R}d\omega'\;\left(g_{0X_+}^{\sigma'\mu'*}(\omega')\ket{0}\bra{0}\otimes \hat{a}_{\sigma'\mu'}^\dagger(\omega')\hat{a}_{\sigma \mu }(\omega) + g_{X_+2X}^{\sigma'\mu'}(\omega')\ket{2X}\bra{0}\otimes \hat a_{\sigma'\mu'}(\omega')\hat{a}_{\sigma \mu }(\omega) \right)\right)(t),
    \end{split}
\end{equation}
where as in Eq.~\eqref{Appendix Equation free Hamiltonian} we dropped out the identity operators and therefore the tensor products for readability. The joint operator $\hat{\xi}_{0X_-}^{\sigma \mu }(\omega) \equiv \ket{X_-}\bra{0}\otimes \hat a_{\sigma \mu }(\omega)$ is defined in the same way as $\hat{\xi}_{0X_+}^{\sigma \mu }(\omega)$. This operator physically represents an optical transition from the ground state to the excitonic state $\ket{X_-}$ with a $\sigma$-polarized photon by means of the fine-structure splitting $S$ coupling -- thus bypassing the optical selection rules which do not allow the $\ket{0}\leftrightarrow\ket{X_\pm}$ transitions for the same polarization $\sigma$ and direction of propagation $\mu$. Injecting Eq.~\eqref{Eq Appendix first dynamical equation for the excitonic joint operators} and Eq.~\eqref{Eq Appendix second dynamical equation for the excitonic joint operators} in the transition operator equation of motion Eq.~\eqref{Appendix Equation equation of motion} 
\begin{equation}
    \begin{split}
        i\frac{d \hat{\xi}_{0X_+}^{\sigma \mu}(\omega,t)}{dt} &= -\omega_X\hat{\xi}_{0X_+}^{\sigma \mu}(\omega,t) + \left(\ket{X_+}\bra{0}\otimes \sum_{\substack{\sigma' \in \{R, L\} \\ \mu' \in \{\pm\}}}\int_{\mathbb R}d\omega'\;\omega'\left[\hat a_{\sigma \mu}(\omega),\hat a_{\sigma'\mu'}^\dagger(\omega')\hat a_{\sigma' \mu'}(\omega') \right]\right)(t) -S\hat \xi_{0X_-}^{\sigma \mu}(\omega,t) \\ 
        &+\left(\sum_{\substack{\sigma' \in \{R, L\} \\ \mu' \in \{\pm\}}}\int_{\mathbb R}d\omega'\;\left(g_{0X_+}^{\sigma'\mu'*}(\omega')\ket{X_+}\bra{X_+}\otimes \hat a_{\sigma \mu}(\omega)\hat a_{\sigma'\mu'}^\dagger(\omega')-g_{0X_+}^{\sigma'\mu'*}(\omega')\ket{0}\bra{0}\otimes \hat a_{\sigma'\mu'}^\dagger(\omega')\hat a_{\sigma \mu}(\omega) \right)\right)(t) \\ 
        &+\left(\sum_{\substack{\sigma' \in \{R, L\} \\ \mu' \in \{\pm\}}}\int_{\mathbb R}d\omega'\;\left(g_{0X_-}^{\sigma'\mu'*}(\omega')\ket{X_+}\bra{X_-}\otimes \hat a_{\sigma \mu}(\omega)\hat a^\dagger_{\sigma'\mu' }(\omega')-g_{X_+2X}^{\sigma'\mu'}(\omega')\ket{2X}\bra{0}\otimes \hat a_{\sigma'\mu'}(\omega')\hat a_{\sigma \mu}(\omega) \right)\right)(t).
    \end{split}
\end{equation}
The bosonic commutation relations lead to
\begin{equation}
    \left[\hat a_{\sigma \mu}(\omega),\hat a_{\sigma'\mu'}^\dagger(\omega')\hat a_{\sigma' \mu'}(\omega')\right] = \delta_{\mu\mu'}\delta_{\sigma\sigma'}\delta(\omega-\omega')\hat a_{\sigma' \mu'}(\omega),
\end{equation}
in such a way that 
\begin{equation}
\label{Eq Appendix excitonic transition operator final differential equation}
    \begin{split}
        i\frac{d \hat{\xi}_{0X_+}^{\sigma \mu}(\omega,t)}{dt} &= \left(\omega-\omega_X\right)\hat{\xi}_{0X_+}^{\sigma \mu}(\omega,t) -S\hat \xi_{0X_-}^{\sigma \mu}(\omega,t) \\ 
        &+\left(\sum_{\substack{\sigma' \in \{R, L\} \\ \mu' \in \{\pm\}}}\int_{\mathbb R}d\omega'\;\left(g_{0X_+}^{\sigma'\mu'*}(\omega')\ket{X_+}\bra{X_+}\otimes \hat a_{\sigma \mu}(\omega)\hat a_{\sigma'\mu'}^\dagger(\omega')-g_{0X_+}^{\sigma'\mu'*}(\omega')\ket{0}\bra{0}\otimes \hat a_{\sigma'\mu'}^\dagger(\omega')\hat a_{\sigma \mu}(\omega) \right)\right)(t) \\ 
        &+\left(\sum_{\substack{\sigma' \in \{R, L\} \\ \mu' \in \{\pm\}}}\int_{\mathbb R}d\omega'\;\left(g_{0X_-}^{\sigma'\mu'*}(\omega')\ket{X_+}\bra{X_-}\otimes \hat a_{\sigma \mu}(\omega)\hat a^\dagger_{\sigma'\mu' }(\omega')-g_{X_+2X}^{\sigma'\mu'}(\omega')\ket{2X}\bra{0}\otimes \hat a_{\sigma'\mu'}(\omega')\hat a_{\sigma \mu}(\omega) \right)\right)(t).
    \end{split}
\end{equation}
A solution to this differential equation is 
\begin{equation}
\label{Eq Appendix General solution to the differential equation excitonic joint operator}
    \hat \xi_{0X_+}^{\sigma \mu}(\omega,t) = e^{-i(\omega-\omega_X)t}\hat \xi_{0X_+}^{\sigma \mu}{(\omega,0)}+\hat l^{\sigma \mu}_{0X_+}(\omega,t),
\end{equation}
where the first term $e^{-i(\omega-\omega_X)t} \hat \xi_{0X_+}^{\sigma \mu}{(\omega,0)}$ comes from the homogeneous differential equation 
\begin{equation}
     i\frac{d \hat{\xi}_{0X_+}^{\sigma \mu}(\omega,t)_{\text{homogeneous}}}{dt} = \left(\omega-\omega_X\right)\hat{\xi}_{0X_+}^{\sigma \mu}(\omega,t)_{\text{homogeneous}}
\end{equation}
and $\hat l^{\sigma \mu}_{0X_+}(\omega,t)$ is a particular solution to the differential equation~\eqref{Eq Appendix excitonic transition operator final differential equation} and is chosen as $\hat l^{\sigma \mu}_{0X_+}(\omega,t) = e^{-i(\omega-\omega_X)t}\hat k^{\sigma \mu}_{0X+}(\omega,t)$ where $\hat k^{\sigma \mu}_{0X_+}(\omega,t)$ is a differentiable operator fulfilling 
\begin{equation}
\label{Appendix equation for the particular solution up to second order}
\begin{split}
    &ie^{-i(\omega-\omega_X)t}\frac{d\hat k^{\sigma \mu}_{0X_+}(\omega,t)}{dt} = -S\hat \xi_{0X_-}^{\sigma \mu}(\omega,t) \\ 
    &+\left(\sum_{\substack{\sigma' \in \{R, L\} \\ \mu' \in \{\pm\}}}\int_{\mathbb R}d\omega'\;\left(g_{0X_+}^{\sigma'\mu'*}(\omega')\ket{X_+}\bra{X_+}\otimes \hat a_{\sigma \mu}(\omega)\hat a_{\sigma'\mu'}^\dagger(\omega')-g_{0X_+}^{\sigma'\mu'*}(\omega')\ket{0}\bra{0}\otimes \hat a_{\sigma'\mu'}^\dagger(\omega')\hat a_{\sigma \mu}(\omega) \right)\right)(t) \\ 
    &+\left(\sum_{\substack{\sigma' \in \{R, L\} \\ \mu' \in \{\pm\}}}\int_{\mathbb R}d\omega'\;\left(g_{0X_-}^{\sigma'\mu'*}(\omega')\ket{X_+}\bra{X_-}\otimes \hat a_{\sigma \mu}(\omega)\hat a^\dagger_{\sigma'\mu' }(\omega')-g_{X_+2X}^{\sigma'\mu'}(\omega')\ket{2X}\bra{0}\otimes \hat a_{\sigma'\mu'}(\omega')\hat a_{\sigma \mu}(\omega) \right)\right)(t).
\end{split}
\end{equation}
Formally integrating Eq.~\eqref{Appendix equation for the particular solution up to second order}
\begin{equation}
\label{Appendix equation particular solution first integral equation}
    \begin{split}
        &\hat k^{\sigma \mu}_{0X+}(\omega,t) = i\int_{0}^t d\tau\; e^{i(\omega-\omega_X)\tau}S\hat \xi_{0X_-}^{\sigma \mu}(\omega,\tau) \\ 
        &-i\int_{0}^t d\tau\; e^{i(\omega-\omega_X)\tau}\left(\sum_{\substack{\sigma' \in \{R, L\} \\ \mu' \in \{\pm\}}}\int_{\mathbb R}d\omega'\;\left(g_{0X_+}^{\sigma'\mu'*}(\omega')\ket{X_+}\bra{X_+}\otimes \hat a_{\sigma \mu}(\omega)\hat a_{\sigma' \mu'}^\dagger(\omega')-g_{0X_+}^{\sigma'\mu'*}(\omega')\ket{0}\bra{0}\otimes \hat a_{\sigma'\mu'}^\dagger(\omega')\hat a_{\sigma \mu}(\omega) \right)\right)(\tau) \\ 
        &-i\int_{0}^t d\tau\; e^{i(\omega-\omega_X)\tau}\left(\sum_{\substack{\sigma' \in \{R, L\} \\ \mu' \in \{\pm\}}}\int_{\mathbb R}d\omega'\;\left(g_{0X_-}^{\sigma'\mu'*}(\omega')\ket{X_+}\bra{X_-}\otimes \hat a_{\sigma \mu}(\omega)\hat a^\dagger_{\mu' \sigma'}(\omega')-g_{X_+2X}^{\sigma'\mu'}(\omega')\ket{2X}\bra{0}\otimes \hat a_{\sigma'\mu'}(\omega')\hat a_{\sigma \mu}(\omega) \right)\right)(\tau).
    \end{split}
\end{equation}
From here, the adiabatic elimination consists in identifying and comparing evolution time scales. The exponential term $e^{i(\omega-\omega_X)\tau}$ oscillates at frequency $|\omega-\omega_X|$ with $\omega$ being the frequency of the photon driving the excitonic transition. On the other hand, up to the zeroth-order in the coupling terms $g_{..}^{..}(.)$ and $S$ -- which again drives the $\ket{X_+}\leftrightarrow \ket{X_-}$ transition -- in Eq.~\eqref{Appendix equation particular solution first integral equation}, the time-dependent two-photon joint transition operators in the Heisenberg picture $\left(\ket{X_+}\bra{X_+}\otimes\hat a_{\sigma \mu}(\omega)\hat a_{\sigma'\mu'}^\dagger(\omega')\right)(\tau)$, $\left(\ket{0}\bra{0}\otimes\hat a^\dagger_{\sigma'\mu' }(\omega')\hat a_{\sigma \mu}(\omega)\right)(\tau)$ and $\left(\ket{X_+}\bra{X_-}\otimes\hat a_{\sigma \mu}(\omega)\hat a^\dagger_{\sigma'\mu'}(\omega')\right)(\tau)$ evolve according to the time-evolution equation at a frequency $|\omega-\omega'|$ with $\omega$ and $\omega'$ matching the excitonic photon spectral distribution while $\left(\ket{2X}\bra{0}\otimes \hat a_{\sigma'\mu' }(\omega')\hat a_{\sigma \mu}(\omega)\right)(\tau)$ evolves at a frequency $|\omega_{2X}-(\omega+\omega')|$ with $\omega$ and $\omega'$ matching the excitonic and biexcitonic photons spectral distributions respectively. Subsequently, provided the detuning $\delta_e(\omega) \equiv \omega-\omega_X$ is large in absolute value compared to $|\omega-\omega'|$ -- representing the excitonic photon spectral bandwidth $\Delta\omega_e$ -- and compared to the two-photon transition $\ket{2X}\leftrightarrow \ket{0}$ detuning $|\omega_{2X}-(\omega+\omega')| \approx |\omega_{2X}-(\omega_e+\omega_b)|$ where $\omega_e$ and $\omega_b$ denote the central frequencies of the excitonic and biexcitonic photons respectively, the previous solution reads 

    \begin{equation}
    \label{Appendix equation particular solution before dropping the S term}
    \begin{split}
        &\hat k^{\sigma \mu}_{0X+}(\omega,t) = i\int_{0}^t d\tau\; e^{i(\omega-\omega_X)\tau}S\hat \xi_{0X_-}^{\sigma \mu}(\omega,\tau) \\ 
        &+\left(\frac{1-e^{i(\omega-\omega_X)t}}{\omega-\omega_X}\right)\left(\sum_{\substack{\sigma' \in \{R, L\} \\ \mu' \in \{\pm\}}}\int_{\mathbb R}d\omega'\;\left(g_{0X_+}^{\sigma'\mu'*}(\omega')\ket{X_+}\bra{X_+}\otimes \hat a_{\sigma \mu}(\omega)\hat a_{\sigma' \mu'}^\dagger(\omega')-g_{0X_+}^{\sigma'\mu'*}(\omega')\ket{0}\bra{0}\otimes \hat a_{\sigma'\mu'}^\dagger(\omega')\hat a_{\sigma \mu}(\omega) \right)\right) \\ 
        &+\left(\frac{1-e^{i(\omega-\omega_X)t}}{\omega-\omega_X}\right)\left(\sum_{\substack{\sigma' \in \{R, L\} \\ \mu' \in \{\pm\}}}\int_{\mathbb R}d\omega'\;\left(g_{0X_-}^{\sigma'\mu'*}(\omega')\ket{X_+}\bra{X_-}\otimes \hat a_{\sigma \mu}(\omega)\hat a^\dagger_{\sigma'\mu' }(\omega')-g_{X_+2X}^{\sigma'\mu'}(\omega')\ket{2X}\bra{0}\otimes \hat a_{\sigma'\mu'}(\omega')\hat a_{\sigma \mu}(\omega) \right)\right)
    \end{split}
\end{equation}
for a time $t$ such that 
\begin{equation}
   1/|\delta_e(\omega)|\ll t \ll  1/\Delta\omega_e, 1/|\omega_{2X}-(\omega_e+\omega_b)|,
\end{equation} 
so that the transition operators are essentially constant during the time of evolution during which the exponential term had had the time to average out to zero. Note that this amounts to a Markovian approximation where memory effects are disregarded~\cite{Adiabaticeliminationbeyondpaulisch2014beyond}. Nonetheless, the last transition operator $\hat \xi_{0X_-}^{\sigma \mu}(\omega,t)$ evolves as the same frequency as $\hat \xi_{0X_+}^{\sigma \mu}(\omega,t)$ that is to say $|\omega-\omega_{X}|$ with $\omega$ matching the excitonic photon spectral distribution. This operator thus cannot be considered constant during the interaction time $t$. However, the first term in Eq.~\eqref{Appendix equation particular solution before dropping the S term} holding the FSS $S$ can -- in the similar fashion to Eq.~\eqref{Eq Appendix General solution to the differential equation excitonic joint operator} -- be expressed as 
\begin{equation}
\begin{split}
    i\int_{0}^t\;d\tau e^{i(\omega-\omega_X)\tau}S\hat{\xi}^{\sigma \mu}_{0X_-}(\omega,\tau) &= i\int_{0}^t\;d\tau e^{i(\omega-\omega_X)\tau}S\times\left(e^{-i(\omega-\omega_X)\tau}\hat{\xi}_{0X_-}^{\sigma \mu}(\omega,0)+\hat l_{0X_-}^{\sigma \mu}(\omega,\tau) \right) \\ 
    &= iSt\hat{\xi}_{0X_-}^{\sigma \mu}(\omega,0) +iS\int_{0}^t\;d\tau e^{i(\omega-\omega_X)\tau}\hat l_{0X_-}^{\sigma \mu}(\omega,\tau).
\end{split}
\end{equation}
As for $\hat l_{0X_+}^{\sigma\mu}(t)$, $\hat l_{0X_-}^{\sigma\mu}(\tau)$ can be expressed as $e^{-i(\omega-\omega_X)\tau}\hat k_{0X_-}^{\sigma\mu}(\omega,\tau)$ with $\hat k_{0X_-}^{\sigma \mu}(\omega,\tau)$ fulfilling an equation identical to Eq.~\eqref{Appendix equation particular solution first integral equation}
\begin{equation}
    \begin{split}
        &\hat k^{\sigma \mu}_{0X_-}(\omega,t) = i\int_{0}^t d\tau\; e^{i(\omega-\omega_X)\tau}S\hat \xi_{0X_+}^{\sigma \mu}(\omega,\tau) \\ 
        &-i\int_{0}^t d\tau\; e^{i(\omega-\omega_X)\tau}\left(\sum_{\substack{\sigma' \in \{R, L\} \\ \mu' \in \{\pm\}}}\int_{\mathbb R}d\omega'\;\left(g_{0X_-}^{\sigma'\mu'*}(\omega')\ket{X_-}\bra{X_-}\otimes \hat a_{\sigma \mu}(\omega)\hat a_{\sigma' \mu'}^\dagger(\omega')-g_{0X_-}^{\sigma'\mu'*}(\omega')\ket{0}\bra{0}\otimes \hat a_{\sigma'\mu'}^\dagger(\omega')\hat a_{\sigma \mu}(\omega) \right)\right)(\tau) \\ 
        &-i\int_{0}^t d\tau\; e^{i(\omega-\omega_X)\tau}\left(\sum_{\substack{\sigma' \in \{R, L\} \\ \mu' \in \{\pm\}}}\int_{\mathbb R}d\omega'\;\left(g_{0X_+}^{\sigma'\mu'*}(\omega')\ket{X_-}\bra{X_+}\otimes \hat a_{\sigma \mu}(\omega)\hat a^\dagger_{\mu' \sigma'}(\omega')-g_{X_-2X}^{\sigma'\mu'}(\omega')\ket{2X}\bra{0}\otimes \hat a_{\sigma'\mu'}(\omega')\hat a_{\sigma \mu}(\omega) \right)\right)(\tau).
    \end{split}
\end{equation}
This term is multiplied by $S$ and is therefore a second order term in coupling terms $S$ and $g_{..}^{..}(.)$ that we neglect as done above. The fast-oscillating contribution in $\hat{\xi}_{0X_-}^{\sigma\mu}(\omega,\tau)$ can thus be ignored in Eq.~\eqref{Appendix equation particular solution before dropping the S term} and the particular solution to the differential equation Eq.~\eqref{Eq Appendix excitonic transition operator final differential equation} $\hat l^{\sigma \mu}_{0X_+}(\omega,t) = e^{-i(\omega-\omega_X)t}\hat k^{\sigma \mu}_{0X_+}(\omega,t)$ thus yields 
  \begin{equation}
  \label{Appendix Equation particular solution B20}
    \begin{split}
        &\hat l^{\sigma \mu}_{0X_+}(\omega,t) = e^{-i(\omega-\omega_X)t}\times iSt\hat{\xi}_{0X_-}^{\sigma \mu}(\omega,0) \\
        &\left(\frac{e^{-i(\omega-\omega_X)t}-1}{\omega-\omega_X}\right)\left(\sum_{\substack{\sigma' \in \{R, L\} \\ \mu' \in \{\pm\}}}\int_{\mathbb R}d\omega'\;\left(g_{0X_+}^{\sigma'\mu'*}(\omega')\ket{X_+}\bra{X_+}\otimes \hat a_{\sigma \mu}(\omega)\hat a_{\sigma' \mu'}^\dagger(\omega')-g_{0X_+}^{\sigma'\mu'*}(\omega')\ket{0}\bra{0}\otimes \hat a_{\sigma'\mu'}^\dagger(\omega')\hat a_{\sigma \mu}(\omega) \right)\right) \\ 
        &+\left(\frac{e^{-i(\omega-\omega_X)t}-1}{\omega-\omega_X}\right)\left(\sum_{\substack{\sigma' \in \{R, L\} \\ \mu' \in \{\pm\}}}\int_{\mathbb R}d\omega'\;\left(g_{0X_-}^{\sigma'\mu'*}(\omega')\ket{X_+}\bra{X_-}\otimes \hat a_{\sigma \mu}(\omega)\hat a^\dagger_{\sigma'\mu' }(\omega')-g_{X_+2X}^{\sigma'\mu'}(\omega')\ket{2X}\bra{0}\otimes \hat a_{\sigma'\mu'}(\omega')\hat a_{\sigma \mu}(\omega) \right)\right).
    \end{split}
\end{equation}
Putting Eq.~\eqref{Appendix Equation particular solution B20} in the solution to the differential equation Eq.~\eqref{Eq Appendix General solution to the differential equation excitonic joint operator} and taking out the $e^{-i(\omega-\omega_X)t}$ fast-oscillating term that averages out to zero over the interaction time $t$ 
\begin{equation}
\begin{split}
\label{Eq Appendix solution to the differential equation for the exitonic transition operator after adiabatic elimination}
    \hat \xi_{0X_+}^{\sigma \mu}(\omega,t) &= -\frac{1}{\omega-\omega_X}\sum_{\substack{\sigma' \in \{R, L\} \\ \mu' \in \{\pm\}}}\int_{\mathbb R}d\omega'\;\bigg(g_{0X_+}^{\sigma'\mu'*}(\omega')\ket{X_+}\bra{X_+}\otimes \hat a_{\sigma \mu}(\omega)\hat a_{\sigma'\mu'}^\dagger(\omega')-g_{0X_+}^{\sigma'\mu'*}(\omega')\ket{0}\bra{0}\otimes \hat a_{\sigma'\mu'}^\dagger(\omega')\hat a_{\sigma \mu}(\omega) \\ 
    & +g_{0X_-}^{\sigma'\mu'*}(\omega')\ket{X_+}\bra{X_-}\otimes \hat a_{\sigma \mu}(\omega)\hat a^\dagger_{\sigma'\mu' }(\omega')-g_{X_+2X}^{\sigma'\mu'}(\omega')\ket{2X}\bra{0}\otimes \hat a_{\sigma'\mu'}(\omega')\hat a_{\sigma \mu}(\omega)\bigg),
\end{split}
\end{equation}
which no longer depends on the interaction time $t$ and can therefore be taken as time-independent throughout the evolution. Their Schrödinger and Heisenberg representations are thus identical for this interaction time.  With exactly the same reasoning, the biexcitonic transition operator $\hat \xi_{X_+2X}^{\sigma \mu}(\omega,t)$ can be shown to read 
\begin{equation}
    \begin{split}
    \label{Eq Appendix solution to the differential equation for the biexcitonic transition operator after adiabatic elimination}
         \hat \xi_{X_+2X}^{\sigma \mu }(\omega,t) &= -\frac{1}{\omega-({\omega_{2X}-\omega_X})}\sum_{\substack{\sigma' \in \{R, L\} \\ \mu' \in \{\pm\}}}\int_{\mathbb R}d\omega'\;\bigg(g_{X_+2X}^{\sigma'\mu'*}(\omega')\ket{2X}\bra{2X}\otimes \hat a_{\sigma \mu }(\omega)\hat a_{\sigma'\mu'}^\dagger(\omega')\\ 
         &-g_{X_+2X}^{\sigma'\mu'*}(\omega')\ket{X_+}\bra{X_+}\otimes \hat a_{\sigma'\mu'}^\dagger(\omega')\hat a_{\sigma \mu }(\omega) \\
    & -g_{X_-2X}^{\sigma'\mu'*}(\omega')\ket{X_-}\bra{X_+}\otimes \hat a_{\sigma'\mu' }^\dagger(\omega')\hat a_{\sigma\mu }(\omega)+g_{0X_+}^{\sigma'\mu'}(\omega')\ket{2X}\bra{0}\otimes \hat a_{\sigma \mu }(\omega)\hat a_{\sigma'\mu' }(\omega')\bigg),
    \end{split}
\end{equation}
which is again time-independent for an interaction time $t$ such that 
\begin{equation}
   1/|\delta_{b}(\omega)|\ll t \ll  1/\Delta \omega_b, 1/|\omega_{2X}-(\omega_e+\omega_b)|,
\end{equation}
with $\delta_{b}(\omega) \equiv \omega-({\omega_{2X}-\omega_X})$ and $\Delta \omega_b$ the biexcitonic photon spectral bandwidth. 
\\ Injecting back these two joint operators $\hat \xi_{0X_+}^{\sigma \mu}(\omega)$ Eq.~\eqref{Eq Appendix solution to the differential equation for the exitonic transition operator after adiabatic elimination} and $\hat \xi_{X_+2X}^{\sigma \mu }(\omega)$ Eq.~\eqref{Eq Appendix solution to the differential equation for the biexcitonic transition operator after adiabatic elimination} in the $\ket{0}\leftrightarrow \ket{X_+}\leftrightarrow \ket {2X}$ branch of the interaction Hamiltonian Eq.~\eqref{Appendix Equation interaction Hamiltonian before adiabatic elimination} 
\begin{equation}
    \begin{split}
        \left(\hat H_{\text{int}}\right)_{\ket{0}\leftrightarrow \ket{X_+}\leftrightarrow \ket {2X}} &= \hbar\sum_{\substack{\sigma'\in \{R,L\} \\ \mu' \in \{\pm\}}} \sum_{\substack{\sigma \in \{R,L\} \\ \mu \in \{\pm\}}}\int_{\mathbb R^2}d\omega d\omega'\;\bigg[-\frac{1}{\omega-\omega_X}\bigg(g_{0X_+}^{\sigma\mu}(\omega)g_{0X_+}^{\sigma'\mu'*}(\omega')\big[\ket{X_+}\bra{X_+}\otimes \hat a_{\sigma \mu}(\omega)\hat a_{\sigma'\mu'}^\dagger(\omega') \\ 
        &- \ket{0}\bra{0}\otimes \hat a_{\sigma'\mu' }^\dagger(\omega')\hat a_{\sigma \mu}(\omega)\big] + g_{0X_+}^{\sigma \mu}(\omega)g_{0X_-}^{\sigma'\mu'*}(\omega')\ket{X_+}\bra{X_-}\otimes \hat a_{\sigma \mu}(\omega)\hat a_{\sigma'\mu'}^\dagger(\omega') \\
        &-g_{X_+2X}^{\sigma'\mu'}(\omega')g_{0X_+}^{\sigma\mu}(\omega)\ket{2X}\bra{0}\otimes \hat a_{\sigma'\mu' }(\omega')\hat a_{\sigma \mu}(\omega) \bigg)  \\
        &-\frac{1}{\omega-(\omega_{2X}-\omega_X)}\bigg(g_{X_+2X}^{\sigma \mu}(\omega)g_{X_+2X}^{\sigma'\mu'*}(\omega')\big[\ket{2X}\bra{2X}\otimes \hat a_{\sigma\mu }(\omega)\hat a_{\sigma'\mu'}^\dagger(\omega')\\
        &-\ket{X_+}\bra{X_+}\otimes \hat a_{\sigma'\mu' }^\dagger(\omega')\hat a_{\sigma \mu }(\omega) \big]\\ &-g_{X_-2X}^{\sigma'\mu'*}(\omega')g_{X_+2X}^{\sigma \mu}(\omega)\ket{X_-}\bra{X_+}\otimes \hat a^\dagger_{\sigma'\mu'}(\omega')\hat a_{\sigma \mu}(\omega) \\
        &+ g_{X_+2X}^{\sigma \mu}(\omega)g_{0X_+}^{\sigma'\mu'}(\omega')\ket{2X}\bra{0}\otimes \hat a_{\sigma \mu }(\omega)\hat a_{\sigma'\mu' }(\omega')\bigg)\bigg] + \text{h.c.}
    \end{split}
\end{equation}
These expressions are similar to those found in \cite{adiabaticeliminationatomicphysicsalexanian1995unitary} wherein the derivation was performed for two frequency modes of the electromagnetic field applying a Schrieffer-Wolff transformation~\cite{SchriefferWolffbravyi2011schrieffer}. The diagonal terms related to $\ket{X_+}\bra{X_+}$, $\ket{0}\bra{0}$ and $\ket{2X}\bra{2X}$ correspond to the intensity-dependent Stark shifts and act as energy renormalization terms in the free Hamiltonian that we thus drop out in the equations for readability. The terms in $\ket{X_+}\bra{X_-}$ are two-photon transition terms arising from the excitonic joint transition operator. Indeed, for the terms in $g_{0X_+}^{\sigma \mu}(\omega)g_{0X_-}^{\sigma'\mu'*}(\omega')$ (respectively $g_{X_-2X}^{\sigma'\mu'*}(\omega')g_{X_+2X}^{\sigma \mu}(\omega)$), since $\hat \xi_{0X_+}^{\sigma \mu}(\omega,t)$ and $\hat \xi_{0X_-}^{\sigma'\mu'  \dagger}(\omega',t)$ evolve rapidly in the adiabatic elimination regime, for a QD that goes from its excitonic state $\ket{X_-}$ (respectively $\ket{X_+}$) to its ground state $\ket{0}$ (respectively excited state $\ket{2X}$) before being promoting back up (respectively decaying down to) to its other excitonic state, all appears as if it went from $\ket{X_-}$ to $\ket{X_+}$ (respectively $\ket{X_+}$ to $\ket{X_-}$) by first emitting (respectively absorbing) a photon and then absorbing (respectively emitting) a photon. Nonetheless, this transition only occurs provided one of the two excitonic states is populated which they are not on average over the interaction time $t$ given that the transition operators that promote the QD to these states have been adiabatically eliminated. Therefore, the $\ket{0}\leftrightarrow \ket{X_+}\leftrightarrow \ket{2X}$ branch of the interaction Hamiltonian now yields 
\begin{equation}
    \begin{split}
        \left(\hat H_{\text{int}}\right)_{\ket{0}\leftrightarrow \ket{X_+}\leftrightarrow \ket {2X}} &= \hbar\sum_{\substack{\sigma'\in \{R,L\} \\ \mu' \in \{\pm\}}} \sum_{\substack{\sigma \in \{R,L\} \\ \mu \in \{\pm\}}}\int_{\mathbb R^2}d\omega d\omega'\;\bigg[-\frac{1}{\omega-\omega_X}\bigg(
        -g_{0X_+}^{\sigma \mu}(\omega)g_{X_+2X}^{\sigma'\mu'}(\omega')\ket{2X}\bra{0}\otimes \hat a_{\sigma'\mu'}(\omega')\hat a_{\sigma \mu}(\omega) \bigg)  \\
        &-\frac{1}{\omega-(\omega_{2X}-\omega_X)}\bigg(
         g_{X_+2X}^{\sigma \mu}(\omega)g_{0X_+}^{\sigma'\mu'}(\omega')\ket{2X}\bra{0}\otimes \hat a_{\sigma \mu }(\omega)\hat a_{\sigma'\mu' }(\omega')\bigg)\bigg] + \text{h.c.}
    \end{split}
\end{equation}
Reordering the dumb indices $\omega$ and $\omega'$ 
\begin{equation}
    \begin{split}
        \left(\hat H_{\text{int}}\right)_{\ket{0}\leftrightarrow \ket{X_+}\leftrightarrow \ket {2X}}  &=\hbar\sum_{\substack{\sigma'\in \{R,L\} \\ \mu' \in \{\pm\}}} \sum_{\substack{\sigma \in \{R,L\} \\ \mu \in \{\pm\}}}\int_{\mathbb R^2}d\omega d\omega'\;g_{0X_+}^{\sigma \mu}(\omega)g_{X_+2X}^{\sigma'\mu'}(\omega')\left[\frac{1}{\omega-\omega_X}-\frac{1}{\omega'-(\omega_{2X}-\omega_X)} \right]\ket{2X}\bra{0} \\ 
        &\otimes \hat a_{\sigma'\mu' }(\omega')\hat a_{\sigma \mu}(\omega)+\text{h.c.}
    \end{split}
\end{equation}
Adding the $\ket{0} \leftrightarrow \ket{X_-}\leftrightarrow \ket{2X}$ branch, the total effective interaction Hamiltonian reads 
\begin{equation}
\label{Appendix Equation final effective Hamiltonian}
    \begin{split}
        \hat H_{\text{int,eff}}&= \hbar\sum_{\substack{\sigma'\in \{R,L\} \\ \mu' \in \{\pm\}}}\sum_{\substack{\sigma \in \{R,L\} \\ \mu \in \{\pm\}}}\int_{\mathbb R^2}d\omega' d\omega\; \ket{2X}\bra{0}\otimes \left[g_{X_+}^{\sigma'\mu'\sigma \mu }(\omega',\omega)\hat a_{\sigma'\mu'}(\omega')\hat a_{\sigma \mu}(\omega) + g_{X_-}^{\sigma'\mu'\sigma \mu}(\omega',\omega)\hat a_{\sigma'\mu'}(\omega')\hat a_{\sigma \mu}(\omega)\right] + \text{h.c},
    \end{split}
\end{equation}
with 
\begin{align}
\label{Appendix two-photon coupling term general expression}
    g_{X_\pm}^{\sigma'\mu'\sigma \mu }(\omega',\omega) &= g_{X_\pm2X}^{\sigma'\mu'}(\omega')g_{0X_\pm}^{\sigma \mu}(\omega)\left[\frac{1}{\omega-\omega_X}-\frac{1}{\omega'-(\omega_{2X}-\omega_X)} \right].
\end{align}
\mohamed{We obtained an effective two-photon Hamiltonian for which the coupling terms are non-separable in the two photons frequencies $\omega$ and $\omega'$. These calculations have been performed up to first order in the coupling terms $S$ and $g_{..}^{..}(.)$ in the one-photon joint transition operators and to the zeroth-order in the two-photon transition operators; under the assumption that the one-photon detunings are large compared to both $S$ and $g_{..}^{..}(\cdot)$ (dispersive regime).} This resulted in a total Hamiltonian Eq.~\eqref{Appendix Equation final effective Hamiltonian} with up to second order coupling terms. In that regard, within the adiabatic elimination, the fine-structrue splitting $S$ would have contributed to a two-photon interaction if we had considered coupling terms up to the third order in the effective Hamiltonian. This stands to reason because this would be driven by $g_{0X_\pm}^{\sigma\mu}(\omega)$ (respectively $g_{X\pm 2X}^{\sigma \mu*}(\omega)$) for the transitions $\ket{0}\rightarrow\ket{X_\pm}$ (respectively $\ket{2X}\rightarrow\ket{X_\pm}$), then a coupling $\ket{X_\pm}\leftrightarrow\ket{X_\mp}$ through $S$ then the transition $\ket{X_\mp}\rightarrow\ket{0}$ or $\ket{X_\mp}\rightarrow\ket{2X}$ (respectively $\ket{X_\mp}\rightarrow \ket{2X}$ or $\ket{X_\mp}\rightarrow\ket{0}$) with the coupling term $g_{0X_\mp}^{\sigma\mu*}(\omega)$ or $g_{X_\mp2X}^{\sigma \mu}(\omega)$ (respectively $g_{X\mp2X}^{\sigma \mu}(\omega)$ or $g_{0X\mp}^{\sigma\mu*}(\omega)$).\\ \mohamed{In summary, the adiabatic elimination regime requires operation at finite one-photon detunings. In the dispersive limit -- where light–matter couplings are much weaker than the detunings -- single-photon transitions are suppressed, whereas two-photon interactions remain resonant.} \\ \\ Considering one-photon joint operators instead of either bare states~\cite{AdiabaticeliminationLambdasystembrion2007adiabatic,Adiabaticeliminationbeyondpaulisch2014beyond} or bare operators~\cite{ArticleracineRamanpassagegerry1990dynamics} has several key benefits. (i) It is more intuitive to adiabatically eliminate operators rather than states since the goal is to obtain an effective Hamiltonian with effective operators, making operators adiabatic elimination the more natural choice. (ii) Because we are working with transition operators in the uniquely defined Heisenberg picture, frequency detuning naturally arises without requiring a shift to an interaction picture, which is not unique and can lead to different predictions of the system's dynamics~\cite{Adiabaticeliminationbeyondpaulisch2014beyond}. (iii) The light-matter coupling depends on the frequency which is continuously distributed. Tracking the evolution of transitions operator joint for each frequency with the field bosonic operators preserves this frequency dependence throughout and yields effective two-photon coupling terms that are functions of both photons' frequencies. (iv) Using bosonic operators instead of bosonic states allows for the explicit application of the bosonic commutation relation, therefore clearly revealing the photonic processes ordering. Note that the previous derivation was performed without taking into account the optical selection rules.   Therefore, the present description can be extended to any four-level system coupled to a bosonic field frequency continuum.

\begin{figure}
    \centering
    \includegraphics[width=0.8\linewidth]{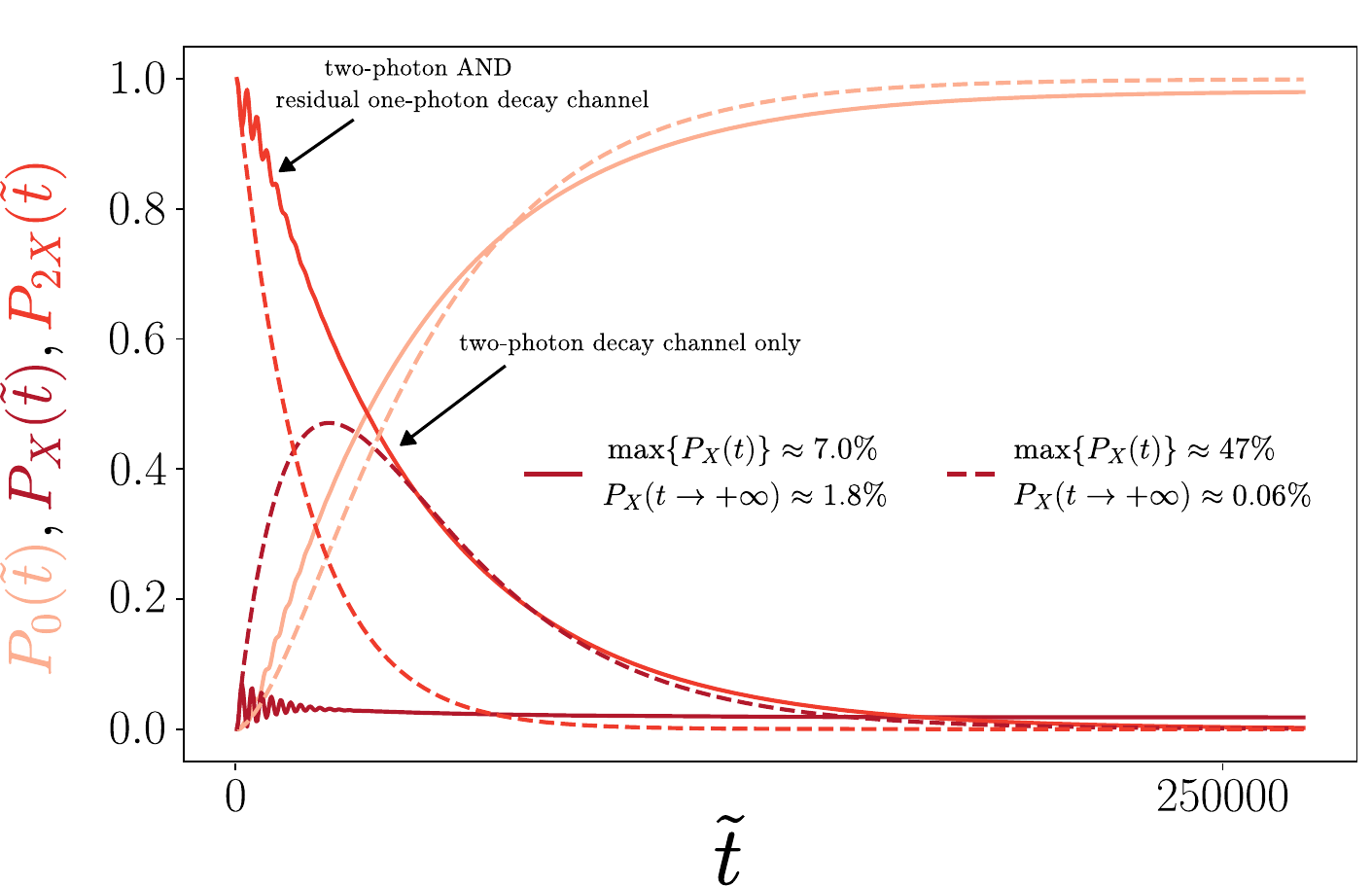}
    \caption{Biexcitonic decay with adiabatic elimination (solid lines) and without adiabatic elimination (dashed lines). All physical constants are normalized to the biexcitonic frequency $\omega_{2X}$ and denoted with a tilde. The emission bandwidth is centered at half the biexcitonic frequency, $\omega_{2X}/2$. In the adiabatic elimination regime, the one-photon transition frequencies lie outside the bandwidth, and the detuning is large compared with $\|g_{..}^{..}\|_{L^2}$. The corresponding bandwidth is $2.5\times 10^{-4}\,\omega_{2X}$. Outside the adiabatic elimination regime, the one-photon transitions overlap with the bandwidth, which is $2.5\times 10^{-5}\,\omega_{2X}$. The coupling functions are chosen as $\sqrt{\omega}\,u(\omega)$, where $u(\omega)$ denotes the waveguide mode centered at $\omega_X-\delta_X/2$ (set a Gaussian for simplicity) and $\sqrt{\omega}$ stems from the waveguide electromagnetic field quantization. We set $\omega_X = 1.005/2\,\omega_{2X} \approx 2\pi \times 190\; \text{THz}$, $\delta_X = 0.005\omega_{2X} \approx 2\pi\times 950\; \text{GHz}$, and $S = 10^{-5}\omega_{2X}\approx 2\pi \times 20\; \text{GHz}$; in accordance with the values indicated in the main text. Using a numerical interpolation, we estimate the biexcitonic decay rate $\Gamma$ in the adiabatic elimination regime to be of order $10^{-5}\,\omega_{2X}$, consistent with the value used in the main text. It is related (in good approximation) to the population of the biexciton at time $t$ as $P_{2X}(t) = \exp(-\Gamma t)$.}

    \label{Appendix Figure: Decay channels}
\end{figure}

\mohamed{\subsection{Bound on fast-oscillating contributions}}
\mohamed{In this section we establish a self-contained bound on the fast-oscillating contributions of the interaction Hamiltonian in Eq.~\eqref{Appendix Equation final effective Hamiltonian}, thereby justifying why they can be neglected. Our analysis will lay emphasis on the terms arising from the time evolution Eq.~\eqref{Eq Appendix General solution to the differential equation excitonic joint operator} of the operator
$\hat \xi_{0X_+}^{\sigma \mu}(\omega,t) 
= \left(\ket{X_+}\bra{0}\otimes \hat a_{\sigma \mu}(\omega)\right)(t)$ noting that the same reasoning applies directly to the other transition operators. Throughout, we restrict to second-order processes and therefore to the at-most-two-photon sector.
 Let $K=L^2(\mathbb R, \mathbb C^2)$ be the one-particle Hilbert space. $\mathbb C^2$ accounts for the polarization and direction of propagation degrees of freedom. Let $\mathcal F_{\leq 2} =\bigoplus_{n=0}^2 K^{\otimes n}$ denote the truncated bosonic Fock space with at most two photons. Denote by $\hat N$ the number operator and by $D(\hat N)$ its domain defined as $D(\hat N) = \{\psi, \psi = \{\psi^{(n)}\}_{2\geq n \geq 0}, \sum_{2\geq n \geq 0}n^2\|\psi^{(n)}\|^2 < \infty\}$. For $\psi \in D(\hat N)$, it acts as $\hat N \psi = \{\hat N \psi^{(n)}\}_{2\geq n \geq 0}$. For $f_t \in K$, the annihilation operator $\hat a [f_t]$ is defined on the finite-particle domain as 
 \begin{equation}
     \hat a[f_t] = \int_{\mathbb R}d\omega\; f_t(\omega)\hat a(\omega),
 \end{equation}
 and satisfies the sectorwise Cauchy-Schwarz bound~\cite{OperatoralgebraBrattelia1987operator,Operatoralgebrarobert2021coherent} 
\begin{equation}
\label{Appendix equation smeared operator bound for the two-photon truncation}
    \|\hat a[f_t]\psi\| \leq \|f_t\|_{L^2}\|\hat N^{1/2}\psi\|,
\end{equation}
for $\psi$ in the form domain of $\hat N$, \textit{i.e.} $\psi \in D(\hat N^{1/2})$. In particular, on the two-photon truncation and assuming a state normalized to one, \textit{i.e.} $\|\psi\| = 1$
\begin{equation}
    \|\hat a[f_t]\psi\| \leq \sqrt{2}\|f_t\|_{L^2}\|\psi\|.
\end{equation}
The norm $\|.\|$ is defined as~\cite{OperatoralgebraBrattelia1987operator} 
\begin{equation}
    \|.\| : \psi \in D(\hat N^{1/2}) \rightarrow \|\psi\|^2 = \|\psi^{(0)}\|^2 + \int d\omega_1|\psi^{(1)}(\omega_1)|^2+\int d\omega_1 d\omega_2|\psi^{(2)}(\omega_1,\omega_2)|^2.
\end{equation}
Note that with the more standard definition from quantum optics, the norm $\|\psi \|$ is equal to $\sqrt{\bra{\psi}\ket{\psi}}$ where $\ket{\psi}$ is the photonic vector state. In the previous adiabatic elimination calculation, we eliminated three kinds of contributions from the interaction Hamiltonian -- up to second order in the coupling terms and the fine-structure splitting $S$ --, all carrying the fast-oscillating $e^{-i(\omega-\omega_X)t}$ term 
\begin{align}
    \hat \xi_1(t) &=\sum_{\substack{\sigma \in \{R, L\} \\ \mu \in \{\pm\}}}\int_{\mathbb{R}}d\omega\;
    e^{-i(\omega-\omega_X)t}g_{0X_+}^{\sigma \mu}(\omega) \hat a_{\sigma \mu}(\omega) \\ 
   \hat \xi_2(t) &= \sum_{\substack{\sigma'\in \{R,L\} \\ \mu' \in \{\pm\}}} \sum_{\substack{\sigma \in \{R,L\} \\ \mu \in \{\pm\}}}\int_{\mathbb R^2}d\omega d\omega'\;e^{-i(\omega-\omega_X)t}\frac{g_{0X_+}^{\sigma \mu}(\omega)g_{X_+2X}^{\sigma' \mu'}(\omega')}{\omega-\omega_X}\hat a_{\sigma'\mu'}(\omega')\hat a_{\sigma \mu}(\omega) \\ 
   \hat \xi_3(t) &=  \sum_{\substack{\sigma \in \{R, L\} \\ \mu \in \{\pm\}}}\int_{\mathbb{R}}d\omega\;te^{-i(\omega-\omega_X)t}Sg_{0X_+}^{\sigma \mu}(\omega) \hat a_{\sigma \mu}(\omega),
\end{align}
where the atomic parts of the transition operators have been disregarded for readibility as they do not alter the operator bounds. $\hat{\xi}_1(t)$ denotes the one-photon transition operator from Eq.~\eqref{Eq Appendix General solution to the differential equation excitonic joint operator}, $ \hat{\xi}_2(t)$ denotes the exponentially time-varying contribution to two-photon processes -- either photon-number-dependent Stark-shift terms (which explicitly involve field annihilation and creation operators) or the direct $\ket{0}\leftrightarrow \ket{2X}$ two-photon transition -- as given in Eq.~\eqref{Appendix Equation particular solution B20}, and $ \hat{\xi}_3(t)$ arises from the coupling between $\ket{X_-}$ and $\ket{X_+}$ induced by the fine-structure splitting (FSS) (see Eq.~\eqref{Appendix Equation particular solution B20}). Let us define the time-averaged operators 
\begin{equation}
    \overline{\hat \xi_i(t)} \equiv \frac{1}{T}\int_{0}^Tdt\; \hat \xi_i(t),
\end{equation}
with $T$ the time after which the interaction started. With this definition, 
\begin{align}
    \overline{\hat \xi_1(t)} &= \frac{1}{T}\sum_{\substack{\sigma \in \{R, L\} \\ \mu \in \{\pm\}}}\int_{\mathbb R}d\omega\; \left[1-e^{-i(\omega-\omega_X)T} \right]\frac{g_{0X_+}^{\sigma \mu}(\omega)}{i(\omega-\omega_X)}\hat a_{\sigma \mu}(\omega) \\
    \overline{\hat \xi_2(t)} &=\frac{1}{iT}\sum_{\substack{\sigma'\in \{R,L\} \\ \mu' \in \{\pm\}}} \sum_{\substack{\sigma \in \{R,L\} \\ \mu \in \{\pm\}}}\int_{\mathbb R}d\omega'\;\hat g_{X_+2X}^{\sigma'\mu'}(\omega')\hat a_{\sigma'\mu'}(\omega')\int_{\mathbb R}d\omega\; \left[1-e^{-i(\omega-\omega_X)T} \right]\frac{g_{0X_+}^{\sigma \mu}(\omega)}{(\omega-\omega_X)^2}\hat a_{\sigma \mu}(\omega) \\ 
    \overline{\hat \xi_3(t)} &= -\sum_{\substack{\sigma \in \{R, L\} \\ \mu \in \{\pm\}}}\int_{\mathbb R}d\omega\; \left(\left[1-e^{-i(\omega-\omega_X)T}\right]\frac{Sg_{0X_+}^{\sigma \mu}(\omega)}{(\omega-\omega_X)^2T}+e^{-i(\omega-\omega_X)T}\frac{Sg_{0X_+}^{\sigma \mu}(\omega)}{i(\omega-\omega_X)} \right).
\end{align}
Employing Eq.~\eqref{Appendix equation smeared operator bound for the two-photon truncation}, the triangular inequality and the excitonic one-photon detuning lower bound $|\omega-\omega_X|\geq \delta_e$
\begin{align}
    \|\overline{\hat \xi_1(t)}\psi\| &\leq 2\sqrt{2}\sum_{\substack{\sigma \in \{R, L\} \\ \mu \in \{\pm\}}}\frac{\|g_{0X_+}^{\sigma\mu}\|_{L^2}}{T\delta_e} \\ 
    \|\overline{\hat \xi_2(t)}\psi\| &\leq 2\sqrt{2}\sum_{\substack{\sigma'\in \{R,L\} \\ \mu' \in \{\pm\}}} \sum_{\substack{\sigma \in \{R,L\} \\ \mu \in \{\pm\}}}\frac{\|g_{0X_+}^{\sigma \mu}\|_{L^2} \|g_{X_+2X}^{\sigma\mu'}\|_{L^2}}{T\delta_e^2} \\ 
    \|\overline{\hat \xi_3(t)}\psi\| &\leq \sqrt{2}\sum_{\substack{\sigma \in \{R, L\} \\ \mu \in \{\pm\}}}\left(2\frac{S\|g_{0X_+}^{\sigma \mu}\|_{L^2}}{\delta_e^2 T} + \frac{S\|g_{0X_+}^{\sigma \mu}\|_{L^2}}{\delta_e} \right).
\end{align}
As $T \gg 1/\delta_e$, $\|\overline{\hat \xi_1(t)}\psi\| \longrightarrow 0$, $\|\overline{\hat \xi_2(t)}\psi\| \longrightarrow 0$ and $\|\overline{\hat \xi_3(t)}\psi\| \leq \sqrt{2}\sum_{\substack{\sigma \in \{R, L\} \\ \mu \in \{\pm\}}}\frac{S\|g_{0X_+}^{\sigma \mu}\|_{L^2}}{\delta_e}$. The fast-oscillating contributions $\hat \xi_1(t)$ and $\hat \xi_2(t)$ can thus be safely discarded in the adiabatic elimination regime with a $O(1/t)$ rate as they do not contribute \textit{on average} to the dynamics for a time $T\gg 1/\delta_e$. Nonetheless, $\hat \xi_3(t)$ \st{do} \mohamed{does} not \textit{a priori} tend to zero on average as $T \gg 1/\delta_e$ because of the secular term $\frac{S\|g_{0X_+}^{\sigma \mu}\|_{L^2}}{\delta_e}$. $\hat \xi_3(t)$ can still be ignored provided $S \ll \|g_{..}^{..}\|_{L^2}$ where $g_{..}^{..}$ is a shorthand notation for any of the involved coupling terms. Indeed, in a similar fashion to before, the constant-in-time resonant two-photon transition operators $\hat \xi_{\text{res.}}$ can be proven to be upper bounded as 
\begin{equation}
    \|\hat \xi_{\text{res.}}\|_{L^2} \leq \sqrt{2}\sum_{\substack{\sigma'\in \{R,L\} \\ \mu' \in \{\pm\}}} \sum_{\substack{\sigma \in \{R,L\} \\ \mu \in \{\pm\}}} \frac{\|g_{..}^{..}\|_{L^2}\|g_{..}^{..}\|_{L^2}}{\delta_e}.
\end{equation}
Thus, provided $S \ll \|g_{..}^{..}\|_{L^2}$, the resonant two-photon transition operators are predominant in the dynamics with compared to the FSS-induced second-order $\ket{0}\leftrightarrow \ket{X_+} \leftrightarrow \ket{X_-}$ transition. This further justifies investigations of means to effectively suppress the FSS~\cite{QDFSSmano2010self,QDFSSwang2012eliminating,QDFSSfognini2018universal,QDFSSlettner2021strain}. \\
A conservative upper bound for the time-averaged one-photon transition operator $\hat\xi_1(t)$ is given by Eq.~\eqref{Appendix equation smeared operator bound for the two-photon truncation} as $\|\hat \xi_1(t)\psi\| \leq \|g_{0X_+}^{\sigma \mu}\|_{L^2}$. This reflects that, at very short times, all modes contribute coherently. At later times $T \gg 1/\delta_e$, the fast oscillations lead to effective suppression as quantified by the time-averaged operator $\overline{\hat \xi_1(t)}$. \\ 
For transparency we integrate the mode index $\omega$ over the entire real line $\mathbb R$ but assume the coupling profiles $g_{0X_+}^{\sigma \mu}(\omega)$ are either (i) compactly supported away from the excitonic frequency $\omega_X$ \textit{i.e.} $\text{supp}\{ g_{0X_+}^{\sigma \mu}\}\subset \{\omega: \; |\omega-\omega_X|\geq \delta_e > 0\}$ or (ii) admit a smooth cutoff which removes a small neighbourhood of $\omega_X$.
Under either assumption the energy denominators $|(\omega-\omega_X)^{-1}|$ are uniformly bounded by $\delta_e$ and the estimates in this subsection hold. If one does not wish to assume that the coupling vanishes exactly at resonance, the integration domain can be separated into a near-resonant region and an off-resonant region. The off-resonant part is straightforward to control, since the energy denominators remain uniformly bounded there. The near-resonant contribution, on the other hand, can be made arbitrarily small by restricting the size of the region under consideration, for instance using a Cauchy–Schwarz estimate, or it can be treated by introducing the physically natural linewidth regularization. For notational clarity we adopt the compact-support formulation above; the alternative splitting produces identical conclusions up to errors that can be made arbitrarily small. \\
Our reasoning focused on the terms arising from the time evolution, Eq.~\eqref{Eq Appendix General solution to the differential equation excitonic joint operator}, of the operator
$\hat\xi_{0X_+}^{\sigma\mu}(\omega,t)=\big(\ket{X_+}\bra{0}\otimes\hat a_{\sigma\mu}(\omega)\big)(t)$ but the same arguments apply straightforwardly to the other transition operators subject to adiabatic elimination.}
\mohamed{\subsection{Numerical simulation of the biexciton decay}}
\mohamed{In this subsection, we run a numerical simulation of the biexcitonic decay within the adiabatic elimination regime previously detailed in order to assess its validity. For simplicity, the polarization and direction of propagation auxiliary degrees of freedom are omitted, as they do not modify the overall dynamics.  Consider a Wigner-Weisskopf ansatz with at most two weighted excitations. The light-matter vector state can be cast as 
\begin{equation}
    \ket{\psi(t)} = \int_{\mathbb R^2}d\omega d\omega'\;C_{0}(\omega',\omega;t)\ket{0;\omega',\omega} + \int_{\mathbb R}d\omega\; C_{X_+}(\omega;t)\ket{X_+;\omega} + \int_{\mathbb R}d\omega\; C_{X_-}(\omega;t)\ket{X_-;\omega} + C_{2X}(t)\ket{2X;\text{vac}},
\end{equation}
and its time-dynamics is driven by the Schrödinger equation $i\hbar \partial_t \ket{\psi(t)} = \hat H\ket{\psi(t)}$. In the reduced subspace with at most two weighted excitations, the full (not the effective one) Hamiltonian $\hat H = \hat H_{\text{free}}+\hat H_{\text{int}}$ (with $\hbar =1$) constituted of $\hat H_{\text{free}}$ Eq.~\eqref{Appendix Equation free Hamiltonian} and $\hat H_{\text{int}}$ Eq.~\eqref{Appendix Equation interaction Hamiltonian before adiabatic elimination} can be expressed within the $\{\omega_i,\omega_i'\}$ sector as 
\begin{equation}
    \hat H_{\{\omega_i,\omega_i'\}} = \begin{bmatrix} \omega_i+\omega_i' & g_{0X_+}(\omega_i) & g_{0X_-}(\omega_i) & g_{0X_+}(\omega_i') & g_{0X_-}(\omega_i') & 0 \\
    g_{0X_+}^*(\omega_i) & \omega_i+\omega_X & S & 0 & 0 & g_{X_+2X}(\omega_i) \\ 
    g_{0X_-}^*(\omega_i) & S & \omega_i +\omega_X & 0 & 0 & g_{X_-2X}(\omega_i) \\ 
    g_{0X_+}^*(\omega_i') & 0 & 0 & \omega_i'+\omega_X & S & g_{X_+2X}(\omega_i') \\
    g_{0X_-}^*(\omega_i') & 0 & 0 & S & \omega_i'+\omega_X & g_{X_-2X}(\omega_i') \\ 
    0 & g_{X_+2X}^*(\omega_i) & g_{X_-2X}^*(\omega_i) & g_{X_+2X}^*(\omega_i') & g_{X_-2X}^*(\omega_i') & 2\omega_X - \delta_X 
    \end{bmatrix},
\end{equation}
in the basis $\{\ket{0;\omega_i,\omega_i'},\ket{X_+;\omega_i}, \ket{X_-;\omega_i},\ket{X_+;\omega_i'}, \ket{X_-;\omega_i'}, \ket{2X;\text{vac}}\}$. We recall that $\omega_X$, $\delta_X$ and $S$ denotes the excitonic frequency, the binding frequency and the FSS, respectively. The numerical simulations are carried out using a discretization of the frequency continuum on a $400\times 400$ frequency grid. In Fig.~\ref{Appendix Figure: Decay channels}, we display the populations $P_0(t) =\int_{\mathbb R^2}d\omega d\omega'\;|C_{0}(\omega',\omega;t)|^2$, $P_{X}(t) = \int_{\mathbb R}d\omega\;\left(|C_{X_+}(\omega;t)|^2+|C_{X_+}(\omega;t)|\right)$ and $P_{2X}(t)=|C_{2X}(t)|^2$ as a function of time in the adiabatic elimination regime and when the latter does not hold. Since we investigate the biexcitonic state decay, $P_{2X}(t=0) = 1$.}
\mohamed{As expected in the adiabatic elimination regime, the excitonic population remains small \textit{on average} ($\max\{P_X(t)\} \approx 7\%$) and can therefore be safely ignored, while the biexcitonic state decays directly to the ground state. This confirms the usual expectation that in the dispersive regime the exciton acts only as a virtual intermediate, contributing weakly to the dynamics. This behavior reflects the absence of a one-photon decay channel when all one-photon detunings are bounded away from zero: the biexciton then decays exclusively through the available two-photon channel. In contrast, when the waveguide spectrum overlaps with the one-photon resonant frequencies, the first-order one-photon decay channel becomes available in addition to the second-order two-photon decay channel. In that case the excitonic population rises significantly ($\max\{P_X(t)\} \approx 47\%$) before decaying more slowly, as the exciton itself eventually leaks to the ground state $(P_X(t\rightarrow+\infty) \approx 0.05\%)$.
Even within the adiabatic elimination regime, one observes rapid oscillations of the excitonic population that decay quickly to a small plateau ($P_X(t \rightarrow +\infty) \approx 1.8\% \neq 0\%$). Because this residual population is minor, it does not alter the dispersive dynamics, and the interaction Hamiltonian Eq.~\eqref{Appendix Equation final effective Hamiltonian} remains a valid approximation. \\
In summary, these simulations show that, in the adiabatic elimination regime, one-photon decay channels can be neglected in favor of the two-photon transition pathways. As the detuning decreases and the system departs from the dispersive regime, the one-photon processes -- being of lower order -- start to dominate, and the interaction Hamiltonian deviates from the effective form of Eq.~\eqref{Appendix Equation final effective Hamiltonian}. This reduces the efficiency of the frequency-entangling gate, as it becomes contaminated by unwanted single-photon events. Achieving the adiabatic elimination regime therefore requires careful waveguide engineering: one-photon transition frequencies must be suppressed, while allowing photons at the two-photon resonance to propagate.} \\
\section{Scattering theory derivation}
\label{Appendix scattering theory derivation}
In this section, we present the main computational steps of the scattering theory method employed by Alushi \textit{et al}~\cite{ArticleracineSimonealushi2023waveguide} adapted to our physical system. Further details can be found in~\cite{ArticleracineSimonealushi2023waveguide} and Sec.~7.2 and Appendix~B.3 of~\cite{LivreJuanjoripoll2022quantum}.
\subsubsection{Wigner-Weisskopf ansatz}
Let us first define the bare transition operator $\hat \xi_{2X0} = \hat \xi_{02X}^\dagger \equiv \ket{2X}\bra{0}$, the following $\hat N$ operator 
\begin{equation}
    \hat N = 2\hat \xi_{02X}\hat \xi_{2X0} + \sum_{\sigma\in\{R,L\}}\sum_{\mu\in\{\pm\}}\int_{\mathbb R}d\omega\; \hat a_{\sigma \mu}^{\dagger}(\omega)\hat a_{\sigma \mu}(\omega),
\end{equation}
accounting of the number of weighted excitations in the system, commutes with the total effective Hamiltonian. Consequently, the system's dynamics can be effectively described using a normalized Wigner-Weisskopf ansatz $\ket{\psi(t)}$ with a fixed number of weighted excitations. Given the expression of the effective interaction Hamiltonian Eq.~\eqref{Appendix Equation final effective Hamiltonian}, the photonic states are decoupled from the emitter in the vacuum and individual photon sectors. As a result, we can focus on solving the scattering problem for input states consisting either of two photons with the emitter in the ground state or a single excited emitter in the vacuum state 
\begin{equation}
 \label{Appendix Equation Wigner-Weisskopf ansatz} 
\ket{\psi(t)} = C_{2X}(t)\hat \xi_{2X0}\ket{\boldsymbol 0}+
   \sum_{\substack{\sigma'\in \{R,L\} \\ \mu' \in \{\pm\}}} \sum_{\substack{\sigma \in \{R,L\} \\ \mu \in \{\pm\}}}\int_{\mathbb R^2}d\omega' d\omega C^{\sigma'\sigma}_{\mu'\mu}(\omega',\omega;t)\hat a_{\sigma'\mu'}^\dagger(\omega')\hat a_{\sigma \mu}^\dagger(\omega)\ket{\boldsymbol 0},
\end{equation}
where $C_{2X}(t)$ and $C^{\sigma'\sigma}_{\mu'\mu}(\omega',\omega;t)$ are the probability amplitudes to have the system respectively in the QD's biexcitonic state without photons and in the QD's ground state with two-photons at frequency $\omega$ and $\omega'$, polarization $\sigma$ and $\sigma'$, and direction of propagation $\mu$ and $\mu'$.  The state $\ket{\boldsymbol 0} \equiv \ket{0}\otimes\ket{\text{vac}}$ denotes the global vacuum state for both light and matter. The entire QD-waveguide isolated system is enclosed within the Wigner-Weisskopf ansatz, whose dynamics is governed by the Schrödinger equation $i\hbar \partial_t \ket{\psi(t)} = \hat H\ket{\psi(t)}$ giving rise to a system of linear, coupled ordinary differential equations 
\begin{subequations}
\label{Appendix Equation system of linear ODEs}
\begin{align}
\label{Appendix Equation linear ordinary differential equation for biexciton}
    &i\dot C_{2X}(t) = \omega_{2X}C_{2X}(t) 
    +\sum_{\mu',\mu\in\{\pm\}}\int_{\mathbb R^2}d\omega'd\omega\; g^{\mu'\mu}(\omega',\omega)C^{\sigma'\sigma}_{\mu'\mu}(\omega',\omega;t) \\
\label{Appendix Equation linear ordinary differential equation for two-photon state}
    &i\dot C^{\sigma'\sigma}_{\mu'\mu}(\omega',\omega;t) = (\omega+\omega')C^{\sigma'\sigma}_{\mu'\mu}(\omega',\omega;t)
    +\left(g^{\mu'\mu}(\omega',\omega) \right)^*C_{2X}(t),
\end{align}
\end{subequations}
with the polarization $\sigma$ and $\sigma'$ left unspecified but implicitly enforced by the coupling terms selection rules. The two-photon state linear differential equation Eq.~\eqref{Appendix Equation linear ordinary differential equation for two-photon state} can be formally integrated as 
\begin{equation}
\label{Appendix Equation probability amplitude two-photon formally integrated}
   C_{\mu'\mu}^{\sigma'\sigma}(\omega',\omega;t_1) = C_{\mu'\mu}^{\sigma'\sigma}(\omega',\omega;t_0)e^{-i(\omega+\omega')(t_1-t_0)}
    -i\left(g^{\mu'\mu}(\omega',\omega)\right)^* \int_{t_0}^{t_1} d\tau\;C_{2X}(\tau)e^{-i(\omega+\omega')(t_1-\tau)},
\end{equation}
which can be substituted back into the differential equation for the biexciton probability amplitude 
\begin{equation}
\label{Appendix Equation differential equations of beixcitonic amplitude 2}
    \begin{split}
        i\dot C_{2X}(t_1) &= \omega_{2X}C_{2X}(t_1) +\sum_{\mu',\mu\in\{\pm\}}\int_{\mathbb R^2}d\omega'd\omega\; \bigg(g^{\mu'\mu}(\omega',\omega)C_{\mu'\mu}^{\sigma'\sigma}(\omega',\omega;t_0)e^{-i(\omega+\omega')(t_1-t_0)} \\ 
        &-i\int_{t_0}^{t_1} d\tau\;|g^{\mu'\mu}(\omega',\omega)|^2 C_{2X}(\tau)e^{-i(\omega+\omega')(t_1-\tau)}\bigg).
    \end{split}
\end{equation}
The first term corresponds to the QD's free evolution, the second to the photons absorbed by the QD and the third to the photons emitted by the QD in the past that got reabsorbed.

\subsubsection{Markovian and weak coupling approximation}
Let us toggle to the collective variables $\omega_\Sigma = \omega+\omega'$ and $\omega_\Delta = \omega-\omega'$. This choice is driven by the forthcoming Markovian approximation and the nature of the frequency entangling gate , which transforms the reference axes of the two-photon distribution from the individual frequencies $\omega$ and $\omega'$ to the collective frequencies variables $\omega+\omega'$ and $\omega-\omega'$. The third term in Eq.~\eqref{Appendix Equation differential equations of beixcitonic amplitude 2} can be expressed as 
\begin{equation}
\label{Appendix Equation emitted by the QD in the past that got reabsorbed}
    F(t) = -i\int_{0}^{t_1-t_0}d\tau \;C_{2X}(t_1-\tau)K(\tau),
\end{equation}
where $K(\tau)$ is the memory kernel 
\begin{equation}
    K(\tau) \equiv \int_{\mathbb R}d\omega_\Sigma\; e^{-i\omega_\Sigma\tau}\left( \frac{1}{2}\sum_{\mu',\mu\in\{\pm\}}\int_{\mathbb R} d\Delta\; |g^{\mu' \mu}(\omega_\Sigma,\omega_\Delta)|^2 \right) 
\end{equation}
connected to the spectral function $J(\omega_\Sigma)$~\cite{LivreJuanjoripoll2022quantum} through a Fourier transform 
\begin{equation}
\label{Appendix Equation spectral function expression}
    J(\omega_\Sigma) = \pi\sum_{\mu',\mu\in\{\pm\}}\int_{\mathbb R} d\omega_\Delta\; |g^{\mu' \mu}(\omega_\Sigma,\omega_\Delta)|^2.
\end{equation}
The spectral function can be interpreted as a function counting the coupling
function contributions in the frequency domain to the term $F(t_1)$ [defined in Eq.~\eqref{Appendix Equation emitted by the QD in the past that got reabsorbed}], in particular here with respect to the sum of photons frequencies. This stands to reason given that the resonance condition for this two-photon process is $\omega_\Sigma = \omega+\omega'=\omega_{2X}$. The Markovian and weak coupling approximations now consist in: (i) Assuming that the coupling strength is negligible with respect to the biexcitonic energy $\omega_{2X}$ so that the probability amplitude $C_{2X}(\tau)$ can be expressed as the product of a free-evolving term and a slowly time-varying function of time $C_{2X}(\tau) = e^{-i\omega_{2X}\tau} S_{2X}(\tau)$. (ii) Supposing that the spectral function Eq.~\eqref{Appendix Equation spectral function expression} -- and hence the coupling terms $g^{\mu'\mu}(\omega_\Sigma, \omega_\Delta)$ -- has a minimal dependence on the collective frequency $\omega_\Sigma$, so that the memory kernel $K(\tau)$ is sharply peaked at $\tau=0$. The term $F(t)$ can thus be approximated as 
\begin{equation}
    F(t) \approx -iC_{2X}(t)\int_{0}^{t_1-t_0}d\tau \; e^{i\omega_{2X}\tau}K(\tau),
\end{equation}
where the envelope was treated as a constant over the short time interval where the memory kernel is nonzero. From this point, $F(t_1)$ can be directly substituted back into the differential equation Eq.~\eqref{Appendix Equation differential equations of beixcitonic amplitude 2} or can be simplified by expanding the time $t_1$ in the integral to infinity. This approach is valid as long as the memory kernel closely approximates a Dirac $\delta$ function in time and aligns with scattering states, where $t_0$ and $t_1$ are extended to $+\infty$ and $-\infty$, respectively. In that case, using the Sokhotski–Plemelj theorem, $F(t_1)$ yields
\begin{equation}
    F(t_1) = -i\left(\frac{\Gamma}{2}-i\delta_{\text{Lamb}}\right)C_{2X}(t_1),
\end{equation}
with $\Gamma$ and $\delta_{\text{Lamb}}$ the decay rate and the frequency Lamb shift respectively 
\begin{align}
    \Gamma &=J(\omega_{2X}) \\
    \delta_{\text{Lamb}} &=\mathcal{PV}\int_{\mathbb R}d\omega_\Sigma\;\left(\frac{J(\omega_\Sigma)}{\omega_{2X}-\omega_\Sigma}\right),
\end{align}
where $\mathcal{PV}$ labels the Cauchy principal value. The Lamb shift is recast in the definition of the biexcitonic frequency $\omega_{2X}\rightarrow \omega_{2X}+\delta_{\text{Lamb}}$. In most cases, the experimental measurement of the emitter's frequency already accounts for the Lamb shift regardless \cite{LivreJuanjoripoll2022quantum}. The extension $t_1\rightarrow +\infty$ and $t_0\rightarrow -\infty$ can now be clarified by defining the scattering times as $(t_1-t_0)\gg 1/\Gamma$, that is to say times such that the emitter has fully decayed before the output state is formally established. Writing the second term from the two-photon probability amplitude Eq.~\eqref{Appendix Equation probability amplitude two-photon formally integrated}
\begin{equation}
\label{Appendix Equation biexcitonic amplitude Fourier transform}
    \int_{t_0}^{t_1}d\tau\;C_{2X}(\tau)e^{-i\omega_\Sigma(t_1-\tau)} \approx e^{-i\omega_\Sigma t_1}\widetilde{C}_{2X}(\omega_\Sigma),
\end{equation}
with $\widetilde{C}_{2X}(\omega_\Sigma)$ the Fourier transform of $C_{2X}(\tau)$ that can obtained from Eq.~\eqref{Appendix Equation differential equations of beixcitonic amplitude 2} using $\int_{\mathbb R}dt\;e^{it(\omega_\Sigma-\omega_\Sigma')}=2\pi\delta(\omega_\Sigma-\omega_\Sigma')$. The two-photon probability amplitude can then be expressed as 
\begin{equation}
\label{Appendix Equation general scattered output states}
\begin{split}
   C^{\sigma'\sigma} _{\mu'\mu}(\omega_\Sigma,\omega_\Delta;t_1) &= e^{-i\omega_\Sigma(t_1-t_0)}\bigg[C^{\theta'\theta} _{\nu'\nu}(\omega_\Sigma,\omega_\Delta;t_0)\delta_{\sigma'\theta'}\delta_{\mu'\nu'}\delta_{ \sigma \theta }\delta_{\mu \nu} \\
   &-\frac{\pi\left(g^{\mu'\mu}(\omega_\Sigma,\omega_\Delta)\right)^*}{\frac{\Gamma}{2}+i(\omega_{2X}-\omega_\Sigma)}\sum_{\nu',\nu\in\{\pm\} }\int_{\mathbb R}d\omega_\Delta'\;g^{\nu'\nu}(\omega_\Sigma, \omega_\Delta')C _{\mu'\mu}(\omega_\Sigma,\omega_\Delta;t_0) \bigg],
\end{split}
\end{equation}
where again, the polarization degrees of freedom indices are omitted whenever there is a two-photon coupling term as it implicitly dictates them due to the selection rules. The two-photon output state, given in Eq.~\eqref{Appendix Equation general scattered output states}, consists of two contributions: one term represents photons that did not interact with the (QD) and are only time-shifted, while the other term corresponds to photons that interacted with the QD, as indicated by the coupling terms. 
\section{Schmidt decomposition for continuous variables}
\label{Appendix Schmidt decomposition for continous variables} 
In this section, we provide additional details on the Schmidt decomposition discussed in the main text. For a comprehensive treatment refer to~\cite{CVSchmidtdecompositionParkerPhysRevA.61.032305,CVSchmidtdecompositionlucaslamata2005dealing,bogdanov_schmidt_2006}. The pure \st{renomarlized} renormalized output state 
\begin{equation}
\label{Appendix Equation two-photon scattered output state before Schmidt decomposition}
    \ket{\psi_{\mu'\mu}^{\sigma'\sigma}(t_1)} = \int_{\mathbb R^2}d\omega'd\omega\; C_{\mu'\mu}^{\sigma'\sigma }(\omega',\omega;t_1)\hat a_{\sigma'\mu'}^\dagger(\omega')\hat a_{\sigma \mu}^\dagger(\omega)\ket{\text{vac}}
\end{equation}
is expressed by means of a discrete expansion 
\begin{equation}
\label{Appendix Equation Schmidt decomposition discrete expansion of amplitude}
    C_{\mu'\mu}^{\sigma'\sigma }(\omega',\omega;t_1) = \sum_{n,m \in\mathbb N}C_{\mu'\mu}^{\sigma' \sigma}(n,m;t_1)O_n^{(2)}(\omega')O_m^{(1)}(\omega),
\end{equation}
where $\{O_m^{(1)}(\omega)\}_{m\in \mathbb N}$ and $\{O_n^{(2)}(\omega')\}_{n\in \mathbb N}$ are two set of orthogonal $L^2(\mathbb R)$ functions associated to the first and second photon respectively. These functions verify 
\begin{align}
    &\int_{\mathbb R}d\nu\;O_k^{(i)}(\nu)O_{k'}^{(i)}(\nu) = \delta_{kk'} \\ 
    &\sum_{k\in\mathbb N} O_k^{(i)}(\nu)O_k^{(i)}(\nu') = \delta(\nu-\nu'),
\end{align}
for $i=1,2$. Here for instance, we choose the Hermite-Gauss orthogonal functions given the Gaussianity of the two-photon distribution. The elements $C_{\mu'\mu}^{\sigma' \sigma}(n,m;t_1)$ of the discrete expansion Eq.~\eqref{Appendix Equation Schmidt decomposition discrete expansion of amplitude} are given by 
\begin{equation}
    C_{\mu'\mu}^{\sigma' \sigma}(n,m;t_1) = \int_{\mathbb R^2}d\omega'd\omega\;  C_{\mu'\mu}^{\sigma'\sigma }(\omega',\omega;t_1)O_n^{(2)*}(\omega')O_m^{(1)*}(\omega).
\end{equation}
By discretizing the continuous problem, we work with the discrete coefficients of the linear combination while preserving the continuous nature of the state through the $\omega$- and $\omega'$-dependence of the set of orthogonal functions $\{O_m^{(1)}(\omega)\}_{m\in \mathbb N}$ and $\{O_n^{(2)}(\omega')\}_{n\in \mathbb N}$. The next step consists in performing a numerical matrix singular-value decomposition to the  $C_{\mu'\mu}^{\sigma'\sigma}(t_1)$ matrix with elements $C_{\mu'\mu}^{\sigma' \sigma}(n,m;t_1)$. In order to perform the numerical singular value decomposition, we first truncate the discrete sums. Throughout the computations, we ensure that these truncations are sufficiently large to maintain an approximation close to the exact solution. This is assessed by the computation of a distance -- \textit{e.g.} the mean squared error -- to the actual JSA $C_{\mu'\mu}^{\sigma'\sigma}(\omega',\omega;t_1)$. The amplitude Eq.~\eqref{Appendix Equation Schmidt decomposition discrete expansion of amplitude} can be rewritten as 
\begin{equation}
   C_{\mu'\mu}^{\sigma'\sigma}(\omega',\omega;t_1) = \sum_{k=1}^{\text{min}(m_0,n_0)} \lambda_k^{\sigma'\mu'\sigma\mu}\Theta_k^{(2)}(\omega')\Theta_k^{(1)}(\omega),
\end{equation}
where the discrete expansion has been truncated for the singular-value decomposition to be applied \begin{equation}
     C_{\mu'\mu}^{\sigma'\sigma }(\omega',\omega;t_1) = \sum_{n=1}^{n_0}\sum_{m=1}^{m_0} C_{\mu'\mu}^{\sigma' \sigma}(n,m;t_1)O_n^{(2)}(\omega')O_m^{(1)}(\omega).
\end{equation}
 The $\Theta_k^{(1)}(\omega)$ and $\Theta_k^{(2)}(\omega')$ functions are linear combinations of the orthogonal functions $\{O_m^{(1)}(\omega)\}_{m\in  \llbracket 1,m_0 \rrbracket}$ and $\{O_n^{(2)}(\omega')\}_{n\in  \llbracket 1,n_0 \rrbracket}$ respectively. The $\lambda_k^{\sigma'\mu'\sigma\mu}$ Schmidt coefficients are real, non-negative, unique up to reordering and obey $\sum_{k=1}^{\text{min}(m_0,n_0)}\left(\lambda_k^{\sigma'\mu'\sigma\mu}\right)^2=1$ provided $C_{\mu'\mu}^{\sigma'\sigma}(\omega',\omega;t_1)$ is normalized to one.  The two-photon scattered output state $\ket{\psi_{\mu'\mu}^{\sigma'\sigma}(t_1)}$ in Eq.~\eqref{Appendix Equation two-photon scattered output state before Schmidt decomposition} can be expressed as a discrete sum 
\begin{equation}
    \ket{\psi_{\mu'\mu}^{\sigma'\sigma}(t_1)} = \sum_{k=1}^{\text{min}(m_0,n_0)} \lambda_k^{\sigma'\mu'\sigma\mu} \hat b^{(2)\dagger}_{\sigma'\mu'}(k)\hat b^{(1)\dagger}_{\sigma\mu}(k)\ket{\text{vac}},
\end{equation}
where $\hat b^{(1)\dagger}_{\sigma\mu}(k)$ and $\hat b^{(2)\dagger}_{\sigma'\mu'}(k)$ are bosonic creation operators which read
\begin{equation}
    \begin{split}
        &\hat b^{(1)\dagger}_{\sigma\mu}(k) = \int_{\mathbb R}d\omega\;\Theta_k^{(1)}(\omega)\hat a_{\sigma\mu}^\dagger(\omega) \\ 
        &\hat b^{(2)\dagger}_{\sigma'\mu'}(k) = \int_{\mathbb R}d\omega'\;\Theta_k^{(2)}(\omega')\hat a_{\sigma'\mu'}^\dagger(\omega').
    \end{split}
\end{equation}
These operators create photons in the Schmidt modes, each labeled by the index $k$. \\

\section{Optimization of the propagation mode} 
\label{Appendix Optimization of the propagation mode}
In this section, we offer numerical insights into optimizing the propagation mode $u(\omega)$, which corresponds to the magnitude of the solution to the waveguide wave equation 
\begin{equation}
\label{Appendix Equation wave equation}
    \boldsymbol \nabla \times \boldsymbol \nabla \times \boldsymbol{\mathcal{E}}_{\sigma\mu} \left(\boldsymbol{r},\beta\right) - \frac{\omega(\beta)^2}{c^2}\epsilon(\boldsymbol r,\mu\beta)\boldsymbol{\mathcal{E}}_{\sigma\mu} \left(\boldsymbol{r},\beta\right) = \boldsymbol 0,
\end{equation}
in such a way for the two-photon coupling term presented in Section II.B.4 of the main text
\begin{equation}
\begin{split}
\label{Appendix Equation two-photon coupling term after assuming same strength for each path}
    g^{\mu'\mu}(\omega',\omega) &= \frac{D}{\hbar^2}u(\omega')u(\omega)\times\bigg[\frac{1}{\omega-\omega_X}-\frac{1}{\omega'-(\omega_{X}-\delta_X)} 
    + \frac{1}{\omega'-\omega_X}-\frac{1}{\omega-(\omega_{X}-\delta_X)}\bigg]
\end{split}
\end{equation}
to be a non-separable Gaussian function of width $\beta$ and defined along the collective variable $\omega_{\Delta} = \omega-\omega'$ 
\begin{equation}
\label{Appendix Equation Gaussian two-photon coupling term}
    g^{\mu'\mu}(\omega_\Delta) = \sqrt{\frac{\gamma^{\mu'\mu}}{\pi}}\left(\frac{1}{2\pi\beta^2}e^{-\frac{\left(\omega_\Delta-(\omega_e-\omega_b)\right)^2}{\beta^2}}\right)^{1/4}.
\end{equation}
\begin{figure}
    \centering
       \centering
        \includegraphics[width=\linewidth]{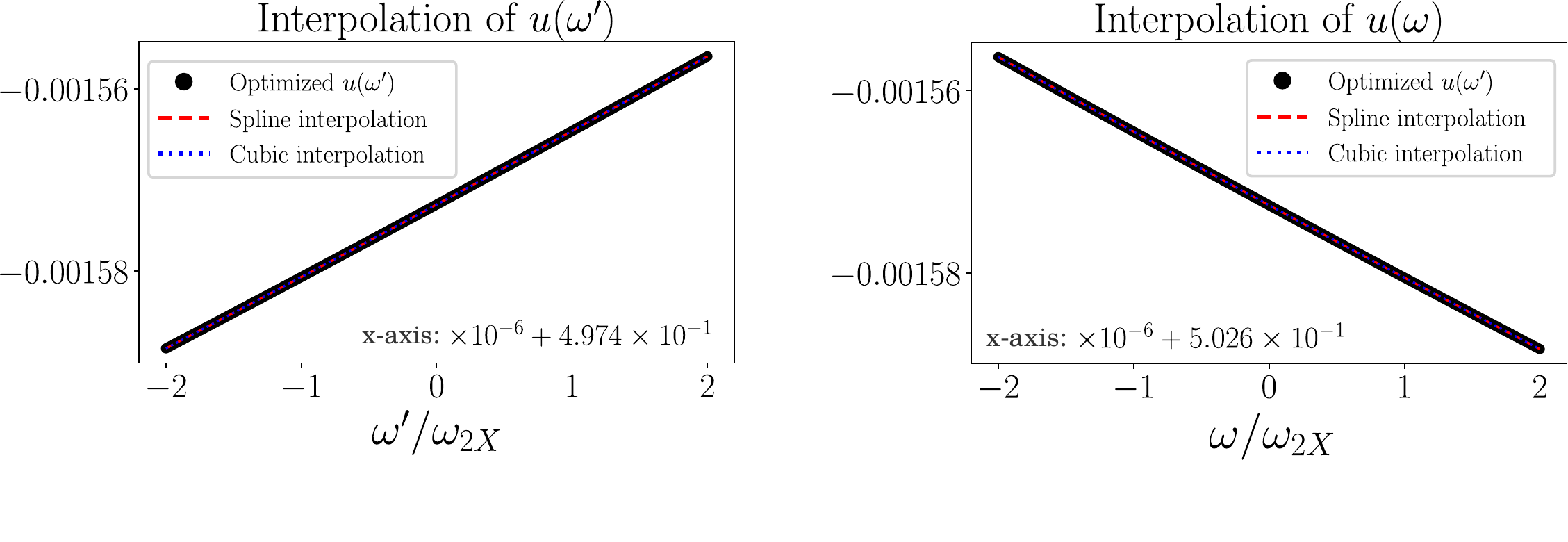}
        \caption{Spline and cubic interpolation of the propagation mode $u$ over the excitonic range $\omega$ and biexcitonic range $\omega'$ normalized to the biexcitonic transition frequency $\omega_{2X}$ for $\alpha=10^{-6}\omega_{2X}$, $\Gamma=10^{-5}\omega_{2X}$ and $\beta\approx10^{-4}\omega_{2X}$, $\omega_X=0.5025\omega_{2X}$, $\delta_X = 0.005\omega_{2X}$, $\omega_e =0.5026\omega_{2X}$, $\omega_b=0.4974\omega_{2X}$. }
        \label{Appendix figure interpolation of u for both frequencies ranges}
\end{figure}
\mohamed{Note that $\sqrt{\omega}$ has been attached to the definition of $u(\omega)$ in accordance to Sec.~\ref{Appendix Field quantization in dielectrics}}. Injecting the constants $D/\hbar^2$ in the definition of the products $u(\omega')u(\omega)$, this problem can be formulated as $u(\omega')u(\omega)h(\omega',\omega) = g(\omega',\omega)$ where $h(\omega',\omega)$ is the non-separable part of Eq.~\eqref{Appendix Equation two-photon coupling term after assuming same strength for each path} and $g(\omega',\omega)$ the target two-variable functions Eq.~\eqref{Appendix Equation Gaussian two-photon coupling term}. Since $g(\omega',\omega)$ is non-separable in the two variables $\omega$ and $\omega'$, \text{i.e.}. it cannot be written as a product $g(\omega',\omega) = g_1(\omega)g_2(\omega')$ with $g_1$ and $g_2$ two functions, the importance of $h(\omega',\omega)$ non-separability becomes apparent. In order to optimize the function $u$ over the ranges of definition of $\omega$ and $\omega'$, let us write 
\begin{equation}
    u(\omega)u(\omega') = \frac{g(\omega',\omega)}{h(\omega',\omega)}
\end{equation}
and express this equation in a matrix form $\boldsymbol{U}=\boldsymbol T$ with
\begin{align}
    &\boldsymbol U \equiv \begin{bmatrix} u(\omega_1)u(\omega'_1) & u(\omega_2)u(\omega'_1) & \dots & u(\omega_n)u(\omega'_1) \\ \vdots & \vdots & \vdots & \vdots \\ u(\omega_1)u(\omega_n') & u(\omega_2)u(\omega_n') & \dots & u(\omega_n)u(\omega_n')
    \end{bmatrix} \\ \\
    &\boldsymbol T \equiv \begin{bmatrix} \frac{g(\omega_1,\omega'_1)}{h(\omega_1,\omega'_2)} & \frac{g(\omega_2,\omega'_1)}{h(\omega_2,\omega'_2)} & \dots & \frac{g(\omega_n,\omega'_1)}{h(\omega_n,\omega'_1)} \\ \vdots & \vdots & \vdots & \vdots \\ \frac{g(\omega_1,\omega'_n)}{h(\omega_1,\omega'_n)} & \frac{g(\omega_2,\omega'_n)}{h(\omega_2,\omega'_n)} & \dots & \frac{g(\omega_n,\omega'_n)}{h(\omega_n,\omega'_n)}
    \end{bmatrix},
\end{align}
where we discretize the frequencies $\boldsymbol{\omega}=[\omega_1 \dots \omega_N]^T$ and $\boldsymbol{\omega'} = [\omega'_1 \dots \omega_N']^T$. The frequencies ranges are chosen to match the initial two-photon distribution used in Section III.B.1 of the main text. The optimization problem is thus equivalent to optimizing the entries of two column vectors $\boldsymbol u \equiv [u(\omega_1)\dots u(\omega_N)]^T$ and $\boldsymbol u' \equiv [u(\omega_1')\dots u(\omega_N')]^T$ by minimizing a distance $\|.\|$ between $\boldsymbol U$ and $\boldsymbol T$. The optimization loop starts with an initial guess that starkly determines the extent of the optimization's reach and must be carefully chosen. A guess motivated by the previous $\boldsymbol U=\boldsymbol T$ matrix formulation of the optimization can be chosen by noticing that $\boldsymbol U = \boldsymbol u^T \boldsymbol{u'}$, and by extension $\boldsymbol T$, should be a rank-1 matrix. Indeed, each row (respectively column) of $\boldsymbol U$ is a multiple of $\boldsymbol{u'}^T$ (respectively $\boldsymbol u$) so all rows (respectively columns) are linearly dependent. Let us now perform the singular value decomposition of $\boldsymbol T$ 
\begin{equation}
    \boldsymbol T = \boldsymbol V \boldsymbol S \boldsymbol W^T,
\end{equation}
where $\boldsymbol V$ and $\boldsymbol W^T$ orthogonal matrix contain the left and right singular vectors respectively and $\boldsymbol S$ is the diagonal matrix of singular values. Any column (respectively row) of $\boldsymbol T$ can be expressed as linear combination of the columns (respectively rows) of $\boldsymbol V$ and $\boldsymbol W^T$. The columns (respectively rows) of $\boldsymbol V$ (respectively $\boldsymbol W^T$) form an orthonormal basis of the column (respectively row) space of $\boldsymbol T$. The leading singular vectors are the first column and row of $\boldsymbol V$ and $\boldsymbol W^T$ respectively and because $\boldsymbol U = \boldsymbol T$ is supposed to be a rank-1 matrix, $\boldsymbol T \approx \boldsymbol S[1]\boldsymbol V[0,:]\boldsymbol W^T[0,:]$ therefore a good initial guess for $\boldsymbol u$ and $\boldsymbol{u'}$ is 
\begin{equation}
    \boldsymbol u = \boldsymbol W^T[0,:] \;\;\; \boldsymbol{u'} = \boldsymbol V[0,:].
\end{equation}
After retrieving the optimized vectors $\boldsymbol u$ and $\boldsymbol{u'}$, we interpolate the functional form of $u$ over the ranges of $\omega$ and $\omega'$ with a spline and cubic interpolation, see Fig.~\ref{Appendix figure interpolation of u for both frequencies ranges}. Remarkably, a form of anticorrelation emerges between the values of $u$ over the two frequency ranges, aligning with the anticorrelations in the target coupling term Eq.~\eqref{Appendix Equation Gaussian two-photon coupling term}.
In Fig.~\ref{Appendix figure comparing target and approximation optimization}, the target Gaussian non-separable coupling Eq.~\eqref{Appendix Equation Gaussian two-photon coupling term} and approximated solution to the optimization problem $u(\omega)u(\omega')h(\omega',\omega)$ are plotted. They demonstrate strong agreement, validating the effectiveness of the previous optimization method. It is important to note that the optimization critically depends on the frequency ranges, which are determined by the two-photon input distribution, the Gaussian coupling width $\beta$, the one-photon detunings and the binding energy $\delta_X$. The parameters that we consider here follow those used in the main text for the output two-photon distribution to be frequency-entangled with Gaussian distributions along both collective variables $\omega_\Sigma = \omega+\omega'$ and $\omega_{\Delta} = \omega-\omega'$. All the preceding optimization was carried out in Python. The present optimization scheme is mostly illustrative. Precisely tailoring the propagation mode $u(\omega)$ starkly depends on the different physical parameters at play, and precise waveguide engineering is paramount to experimentally manufacture the required propagation mode $u(\omega)$. 
\begin{figure}
    \centering
       \centering
        \includegraphics[width=\linewidth]{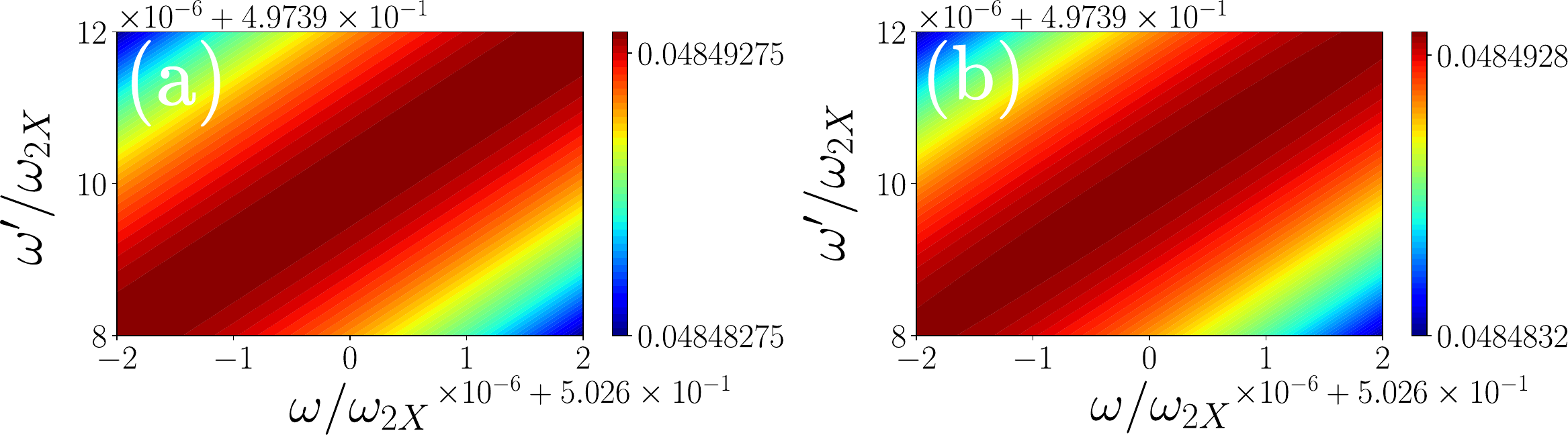}
        \caption{(a) Target Gaussian non-separable coupling Eq.~\eqref{Appendix Equation Gaussian two-photon coupling term}. (b) Approximated Gaussian non-separable coupling drawn from the optimization of $u$ over the two frequencies ranges, see Fig.~\ref{Appendix figure interpolation of u for both frequencies ranges}. }
        \label{Appendix figure comparing target and approximation optimization}
\end{figure}
\section{Frequency qudit states computation details}
\label{Appendix Grid states}
In this section, we provide the computation details leading to the output distribution for the cavity-filtered input $C_{++}^{LR}(\omega_\Sigma,\omega_\Delta;t_0)f_{\text{filter}}(\omega)f_{\text{filter}}(\omega')$. For a \st{non-interefered} non-interfered output, that is an output with auxiliary degrees of freedom $\{\mu,\mu'\}\neq \{+,+\}$, the scattered two-photon distribution reads 
\begin{equation}
     C_{\mu'\mu}^{\sigma'\sigma }(\omega_\Sigma,\omega_\Delta;t_1)=-e^{-i\omega_\Sigma(t_1-t_0)}\frac{\pi\left(g^{\mu'\mu}(\omega_\Sigma,\omega_\Delta)\right)^*}{\frac{\Gamma}{2}+i(\omega_{2X}-\omega_\Sigma)}\int_{\mathbb R}d\omega_\Delta'\;g^{++}(\omega_\Sigma, \omega_\Delta')C_{++}^{LR}(\omega_\Sigma,\omega_\Delta';t_0)f_{\text{filter}}(\omega)f_{\text{filter}}(\omega'),
\end{equation}
with 
\begin{equation}
    g^{\mu'\mu}(\omega_\Delta) = \sqrt{\frac{\gamma^{\mu'\mu}}{\pi}}\left(\frac{1}{2\pi\beta^2}e^{-\frac{\left(\omega_\Delta-(\omega_e-\omega_b)\right)^2}{\beta^2}}\right)^{1/4},
\end{equation}
and 
\begin{equation}
    C_{++}^{LR}(\omega_\Sigma,\omega_\Delta;t_0) = \frac{1}{\sqrt{2\pi\alpha ^2}}e^{-\frac{\left[\omega_\Sigma- (\omega_e+\omega_b)\right]^2}{4\alpha^2}}e^{-\frac{\left[\omega_\Delta-(\omega_e-\omega_b)\right]^2}{4\alpha^2}}.
\end{equation}
Let us focus now on the integral terms. Given that $f_{\text{filter}}(\omega) = \sum_{n\in \mathbb Z}T_{n}(\omega)$ where $T_n(\omega) = \exp(-(\omega-n\overline \omega)^2/(2\delta\omega^2))$, the integral can be expressed as a sum of the following integrals 
\begin{equation}
    \sqrt{\frac{\gamma^{\mu'\mu}}{\pi}}\left(\frac{1}{2\pi\beta^2}\right)^{1/4}\frac{1}{\sqrt{2\pi\alpha^2}}e^{\frac{\left[\omega_\Sigma-(\omega_e+\omega_b)\right]^2}{4\alpha^2}}\int_{\mathbb R}d\omega_{\Delta}'e^{-\left[\omega_{\Delta}'-(\omega_e-\omega_b)\right]^2\left(\frac{1}{4\alpha^2}+\frac{1}{4\beta^2} \right)}e^{-\frac{1}{4\delta\omega^2}\left(\left[\frac{\omega_\Sigma+\omega_{\Delta}'}{2}-n\overline \omega\right]^2+\left[\frac{\omega_\Sigma-\omega_{\Delta}'}{2}-m\overline \omega\right]^2\right)}.
\end{equation}
Rewriting the argument of the second exponential of the integral as
\begin{equation}
    -\frac{1}{2\delta\omega^2}\left(\left[\frac{\omega_\Sigma+\omega_{\Delta}'}{2}-n\overline \omega\right]^2+\left[\frac{\omega_\Sigma-\omega_{\Delta}'}{2}-m\overline \omega\right]^2\right)= -\frac{1}{2\delta\omega^2}\left(A_{\Sigma}\left[\omega_\Sigma-c_{\Sigma,nm}\right]^2+A_{\Delta}'\left[\omega_{\Delta}'-c_{\Delta,nm}' \right]^2+B_{nm}\right),
\end{equation}
with $A_\Sigma, A_\Delta',c_{\Sigma,nm}, c_{\Delta,nm}'$ and $B$ which can be determined \begin{align}
    A_{\Sigma} &= A_{\Delta}'=\frac{1}{2}\\
    c_{\Sigma,nm} &= \overline \omega(n+m) \\
    c_{\Delta',nm} &= \overline \omega (n-m) \\
    B_{nm}&= 0,
\end{align}
the term in $A_\Sigma\left[\omega_\Sigma-c_{\Sigma,nm} \right]^2$ can be taken out of the previous integral, which now reads 
\begin{equation}
    \int_{\mathbb R}d\omega_{\Delta}'e^{-\left(\frac{1}{4\alpha^2}+\frac{1}{4\beta^2} \right)\left[\omega_{\Delta}'-(\omega_e-\omega_b)\right]^2}e^{-\frac{1}{2\delta \omega^2}\left[\omega_{\Delta}'-\overline \omega(n-m)\right]^2}.
\end{equation}
The argument of the exponential can be expressed as 
\begin{equation}
    -\left(\frac{1}{4\alpha^2}+\frac{1}{4\beta^2} \right)\left[\omega_{\Delta}'-(\omega_e-\omega_b)\right]^2 -\frac{1}{4\delta \omega^2}\left[\omega_{\Delta}'-\overline \omega(n-m)\right]^2 = -\frac{\left[\omega_{\Delta}'-\mu_{\Delta,nm}'\right]^2}{\sigma_{\Delta}'^2}+D_{nm}
\end{equation}
with 
\begin{align}
    \sigma_{\Delta}'^2 &= \frac{4\alpha^2\beta^2\delta \omega^2}{\alpha^2\beta^2+\delta \omega^2(\alpha^2+\beta^2)} \\
    \mu_{\Delta,nm}' &= \frac{\delta \omega^2(\alpha^2+\beta^2)(\omega_e-\omega_b)+\alpha^2\beta^2\overline \omega(n-m)}{\alpha^2\beta^2+\left(\alpha^2+\beta^2\right)\delta \omega^2} \\
    D_{nm} &= -\frac{1}{4}\frac{\left(\alpha^2+\beta^2\right)\left((\omega_e-\omega_b)-\overline \omega(n-m) \right)^2}{\alpha^2\beta^2+(\alpha^2+\beta^2)\delta \omega^2}.
\end{align}
The integral can then readily be calculated as 
\begin{equation}
      \int_{\mathbb R}d\omega_{\Delta}'e^{-\left(\frac{1}{4\alpha^2}+\frac{1}{4\beta^2} \right)\left[\omega_{\Delta}'-(\omega_e-\omega_b)\right]^2}e^{-\frac{1}{4\delta \omega^2}\left[\omega_{\Delta}'-\overline \omega(n-m)\right]^2} =e^{D_{nm}}\sqrt{\pi \sigma_{\Delta}'^2}.
\end{equation}
One can therefore compute 
\begin{equation}
\begin{split}
    \int_{\mathbb R}d\omega_\Delta'\;g^{++}(\omega_\Sigma, \omega_\Delta')C_{++}^{LR}(\omega_\Sigma,\omega_\Delta';t_0)f_{\text{filter}}(\omega)f_{\text{filter}}(\omega') &= \sum_{n,m\in \mathbb Z}\sqrt{\frac{\gamma^{\mu'\mu}}{\pi}}\left(\frac{1}{2\pi\beta^2}\right)^{1/4}\sqrt{\frac{\sigma_{\Delta}'^2}{2\alpha^2}}e^{D_{nm}} \\
    &\times e^{-\frac{\left[\omega_\Sigma-(\omega_e+\omega_b)\right]^2}{4\alpha^2}}e^{-\frac{1}{4\delta \omega^2}\left[\omega_\Sigma-\overline \omega(n+m) \right]^2}.
\end{split}
\end{equation}
All in all, for an isotropic emission $\gamma^{\mu'\mu} = \Gamma/4$, the scattered output two-photon distribution is \begin{equation}
     C_{\mu'\mu}^{\sigma'\sigma}(\omega_\Sigma,\omega_\Delta;t_1)= -e^{-i\omega_\Sigma(t_1-t_0)}e^{-\frac{[\omega_\Delta-(\omega_e-\omega_b)]^2}{4\beta^2}} \frac{\Gamma}{\frac{\Gamma}{2}+i(\omega_{2X}-\omega_\Sigma)}e^{-\frac{[\omega_\Sigma-(\omega_e+\omega_b)]^2}{4\alpha^2}}\sqrt{\frac{\sigma_{\Delta}'^2}{64\pi\alpha^2\beta^2}}\sum_{n,m\in \mathbb Z}f_{nm}(\omega_\Sigma),
\end{equation}
with 
\begin{equation}
    f_{nm}(\omega_\Sigma) = e^{D_{nm}}e^{-\frac{1}{4\delta \omega^2}\left[\omega_\Sigma-\overline \omega(n+m) \right]^2},
\end{equation}
where 
\begin{equation}
    D_{nm} = -\frac{1}{4}\frac{\left(\alpha^2+\beta^2\right)\left((\omega_e-\omega_b)-\overline \omega(n-m) \right)^2}{\alpha^2\beta^2+(\alpha^2+\beta^2)\delta \omega^2}.
\end{equation}
Numerically, one cannot extend the sums to infinity and has to truncate the number of peaks. Nonetheless, if the width $\alpha$ is too small, the frequency-filtering function cannot properly scan the one-photon distribution. Thus, we artificially displace the frequency-filtering as $f_{\text{filter}}(\omega) \rightarrow f_{\text{filter}}(\omega-\omega_e)$ and $f_{\text{filter}}(\omega') \rightarrow f_{\text{filter}}(\omega'-\omega_b)$. This changes the different computed coefficients as follows 
\begin{align}
    A_\Sigma &= A_\Delta' = \frac{1}{2}, \\
    c_{\Sigma,nm} &= \overline{\omega}(n + m) + (\omega_e + \omega_b), \\ 
    c_{\Delta',nm} &= \overline{\omega}(n - m) + (\omega_e - \omega_b), \\ 
    B &= 0, \\
    \mu_{\Delta,nm}' &= \frac{(\omega_e-\omega_b)\left[\delta \omega^2(\alpha^2+\beta^2)+\alpha^2\beta^2\right]+\alpha^2\beta^2\overline \omega(n-m)}{\alpha^2\beta^2+\delta\omega^2(\alpha^2+\beta^2)} \\
    \sigma_\Delta '^2 &= \frac{4\alpha^2\beta^2\delta \omega^2}{\alpha^2\beta^2+\delta \omega^2(\alpha^2+\beta^2)} \\
    D_{nm} &= -\frac{1}{4}\frac{(\alpha^2+\beta^2)\overline \omega^2(n-m)^2}{\alpha^2\beta^2+(\alpha^2+\beta^2)\delta\omega^2},
\end{align}
effectively shifting off the center frequency in $D_{nm}$. The scattered output two-photon distribution can be expressed as \begin{equation}
      C_{\mu'\mu}^{\sigma'\sigma}(\omega_\Sigma,\omega_\Delta;t_1)=-e^{-i\omega_\Sigma(t_1-t_0)}e^{-\frac{[\omega_\Delta-(\omega_e-\omega_b)]^2}{4\beta^2}} \frac{\Gamma}{\frac{\Gamma}{2}+i(\omega_{2X}-\omega_\Sigma)}e^{-\frac{[\omega_\Sigma-(\omega_e+\omega_b)]^2}{4\alpha^2}}\sqrt{\frac{\sigma_{\Delta}'^2}{64\pi\alpha^2\beta^2}}\sum_{n,m\in \mathbb Z}f_{nm}(\omega_\Sigma)
\end{equation}
with 
\begin{equation}
    f_{nm}(\omega_\Sigma) = e^{D_{nm}}e^{-\frac{1}{4\delta \omega^2}\left[\omega_\Sigma-(\omega_e+\omega_b)-\overline \omega(n+m) \right]^2}.
\end{equation}

\section*{ACKNOWLEDGMENT}
\mohamed{Mohamed Meguebel acknowledges the support of the Program QuanTEdu-France n° ANR-22-CMAS-0001 France 2030}. Maxime Federico acknowledges funding from European Union’s Horizon Europe research and
innovation programme under the project Quantum Secure Network Partnership (QSNP, grant agreement No 101114043). Simone Felicetti acknowledges financial support from National Recovery and Resilience Plan (PNRR) Extended Partnership (MUR) project PE0000023-NQSTI, financed by the European Union-- Next Generation EU and from the foundation Compagnia di San Paolo, grant vEIcolo no. 121319. We acknowledge discussions with Hélène Ollivier, Eva Maria Gonzalez Ruiz, Hanna le Jeannic, Arne Keller and Pérola Milman for the completion of this manuscript.
\section*{Disclosures}
The authors declare no conflicts of interest.
\section*{Data availability}
Data underlying the results presented in this paper are not publicly available at this time but may be obtained
from the authors upon reasonable request.

\bibliography{biblio_supplement}